\documentclass[10pt,aps,prd,twocolumn,showpacs,superscriptaddress,nofootinbib,nobibnotes,longbibliography,floatfix]{revtex4-1}
\usepackage{aas_macros}
\usepackage{amsmath}
\usepackage{graphicx}
\usepackage{subcaption} 
\usepackage{dcolumn}
\usepackage{bm}

\usepackage{float}
\usepackage{hyperref}
\usepackage{orcidlink} 

\renewcommand\footnotemark{}

\begin{document}

\preprint{APS/123-QED}

\title{\boldmath{Constraints on conformal ultralight dark matter couplings from the European Pulsar Timing Array}}

\author{Clemente Smarra \orcidlink{0000-0002-0817-2830}}
 \email{csmarra@sissa.it}
\affiliation{ SISSA, Via Bonomea 265, 34136 Trieste, Italy and INFN Sezione di Trieste}
\affiliation{ IFPU - Institute for Fundamental Physics of the Universe, Via Beirut 2, 34014 Trieste, Italy}

\author{ Adrien Kuntz \orcidlink{0000-0002-4803-2998}}
 \email{akuntz@sissa.it}
\affiliation{ SISSA, Via Bonomea 265, 34136 Trieste, Italy and INFN Sezione di Trieste}
\affiliation{ IFPU - Institute for Fundamental Physics of the Universe, Via Beirut 2, 34014 Trieste, Italy}

\author{Enrico Barausse \orcidlink{0000-0001-6499-6263}}
\affiliation{ SISSA, Via Bonomea 265, 34136 Trieste, Italy and INFN Sezione di Trieste}
\affiliation{ IFPU - Institute for Fundamental Physics of the Universe, Via Beirut 2, 34014 Trieste, Italy}

\author{Boris Goncharov \orcidlink{0000-0003-3189-5807}}
\affiliation{ Gran Sasso Science Institute (GSSI), I-67100 L'Aquila, Italy}
\affiliation{ INFN, Laboratori Nazionali del Gran Sasso, I-67100 Assergi, Italy}
\affiliation{ Max Planck Institute for Gravitational Physics (Albert Einstein Institute), D-30167 Hannover, Germany}
\affiliation{Leibniz Universität Hannover, D-30167 Hannover, Germany}

\author{Diana L\'opez Nacir \orcidlink{0000-0003-4398-1147}}

\affiliation{ Universidad de Buenos Aires, Facultad de Ciencias Exactas y Naturales, Departamento de F\'isica. Buenos Aires, Argentina.}
\affiliation{ CONICET - Universidad de Buenos Aires, Instituto de Física de Buenos Aires (IFIBA). Buenos Aires, Argentina}

\author{Diego Blas}
\affiliation{Institut de Fisica d’Altes Energies (IFAE), The Barcelona Institute of Science and Technology, Campus UAB, 08193 Bellaterra (Barcelona), Spain}
\affiliation{ Instituci\'o Catalana de Recerca i Estudis Avan\c cats (ICREA), Passeig Llu\'is Companys 23, 08010 Barcelona, Spain}

\author{Lijing Shao \orcidlink{0000-0002-1334-8853}}

\affiliation{Kavli Institute for Astronomy and Astrophysics, Peking University, Beijing 100871, China}
\affiliation{National Astronomical Observatories, Chinese Academy of Sciences, Beijing 100012, China}

\author{J. Antoniadis \orcidlink{0000-0003-4453-3776}}

\affiliation{  FORTH Institute of Astrophysics, N. Plastira 100, 70013, Heraklion, Greece}
\affiliation{ Max-Planck-Institut f\"ur Radioastronomie, Auf dem H\"ugel 69, D-53121 Bonn, Germany}

\author{D. J. Champion \orcidlink{0000-0003-1361-7723}}
\affiliation{ Max-Planck-Institut f\"ur Radioastronomie, Auf dem H\"ugel 69, D-53121 Bonn, Germany}

\author{ I. Cognard \orcidlink{0000-0002-1775-9692}}
\affiliation{ Laboratoire de Physique et Chimie de l'Environnement et de l'Espace, Universit\'e d’Orl\'eans/CNRS, 45071 Orl\'eans Cedex 02, France}
\affiliation{Observatoire Radioastronomique de Nan\c{c}ay, Observatoire de Paris, Universit\'e PSL, Universit\'e d’Orléans, CNRS, 18330 Nan\c{c}ay, France}

\author{L. Guillemot \orcidlink{0000-0002-9049-8716}}
\affiliation{ Laboratoire de Physique et Chimie de l'Environnement et de l'Espace, Universit\'e d’Orl\'eans/CNRS, 45071 Orl\'eans Cedex 02, France}
\affiliation{Observatoire Radioastronomique de Nan\c{c}ay, Observatoire de Paris, Universit\'e PSL, Universit\'e d’Orléans, CNRS, 18330 Nan\c{c}ay, France}

\author{H. Hu \orcidlink{0000-0002-3407-8071}}
\affiliation{ Max-Planck-Institut f\"ur Radioastronomie, Auf dem H\"ugel 69, D-53121 Bonn, Germany}

\author{ M. Keith\orcidlink{0000-0001-5567-5492}}
\affiliation{  Jodrell Bank Centre for Astrophysics, Department of Physics and Astronomy, University of Manchester, Manchester M13 9PL}

\author{M. Kramer \orcidlink{0000-0002-4175-2271}}
\affiliation{ Max-Planck-Institut f\"ur Radioastronomie, Auf dem H\"ugel 69, D-53121 Bonn, Germany}
\affiliation{ Jodrell Bank Centre for Astrophysics, Department of Physics and Astronomy, University of Manchester, Manchester M13 9PL}

\author{ K. Liu \orcidlink{0000-0002-2953-7376}}
\affiliation{ Max-Planck-Institut f\"ur Radioastronomie, Auf dem H\"ugel 69, D-53121 Bonn, Germany}
\affiliation{ Shanghai Astronomical Observatory, Chinese Academy of Sciences, 80 Nandan Road, Shanghai 200030, China}

\author{ D. Perrodin \orcidlink{0000-0002-1806-2483}}
\affiliation{ Affiliation: INAF - Osservatorio Astronomico di Cagliari, via della Scienza 5, 09047 Selargius (CA), Italy}

\author{ S. A. Sanidas \orcidlink{0000-0002-5956-5546}}
\affiliation{  Jodrell Bank Centre for Astrophysics, Department of Physics and Astronomy, University of Manchester, Manchester M13 9PL}

\author{ G. Theureau \orcidlink{0000-0002-3649-276X}}
\affiliation{ Laboratoire de Physique et Chimie de l'Environnement et de l'Espace, Universit\'e d’Orl\'eans/CNRS, 45071 Orl\'eans Cedex 02, France}
\affiliation{ Observatoire Radioastronomique de Nan\c{c}ay, Observatoire de Paris, Universit\'e PSL, Universit\'e d’Orléans, CNRS, 18330 Nan\c{c}ay, France}
\affiliation{ Laboratoire Univers et Th\'eories LUTh, Observatoire de Paris, Universit\'e PSL, CNRS, Universit\'e de Paris, 92190 Meudon, France}


\begin{abstract}
Millisecond pulsars are extremely precise celestial clocks: as they rotate, the beamed radio 
waves emitted along the axis of their magnetic field can be detected with radio telescopes, which allows for tracking subtle changes in the pulsars' rotation periods. 
A possible effect on the period of a pulsar is given by a potential coupling to dark matter, in cases where it is modeled with an ``ultralight'' scalar field. In this paper, we consider a universal conformal coupling of the dark matter scalar to gravity, which in turn mediates an effective coupling between pulsars and dark matter.
If the dark matter scalar field is changing in time, as expected in the Milky Way, this effective coupling produces a periodic modulation of the pulsar rotational frequency. By studying the time series of observed radio pulses collected by the European Pulsar Timing Array experiment, we present constraints on the coupling of dark matter, improving on existing bounds. These bounds can also be regarded as constraints on the parameters of scalar-tensor theories of the Fierz-Jordan-Brans-Dicke and Damour-Esposito-Far\`{e}se types in the presence of a (light) mass potential term. 
\end{abstract}

\maketitle


\section{Introduction}\label{sec:introduction}
\noindent
Elucidating the nature of Dark Matter (DM) remains as one of the most pressing questions of modern physics. 
The widely accepted Cold Dark Matter (CDM) paradigm successfully explains numerous aspects of the Universe's large-scale structure but encounters difficulties in predicting some observations on scales smaller than approximately a kiloparsec ($\sim$ kpc). Notably, observations indicate a flat density profile in the inner regions of galaxies, contradicting the pure CDM prediction of a steep power-law-like behavior (\textit{cusp-core problem})~\cite{Flores_1994, Moore_1994, Karukes_2015}. Additionally, known challenges arise from the mismatch between the observed and expected number of satellites of the Milky Way (MW) (\textit{missing satellite problem})~\cite{Klypin_1999, Moore_1999}, and simulations based on $\Lambda\text{CDM}$ theory suggest that the most massive subhaloes of the MW would be too dense not to host bright satellites (\textit{too-big-to-fail problem})~\cite{Boylan-Kolchin_2011}. Although these issues may be mitigated by considering baryonic feedback mechanisms, such as supernova feedback~\cite{Navarro_1996, Governato_2012, Brooks_2013, Chan_2015, Onorbe_2015, Read_2016},  another possibility is to assume that DM is an ultralight scalar field (with mass $m \sim 10^{-22}~\text{eV}$) with negligible self-interactions \cite{Hu:2000ke,Hui_2017}. In this scenario, the de Broglie wavelength of the scalar field in galaxies can reach $\sim \text{kpc}$, suppressing  power on smaller scales, while retaining all the successes of CDM at large scales. Furthermore, the presence of ultralight scalars is also motivated from a theoretical standpoint by string theory~\cite{Green_1987, Svrcek_2006, Arvanitaki_2010}. In this context, the mass range can be much broader, which motivates considering very light bosons in a large span of ultra-light masses (including, but not restricted to, $m \sim 10^{-22}\,$eV) as natural candidates for theories beyond the Standard Model.

Numerous studies have been conducted to investigate the existence of ultra-light DM (ULDM). Among them, the integrated Sachs-Wolfe effect on the Cosmic Microwave Background (CMB) anisotropies excludes masses $m \lesssim 10^{-24}~\text{eV}$~\cite{Hlozek_2015}. Lyman-$\alpha$ observations strongly suggests a lower limit of $m \gtrsim 10^{-21}~\text{eV}$ for ultra-light candidates accounting for more than $\sim 30\%$ of the DM~\cite{Irsic_2017,Armengaud_2017,Kobayashi_2017,Nori_2018, Rogers_2021}. However, the susceptibility of non-CMB constraints to uncertainties in the modeling of small-scale structure properties~\cite{Schive_2014, Zhang_2019} emphasizes the importance of complementary and independent investigations.
In this context, the rotation curves of  well-resolved nearby galaxies also disfavor masses $m\lesssim 10^{-21}\,$eV \cite{Bar:2018acw}. In addition, measurements of stellar orbit kinematics in ultra-faint dwarf (UFD) galaxies may potentially constrain the scalar field mass to be $m \gtrsim 10^{-19}~\text{eV}$, although this remains a topic of ongoing debate~\cite{Hayashi_2021, Dalal_2022}. Dwarf galaxies can also be used to set robust bounds  $m \gtrsim 10^{-22}~\text{eV}$~\cite{Marsh:2018zyw,Hui_2017}. Given these observations, the current lore is to consider that ULDM of mass below $m\sim 10^{-22}\,$eV can not constitute 100\% of the dark matter, but masses below these bound are certainly possible if they constitute a fraction of the dark matter, see Ref.~\cite{Ferreira:2020fam}. This possibility is natural in the case of the axiverse, where several ULDM candidates coexist at low masses \cite{Arvanitaki:2009fg} and also for ultra-low mass particles that may be cosmologically produced to significant values, see e.g. Ref.~\cite{Hamaide:2022rwi}. 

A completely independent method to probe these small masses was suggested in Ref.~\cite{Khmelnitsky_2014},  where it was pointed out that the oscillating \emph{gravitational potential} induced by the presence of ULDM influences the light travel time of radio signals emitted by pulsars.
Pulsar Timing Array (PTA) experiments~\cite{FosterBacker1990,ManchesterHobbs2013,McLaughlin2013,KramerChampion2013,InPTA, MeerTime} rely on the exquisite predictability of the millisecond pulsar (MSP) spin periods behavior. Each time a MSP magnetic field axis points towards Earth, radio waves are observed by radio telescopes. After measuring and modelling consecutive pulses in decade-long observational campaigns, PTAs search for signals of physical effect that affect all of the observed pulsars, including the ULDM signal.
Based on this principle, previous PTA searches have established 95\% upper limits on the local energy density of ULDM, reaching $\rho\lesssim 0.15$ GeV/cm$^3$ in the mass range  $10^{-24.0}~\text{eV} \lesssim m \lesssim 10^{-23.7}~\text{eV}$~\cite{Smarra_2023, Porayko_2018}.

The ideas of Ref.~\cite{Khmelnitsky_2014}  rely on the universal gravitational coupling of DM to ordinary matter. However, ULDM
may also be coupled to the Standard Model fields~\cite{Kaplan_2022, Afzal_2023}. Indeed, a natural possibility that respects the weak equivalence principle is that ULDM may be
universally (conformally) coupled to gravity, or (equivalently, in the Einstein frame) to the Standard Model. This universal  coupling, together with the oscillations of the scalar field in the MW, would give rise to periodic orbital perturbations in binary systems, which allows to place constraints on the model~\cite{Blas_2017,Blas_2020,Kus:2024vpa}.
In this context, ULDM may be regarded as a scalar-tensor theory of the Fierz-Jordan-Brans-Dicke~\cite{Fierz:1956zz,Jordan:1959eg,Brans_1961,Dicke62} or Damour-Esposito-Far\`{e}se type~\cite{Damour_1992,Damour_1993}, in the presence of a (light) mass potential term~\cite{Alsing:2011er}. As a result of the strong gravitational fields active inside neutron stars, the conformal coupling to gravity gives rise to an effective (gravity-mediated) interaction between neutron stars (and thus pulsars) and the scalar ULDM field~\cite{Nordtvedt68,Eardley1975ApJ,1989ApJ...346..366W,1977ApJ...214..826W,Damour_1992,Will:1993ns}. This effect, known as Nordtvedt effect~\cite{Nordtvedt68}, violates the strong equivalence principle, and it has long been constrained with binary pulsar data~\cite{Damour:1991rd,weisberg2004relativistic,Kramer:2006nb,Ransom_2014, Archibald_2018, Voisin_2020, Kramer_2021,Zhao:2022vig}.

In a recent companion paper~\cite{Kuntz_2024}, following former studies ~\cite{Damour_1994,1989ApJ...346..366W,Will:1993ns}, two of us computed the effect of this effective interaction on the rotational period of an {\it isolated} pulsar. More specifically, Ref.~\cite{Kuntz_2024} found that the effective coupling between ULDM and pulsars produces a change in the moment of inertia (and therefore on the rotational period) of the pulsar. This change is proportional to the rate of change (in time) of the scalar field, which -- as mentioned above -- is expected to oscillate in the MW.
Deriving precise constraints from  current observations on the conformal ULDM coupling based on this effect is the main objective of this work. 

This work is structured as follows. In Section~\ref{sec:nonmin}, we define the Lagrangian of our theory, and we briefly review some general features of ULDM relevant to our analysis. 
Section~\ref{sec:methods} will be devoted to a detailed description of the dataset and the procedure used to carry out the analysis. Finally, in Section~\ref{sec:spin}, we will show how a non-minimally coupled ULDM candidate affects the spin frequency of MSPs.
We constrain the coupling strength, resulting in bounds which are several orders of magnitude better than what obtained from Cassini tests of General Relativity \cite{Bertotti_2003, Blas_2017} or from the pulsar in a triple stellar system \cite{Ransom_2014, Archibald_2018, Voisin_2020} in the mass region which PTAs are sensitive to. 
Our conclusions are presented in Section \ref{sec:conclusions}. The appendices are devoted to technical details and plots.

\section{Conformally coupled ultralight dark matter} \label{sec:nonmin}

Light scalar and pseudoscalar fields emerge naturally from string theory and from theories with pseudo-Goldstone bosons (as the axions introduced to tackle the strong CP problem) \cite{Damour:1994zq,Arvanitaki:2009fg,Kim:1986ax}. These fields are, in principle, coupled to Standard Model fields. Hence it is natural to consider
\emph{non-minimally coupled} scenarios when exploring their phenomenology. In this work, we consider a scalar field $\phi$ with mass $m$.
The action for this field  in the Einstein frame is given by \footnote{Note that our scalar field $\phi$ is not canonically normalized, but appears in the action multiplied by $M_{\text{P}}$. This convention makes comparisons with gravitational phenomena more straightforward.}:
\begin{align}
    S = M_{\text{P}}^2 \int \mathrm{d}^4x \sqrt{-g} \, &\bigg[ \frac{R}{2} - g^{\mu \nu} \partial_\mu \phi \partial_\nu \phi + m^2\phi^2 \bigg] \nonumber\\
    +& S_m[ \psi_m, \tilde g_{\mu \nu}] \; ,
    \label{eq:action}
\end{align}
where $M_{\text{P}}$ is the Planck mass  and the matter action $S_m$ includes a universal 
{\it conformal}
coupling of the scalar field to the matter content $\psi_m$ through the (Jordan) effective metric $\tilde g_{\mu\nu} = \mathcal{A}^2(\phi) g_{\mu \nu}$, normalized such that $\mathcal{A}^2(0)=1$. 
Note that when re-expressed in the Jordan frame, this direct coupling to matter disappears and is replaced by a conformal coupling to gravity (\textit{i.e.}, to the Ricci scalar). In other words, this model satisfies the weak equivalence principle. In particular, the free-fall motion of non-gravitating objects is universal (i.e., independent of the body composition)~\cite{Will_1993}. However, in strongly gravitating objects (such as neutron stars), there appears an effective (tensor-mediated) coupling between matter and the scalar field.

As a first model, we consider the Fierz-Jordan-Brans-Dicke (FJBD) theory~\cite{Fierz:1956zz,Jordan:1959eg,Brans_1961,Dicke62}, in which 
the conformal coupling is linear:
\begin{equation}
    \mathcal{A}(\phi) = e^{\alpha \phi} \sim 1+\alpha \, \phi.
    \label{eq:fjbd}
\end{equation}
The constant \textit{scalar coupling} $\alpha$ is constrained by several observations, notably by the Cassini mission to the level of $\alpha^2\lesssim 10^{-5}$~\cite{Bertotti_2003} and by the triple  system PSR J0337+1715 to the level of $\alpha^2\lesssim 4\times10^{-6}$~\cite{Voisin_2020}. A simple way to evade this constraint is to consider masses generating a Yukawa suppression at scales of order of the typical distances probed by these systems~\cite{Alsing:2011er}. Since our focus is on the PTA, and hence to DM masses with Compton wavelength  larger than $\sim 10^3$ AU, this constraint applies to the models we explore\footnote{The center of mass of the inner binary in the triple system PSR J0337+1715 completes a rotation around the center of mass of the entire system in about 327 days \cite{Voisin_2020}, which corresponds to a distance of $\sim 0.9$ AU.}.

As a second model,  we also consider the Damour-Esposito-Far\`{e}se (DEF) gravity theory~\cite{Damour_1992,Damour_1993}, where the conformal coupling is quadratic:
\begin{equation}
    \mathcal{A}(\phi) = e^{\beta\phi^2/2}.
    \label{eq:def}
\end{equation}
Notice that we have chosen to set the linear coupling of the field in $\mathcal{A(\phi)}$ to zero, so that FJBD cannot be recovered as a special case of DEF, rather the two theories are part of a more general theory where both linear and quadratic couplings are nonzero.
The absence of significant 
deviations from the General Relativity (GR) predictions in binary pulsar data requires $\beta \gtrsim -4.3$ 
(depending on the exact equation of state (EoS) for the neutron star model)
in order to avoid non-perturbative spontaneous scalarization phenomena~\cite{Damour_1993, Shao:2017gwu}, while the value $\phi_0$ of the scalar field on cosmological scales is constrained by the Cassini (pulsar in a triple stellar system) bound to the level $(\beta \phi_0)^2 \lesssim 10^{-5}$ and by the pulsar in a triple stellar system to $(\beta \phi_0)^2 \lesssim 4 \times 10^{-6}$.

The energy momentum tensor of the scalar field sourcing the metric $g_{\mu\nu}$ on cosmological scales follows, in the Einstein frame, from \eqref{eq:action}:
\begin{equation}
    T_{\mu\nu} = M_\text{P}^2\left[2\partial_\mu\phi\partial_\nu\phi -g_{\mu\nu}\left(\left(\partial\phi\right)^2-m^2\phi^2\right)\right].
\end{equation}
In the Jordan frame $\tilde T_{\mu\nu} = \mathcal{A}^{-2}(\phi)T_{\mu\nu}$, it reduces to
\begin{equation}
    \tilde T_{\mu\nu} = \mathcal{A}^{-2}(\phi)T_{\mu\nu}\simeq\left(1-2\,\alpha\left(\phi\right)\right)  T_{\mu\nu},
    \label{eq:relT}
\end{equation}
where the effective scalar coupling is\footnote{In order to avoid confusion, we stress that $\alpha(\phi)$ is generally different from $\alpha$, but reduces to it in  FJBD theory. Binary pulsars have constrained $|\alpha(\phi)|$ to be $\lesssim 10^{-2}$ for neutron stars~\cite{Zhao:2022vig}.} $\alpha(\phi) = \mathrm{d} \log \mathcal{A} / \mathrm{d} \phi$ and we work in the limit $\alpha(\phi) \ll 1$.
The mass of the scalar field can be as small as  $m\sim 10^{-22}~\text{eV}$, and still be a viable very light candidate for CDM. We will refer to models of masses not far from this limit as ULDM. In these models, the distance between particles is much smaller than the corresponding de Broglie wavelength, which implies that they can be treated as a classical superposition of free waves with dynamical properties generated by galactic dynamics. For the MW, this superposition has a very small dispersion velocity ($\sigma_\phi \sim 10^{-3}$), and therefore the ULDM field can be described as a standing wave\footnote{Close to the galactic center, the distribution may condense into a different configuration known as ``soliton" or ``bose star" \cite{Schive_2014}. We will not deal with this situation since the pulsars used by PTA are far from the galactic center.}:
\begin{align}
    &\phi(\boldsymbol{x}, t) = A(\boldsymbol{x})\cos (mt + \theta(\boldsymbol{x})),
    \label{eq:scal}
\end{align}
with $A(\boldsymbol{x})$ determined by\footnote{Notice that, as compared to field theory conventions, there are factors of 2 of difference, arising from the non-canonical normalization of the scalar field in Eq.~\eqref{eq:action}.}
\begin{align}
    &\tilde\rho_\phi \approx \rho_\phi = m^2M_\text{P}^2A(\boldsymbol{x})^2 = \rho\,\hat\phi (\boldsymbol{x})^2,
\end{align}
where we have used Eq.~\eqref{eq:relT} to relate the stress-energy momentum tensor in the Einstein frame to the one in the Jordan frame, and we have neglected the terms $\sim (\partial_i \phi)^2$ which are suppressed by $v_\phi ^2$.
Here, $\rho$ is the average density of the scalar field and $\hat\phi(\boldsymbol{x})$ is a \textit{stochastic} parameter extracted from the Rayleigh distribution ($P(\hat\phi^2) = e^{-\hat\phi^2}$)~\cite{Castillo_2022}. This parametrization 
takes into account the fact the ULDM configurations are built by the superposition of several waves of random phases that interfere. 
For the average density, we take $\rho = \rho_{\text{DM}} = 0.4~\text{GeV/cm}^3$ as a benchmark value for the \textit{average} DM density expected at the Sun position in the Milky Way~\cite{Nesti:2013uwa}. 
As commented in the introduction, masses $m\lesssim 10^{-22}~\text{eV}$ are strongly disfavored to constitute all the dark matter in our Galaxy. As a result, for these masses one needs to focus on scenarios where ULDM is a fraction $f_\text{DM}\equiv \rho/\rho_{\rm DM}<1$. The factor $\hat \phi$ appears from the interference caused by the wavelike nature of the scalar field. It reproduces the expected random local value from the superposition of the waves, and makes the scalar field density $\tilde\rho_\phi$ approach $\rho$ when averaging 
over timescales longer than the ULDM \textit{coherence time}
\begin{equation}
	\tau_c \sim  \frac{2}{mv^2} = 2\times 10^5~\text{yr} \left(\frac{10^{-22}~\text{eV}}{m}\right),
	\label{eq:coh_time}
\end{equation}
or, equivalently, on a length scale larger than the ULDM \textit{coherence length}
\begin{equation}
	l_c \sim \frac{1}{mv} \sim v\cdot\tau_c \sim 0.4~\text{kpc} \left(\frac{10^{-22}~\text{eV}}{m}\right).
	\label{eq:coh_len}
\end{equation}
We have conveniently normalized the mass to values relevant to PTA observations. \\
In the Jordan frame, an oscillating scalar field,  such as the one presented in Eq.~\eqref{eq:scal}, induces a temporal variation of Newton's constant. Its measured value is~\cite{Damour_1992}:
\begin{equation} \label{eq:GJordan}
    G = \big( 8 \pi M_\text{P}^2 \big)^{-1} \mathcal{A}^2(\phi) \big( 1 + \alpha^2(\phi) \big),    
\end{equation}
where $\alpha(\phi) = \mathcal{A}'(\phi) / \mathcal{A}(\phi)$. In order to use this  property of scalar-tensor theories,
we follow Ref.~\cite{Kuntz_2024} and neglect
the scalar field mass
 on the typical length scale of a pulsar, a valid approximation given our mass range for ULDM. 
 In turn, a variation of the local gravitational constant modifies the gravitational mass and the radius of the neutron star~\cite{Damour_1992}.
 This dependence is encoded in the \textit{sensitivities}, which represent the rate of change of these parameters with respect to changes in the scalar field~\cite{Will_1993}.

 To explore this effect, one can recall that the conservation of angular momentum $J$ relates the changes in the moment of inertia $I$ of the neutron star (depending on the local value of $G$) to the observed pulse frequency $\Omega_\mathrm{obs}$ through the relation $J = I \Omega_\mathrm{obs}$. 
Particular attention has to be paid to the frame used for the definition of the angular momentum $J$, as the latter is only conserved in the Einstein-frame (see Ref.~\cite{Kuntz_2024} for more details).
 We use the code presented in Ref.~\cite{Kuntz_2024} to compute the angular momentum sensitivity, defined by:
\begin{equation} \label{eq:angularmomsensitivity}
    s_I = - \frac{1}{2 \alpha(\phi)} \frac{\mathrm{d} \ln I}{\mathrm{d} \phi} \bigg |_{N,J} = \frac{1}{2 \alpha(\phi)} \frac{\mathrm{d} \ln \Omega_\mathrm{obs}}{\mathrm{d} \phi} \bigg |_{N,J},
\end{equation}
where the pulsar's baryon number $N$ and Einstein-frame angular momentum   $J$  are kept constant. With this sensitivity at hand, a change in the scalar field value can be directly related to a change in the frequency of the pulsar, and consequently to a change in the pulsar's pulse time of arrival (TOA). We will use this fact in Section~\ref{sec:spin} to constrain ULDM models.

\section{Dataset and methodology}\label{sec:methods}

In this work, we analyze the second data release (DR2)~\cite{EPTA_I_2023} of the European Pulsar Timing Array collaboration (EPTA)~\cite{KramerChampion2013,DesvignesCaballero2016} . In particular, we use the EPTA-DR2Full dataset\footnote{The dataset can be found at \url{https://gitlab.in2p3.fr/epta/epta-dr2}.}, consisting of a 24.7 years collection of TOAs of radio pulses of 25 millisecond pulsars, surveyed with an approximately biweekly cadence\footnote{The cadence is non-uniform. EPTA combines observations from several telescopes, so sometimes the EPTA-wide cadence can be much shorter.} and observed by five telescopes located in France,
Germany, Italy, the Netherlands, and the United Kingdom.
The relation between the time of emission of a radio pulse and its TOA at the Solar System Barycentre (SSB) is encoded in a pulsar-specific \textit{timing model}~\cite{EdwardsHobbs2006}, which takes into account the pulsar spin down, its position and proper motion, the motion around a binary companion, etc..
Any deviations between the predicted TOAs and the actual measurements, referred to as \textit{timing residuals}, may include contributions from various sources of noise, including stochastic dispersion measure fluctuations and irregularities in pulsar rotation~\cite{GoncharovReardon2021,EPTA_II_2023}. However, these residuals might also be indicative of processes of astrophysical significance, c.f. the recent evidence of a stochastic gravitational wave background (SGWB) in the data of various PTA experiments~\cite{NANOGrav_2023, EPTA_2023, PPTA_2023, CPTA}.

Because the EPTA-DR2Full dataset does not yield a strong evidence in favor of  the hallmark Hellings-Downs (HD)~\cite{HellingsDowns1983} inter-pulsar correlation (in contrast to the 10.3 yr dataset~\cite{EPTA_I_2023, EPTA_II_2023, EPTA_III_2023, EPTA_V_2023}), we only account for possible contributions from the SGWB via a PTA-wide spatially-uncorrelated but temporally-correlated noise term, characterized by an amplitude $A_\text{GWB}$ and a spectral index $\gamma_\text{GWB}$ (see Table~\ref{tab:priors} for more details).
Such a signal appears as a precursor to the SGWB~\cite{nanograv12.5gwb,GoncharovShannon2021,epta2021gwb}, because of the stronger autocorrelation of the Hellings-Downs process.

We utilize Bayesian inference to detect the ULDM-induced \textit{deterministic}\footnote{The signal induced by ULDM is deterministic, as opposed to the stochastic nature of \textit{e.g.} the common red noise process describing the SGWB.} signal, while simultaneously marginalizing over timing model parameters~\cite{EPTA_I_2023} and accounting for all known sources of noise in the data~\cite{EPTA_II_2023}. 
Given the model parameters $\theta$, the likelihood function for the timing residuals, denoted as $\mathcal{L}(\delta t|\theta)$,  is represented as~\cite{vanHaasterenLevin2009,Lentati_2014, Taylor_2017,enterprise,enterprise_ext}: 
\begin{equation}
    \ln \mathcal{L}(\delta t | \theta) \propto  -\frac{1}{2} (\delta t - \mu)^\text{T} C^{-1}(\delta t-\mu).
\end{equation}
In this time-domain Gaussian likelihood, $\delta t$ is a vector with dimension corresponding to the number of observations. The deterministic ULDM contribution, which we derive below in Eqs.~\eqref{eq:dt} and~\eqref{eq:dtbeta}, is taken into account in $\mu$, which includes contributions from both the timing model and noise processes, as analyzed in Ref.~\cite{EPTA_II_2023}.
Temporally-uncorrelated ``white'' noise and other sources of uncertainty in TOA measurements are factored in the diagonal components of the covariance matrix $C$. Off-diagonal elements of the matrix $C$ could, in principle, contain contributions from temporally correlated ``red'' noise; yet they are more commonly incorporated into $\mu$ for computational efficiency, following the approach described in Refs.~\cite{Lentati_2014,Taylor_2017}.
The priors $\pi(\theta)$ of the parameters used for the search are described in Table~\ref{tab:priors}. Parameter estimations are carried out by evaluating posterior distributions, denoted as $\mathcal{P}(\theta|\delta t) \propto \mathcal{L}(\delta t|\theta) \pi(\theta)$, produced with the Parallel-Tempering-Markov-Chain Monte-Carlo sampler~\cite{justin_ellis_2017_1037579} implemented in \textsc{enterprise}~\cite{enterprise} and \textsc{enterprise\_extensions}~\cite{enterprise_ext}, using the
PTArcade~\cite{andrea_mitridate_2023, Mitridate:2023oar} wrapper and adapting it to the EPTA dataset.

In this work, we consider the effect of the scalar field on the TOAs, and we constrain the conformal coupling by looking at its effect on the timing residuals. Since the duration $T_\text{obs} = 24.7~\text{yr}$ of the EPTA-DR2Full dataset is much shorter than the coherence time in Eq.~\eqref{eq:coh_time} for the ULDM considered here, different coherence patches with characteristic dimension $l_c$ will have different $\rho_\phi$. However, notice that if $m$ is sufficiently small so that $l_c > R$, where $R$ is the characteristic radius probed by Galactic rotation curves, we are really observing one single patch of ULDM. We refer to this case as \textit{correlated} scenario.

Based on these premises, we thus distinguish three different regimes~\cite{Smarra_2023}, according to the interplay between the mass of the ULDM candidate and the typical interpulsar separation. In the \textit{uncorrelated} regime, each of the pulsars timed by the EPTA resides in a different coherent patch. Therefore, each pulsar is characterized by its own $\hat\phi(\boldsymbol{x})$ parameter. As the average interpulsar distance is $\mathcal{O}(\text{kpc})$, the \textit{uncorrelated} approximation holds for ULDM masses $m \gtrsim 5 \times 10^{-23}~\text{eV}$. 
For masses $m \lesssim 2 \times 10^{-24}~\text{eV}$ (\textit{correlated} scenario), the  coherence length described by Eq.~\eqref{eq:coh_len} encompasses the inner Galacto-centric $\sim \,20\,\text{kpc}$, which is the benchmark area examined by the most accurate measurements of MW rotation curves~\cite{Nesti:2013uwa}, from which the local DM abundance is inferred. Therefore, as the kinematic tests of the DM halo explore the same coherence patch that hosts all the EPTA pulsars, we can safely absorb the \textit{common} parameter $\hat\phi(\boldsymbol{x})$ into the measured value of the local ULDM abundance. Equation~\eqref{eq:scal} then reads:
\begin{equation}
    \phi(\boldsymbol{x}, t) = \frac{\sqrt{\rho}}{m M_\text{P}} \cos (mt + \theta(\boldsymbol{x})),
\end{equation}
with $\rho$ representing the value of the scalar field density $\rho_\phi$ in our Galaxy. 
Finally, for masses $2 \times 10^{-24}\text{eV} \lesssim m \lesssim 5 \times 10^{-23}\text{eV}$, one ULDM coherence patch can encompass all pulsars, but does not reach the typical radius explored by rotation curves. In this \textit{pulsar-correlated} scenario, the stochastic parameter $\hat\phi(\boldsymbol{x})$ is common across all the pulsars. However, estimates of DM density derived from rotation curves average over different coherence patches. Hence, we maintain $\hat\phi(\boldsymbol{x})$ as an independent parameter and marginalize over it, so that the constraints on $\rho$ derived from the following analysis can be compared to the density estimated through rotation curves.
Regardless of the scenario, we always draw one phase parameter $\theta(\boldsymbol{x})$ per pulsar. This phase encodes the uncertainty on current pulsar distance measurements~\cite{Verbiest_2012, Smarra_2023, Hwang_2024} (see below Eq.~\eqref{eq:alphadep} in Sec.~\ref{sec:spin} for more details).

We focus on the ULDM mass range $m\in [10^{-24}~\text{eV}, 10^{-21}~\text{eV}]$, since this is the interval where PTA constraints are the most compelling. 
Notice that the low-mass end corresponds to a frequency of $f_\text{low} \sim 2.4 \times 10^{-10}~\text{Hz}$, which is far below the inverse of EPTA-DR2Full observation length $f_\text{obs} =  1/T_\text{obs} \sim 1.3~\text{nHz}$. 
In this regime, the ULDM-induced signal (see Eq.~\eqref{eq:dt} later) can still be expanded in powers of ($mt$)~\cite{Kaplan_2022}. The first terms in the expansion are degenerate with the simultaneous fitting of pulsar spin frequency derivatives~\cite{HazbounRomano2019,BlandfordNarayan1984, Ramani_2020}; therefore, the lowest-order term that PTAs are sensitive to is $(mt)^3$.
Although this introduces a sensitivity loss - which is also confirmed \textit{e.g.} by Fig.~\ref{fig:comp_tot} - the ULDM-induced signal amplitude is inversely proportional to the particle mass (see Eqs.~\eqref{eq:dt}-~\eqref{eq:psihere}). Therefore, we can still provide significant bounds at $f \lesssim 1/T_\text{obs}$ ~\cite{Unal_2022}. The upper end of the ULDM mass range, instead, corresponds to $f_\text{up} \sim 2.4 \times 10^{-7}~\text{Hz}$, which is somewhat lower than the observational cadence of $f \sim 1/2\,\text{weeks} \sim 8.3 \times 10^{-7}~\text{Hz}$, after which the PTA data are not sensitive.
Finding no evidence for a signal in the examined mass range, we present 95\% upper limits in the following sections.

\section{Results}\label{sec:spin}
In this section, we apply the theoretical framework laid down in Sec.~\ref{sec:nonmin} to constrain the FJBD and the DEF conformal couplings.
\subsection{FJBD conformal coupling}
Let us focus on  FJBD theory. In this case, by inspecting Eq.~\eqref{eq:fjbd} and recalling the definition of $\alpha(\phi)$, we find $\alpha(\phi) = \alpha$. Moreover, the numerical analysis carried out in Ref.~\cite{Kuntz_2024} shows that the angular momentum sensitivity $s_I$ has a very weak dependence on the scalar coupling $\alpha$ that we neglect in this analysis.  Depending on the context, we conveniently write $s_I(\alpha, M)$ as $s_I (M)$ or $s_I$ to avoid cluttering the notation.

For small scalar fluctuations such as the ones considered in this article, it follows from Eq.~\eqref{eq:angularmomsensitivity} that
\begin{align}
   \Omega_\text{obs} (t) =& \bar\Omega\left(1 + 2\alpha s_I \delta\phi (t)\right)\nonumber\\ 
    =& \bar\Omega\left(1 + 2\alpha s_I \frac{\sqrt{\rho}}{M_\text{P} m} \hat{\phi}(\boldsymbol{x})\cos(mt + \theta(\boldsymbol{x}))\right),\label{eq:domega2}
\end{align}
where we used Eq.~\eqref{eq:scal}  and we denoted the spin frequency of the pulsar \textit{in the absence of} the scalar field by $\bar\Omega$. Notice that Eq.~\eqref{eq:domega2} describes the correct frequency shift only under the assumption that the oscillating timescale $t_\text{osc} \sim 1/m$ is much longer than the timescale 
on which a neutron star  adjusts its internal structure ($t_\text{int}$). This is a reasonable assumption. For instance,  the Vela Pulsar shows a fast core-crust coupling with a timescale $t_\text{int} \sim 10~\text{s}$~\cite{Abney_1996}, which would be larger than $t_\text{osc}$ only for ULDM masses $m\gtrsim 10^{-16}~\text{eV}$, which are not discussed here (see Sec.~\ref{sec:methods}).

To find the TOA change induced by the scalar field, we write Eq.~\eqref{eq:angularmomsensitivity} as 
\begin{equation}
    \frac{\delta \Omega_\text{obs}}{\bar\Omega} = 2\alpha s_I \delta\phi = - \frac{\delta P}{\bar P},
\end{equation}
where $\bar P$ is the pulsar period \textit{in the absence of} the scalar field. The total timing residual $\Delta t(t)$ after a time $t$ is the integral of infinitesimal period variations over time:
\begin{align}
    \Delta t(t) &= -\int \mathrm d t\frac{\delta \Omega_\text{obs}}{\bar\Omega}  \nonumber\\
    &= 2\alpha s_I \left.\frac{\sqrt{\rho}}{M_\text{P} m^2}\hat{\phi}(\boldsymbol{x}) \sin(mt + \theta(\boldsymbol{x}))\right|_{t_{\text {start }}-\frac{d}{c}} ^{t_{\text {end }} -\frac{d}{c}},
    \label{eq:alphadep}
\end{align}
where we have highlighted the dependence on the \textit{retarded} time $t_\textit{i} - d/c$, with $i = \textit{start, end}$ and $d$ referring to the Earth-pulsar distance. As mentioned before, the pulsar distance can be re-absorbed in a redefinition of the phase $\theta(\boldsymbol{x}) \rightarrow \theta(\boldsymbol{x}) + md/c$. 
Present uncertainties on the pulsar distances are $\mathcal{O}(0.1\div 1)~\text{kpc}$~\cite{Verbiest_2012}, implying that this redefinition gives rise to an effective pulsar-dependent random phase. Therefore, as mentioned in Sec.~\ref{sec:methods}, we treat $\theta(\boldsymbol{x})$ as a \textit{pulsar-specific} random parameter, and we neglect the distance in the retarded time from now on.

In analogy with the ULDM search results in Ref.~\cite{Smarra_2023}, where the timing residual for the model considered was written as 
\begin{equation}
    \delta t_\text{DM} = \frac{\Psi_\text{c}}{2m} [\hat{\phi}^2_\text{E}\sin{(2mt + \theta_\text{E} )} - \hat{\phi}^2_\text{P}\sin{(2mt + \theta_\text{P} )} ], 
    \label{eq:st}
\end{equation}
with $E, P$ labeling respectively the Earth and the pulsar\footnote{The stochastic parameter $\hat{\phi}_\text{P}$ corresponds to $\hat{\phi}$ in our notation. }, we can define an \textit{effective} amplitude
\begin{equation}
    \Psi = 2\alpha\frac{\sqrt{\rho}}{M_\text{P} m},
    \label{eq:psihere}
\end{equation}
such that
\begin{equation}
    \Delta t(t) = \left.\frac{\Psi}{m} s_I \hat{\phi}(\boldsymbol{x}) \sin(mt + \theta(\boldsymbol{x}))\right|_{t_{\text {start }}} ^{t_{\text {end }}}.
    \label{eq:dt}
\end{equation}
This form helps to understand what a sensible prior  on $\Psi$ may be. In fact, noticing that Eqs.~\eqref{eq:st} and~\eqref{eq:dt} have the same form (differing only for the presence of the Earth term and some $\mathcal{O}(1)$ factors), we can use the same prior for $\Psi$ and $\Psi_c$, \textit{i.e.} $\text{Log}_{10}-\text{Uniform}\left[-20,-12\right]$. 
In other words, the similarity between the two equations
shows that what we are really testing is whether the PTA data can constrain the presence of a sinusoidal signal.

\begin{figure}[htbp]
\centering
\includegraphics[width=0.5\textwidth]{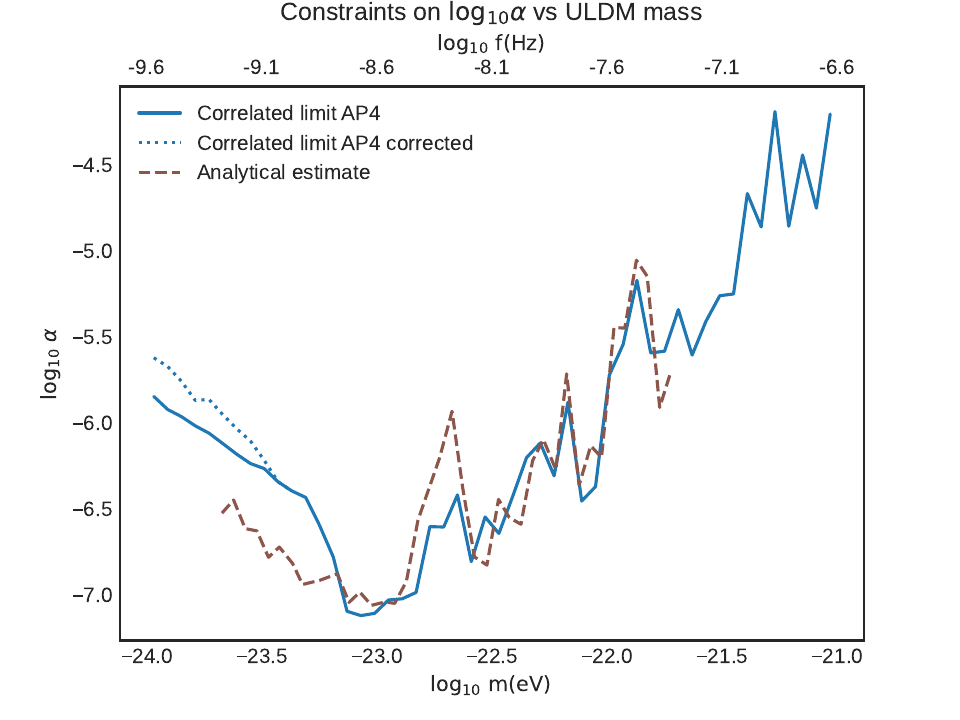}
\caption{Upper limits on $\log_{10}\alpha$ at 95\% credibility compared to the analytical estimate described by Eq.~\eqref{eq:theo}, for the AP4 EoS. The solid line shows the upper limits on $\log_{10}\alpha$ for the \textit{correlated} analysis, assuming that the background DM density is $\rho_\text{DM} = 0.4 \, \text{GeV/cm}^3$, while the brown dashed lines displays the expected behavior described by Eq.~\eqref{eq:theo}. The dotted line shows the degradation of the bounds when choosing $\rho = \text{min}(\bar \rho, \rho_\text{DM})$, optimistically setting $\bar \rho$ to the upper bounds presented in Ref.~\cite{Smarra_2023}. Smaller $\rho$ yield a stronger degradation of the limits.}
\label{fig:comp}
\end{figure}

As previously stated, the sensitivity $s_I$ as a function of the pulsar mass is computed from a fit to the models of Ref.~\cite{Kuntz_2024}. In particular, we consistently utilize the pulsar \textit{gravitational} mass as a parameter of the fit instead of the \textit{inertial} one, because the former is the value measured by experiments. 
We implement the pulsar masses  in the analysis in the following way:
\begin{itemize}
    \item if a pulsar mass is determined from other experiments as $M \pm \delta M$,  we draw the mass  from a normal distribution centered on $M$, with uncertainty $\delta M$ and truncated for masses below $M_\text{min} = 1.1\,M_\odot$ and above $M_\text{max} \sim 2\, M_\odot$. The precise value of $M_\text{max}$ depends on the EoS considered (see Appendix~\ref{sec:app_par} for more details);
    \item if we have no determination of the pulsar mass, we draw it from a uniform distribution (\textit{e.g.} $M \in \left[1.1\,M_\odot, 2.2 M_\odot \right]$ for the AP4 model, see Appendix~\ref{sec:app_par} for more details).
\end{itemize}
At this point, it is worth obtaining some analytical understanding of the results that we expect for $\alpha$. In particular, writing the residual induced by an ULDM candidate of mass $m$ as $\Delta t(m)$, we notice that
\begin{equation}
    \Delta t(m) \sim \delta t _{DM} \left(\frac{m}{2}\right) \rightarrow \Psi \sim \Psi_c \left(\frac{m}{2}\right) ,
\end{equation}
which, through Eq.~\eqref{eq:psihere}, yields
\begin{equation}
    \alpha \sim \frac{m M_\text{P}}{\sqrt{\rho}}\Psi_c\left(\frac{m}{2}\right) .
    \label{eq:theo}
\end{equation}
By substituting the upper limits for $\Psi_c$ found in Fig.~1 of Ref.~\cite{Smarra_2023}, Eq.~\eqref{eq:theo} gives us an approximate estimate of the upper limits that we expect to find.
Fig~\ref{fig:comp} shows the comparison between the \textit{correlated limit} and its theoretical prediction based on Eq.~\eqref{eq:theo}, assuming for reference $\rho = \rho_{\text{DM}} =  0.4~ \text{GeV/cm}^3 $. As  can be seen, our analytical prediction captures the scaling and the overall shape of the constraints, but there are  deviations   caused by the intrinsic difference between the signals in Eq.~\eqref{eq:st} and Eq.~\eqref{eq:dt} (for example, the fact that, in our scenario, the timing residual depends on the mass of the pulsar through the sensitivity).

\begin{figure}[htbp]
\centering
\includegraphics[width=0.5\textwidth]{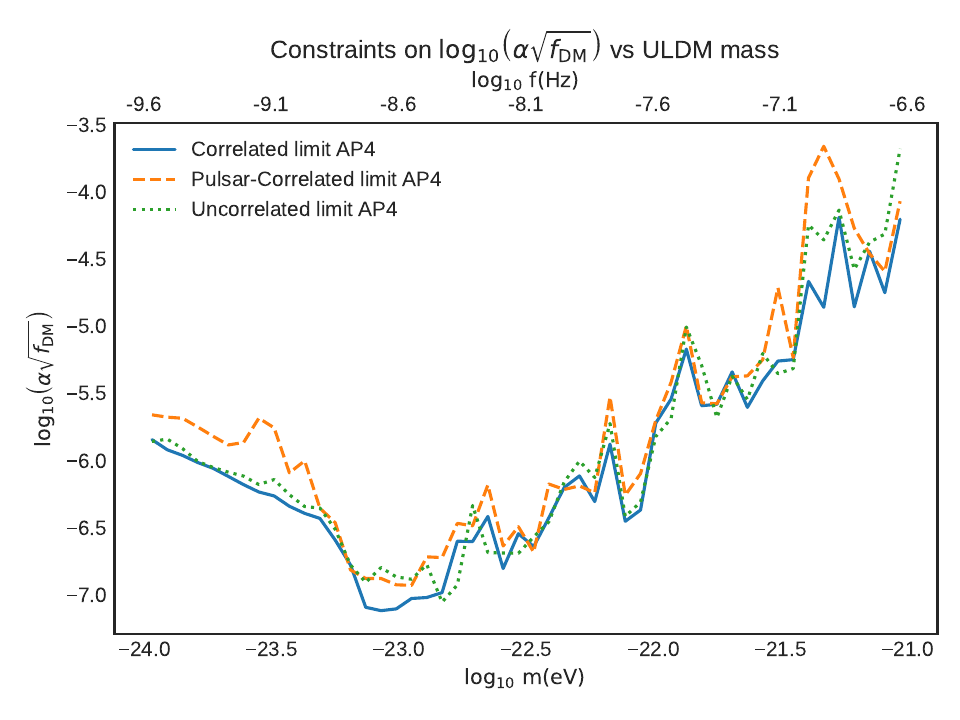}
\caption{Upper limits on $\log_{10}\left(\alpha\sqrt{f_\text{DM}}\right)$ at 95\% credibility versus the ULDM mass. We compare results for the \textit{correlated, pulsar-correlated} and \textit{uncorrelated} scenarios in solid, dashed and dotted lines, respectively. The results are obtained using the AP4 EoS, and the priors on the parameters of the search are presented in Table~\ref{tab:priors}. Bounds from the Cassini mission and from the pulsar in a triple stellar system constrain  $\alpha^2\lesssim 10^{-5}$ and $\alpha^2\lesssim 4\times10^{-6}$, respectively.}\label{fig:comp_tot}
\end{figure}

In the following, 
we present results in terms of $\alpha \sqrt{f_\text{DM}}$ to take into account the relative energy density  of this ULDM candidate $f_\text{DM}$. This is indeed the quantity constrained by Eq.~\eqref{eq:psihere}, and allows for rapidly obtaining the relevant bound on $\alpha$ once a value for the scalar field density $\rho$ is chosen.
Fig.~\ref{fig:comp_tot} displays the upper limits for the \textit{correlated, pulsar correlated} and \textit{uncorrelated} analyses. 

We detect the existence of additional signal power above the common red noise background for masses around $m \sim 10^{-22.7}~\text{eV}$ and $m \sim 10^{-21.4}~\text{eV}$ across all the three analyses. The first excess is consistent with what was observed in recent searches \cite{Smarra_2023, Afzal_2023}, while the second one is thought to be associated with unmodeled perturbations in the orbital elements of Mercury, whose synodic period matches the detected excess frequency \cite{Porayko_2018}. While both of them could in principle be interpreted as evidence of non-minimally coupled ULDM candidates, the fact that they can be accounted for by different physical models makes us more cautious in drawing definitive conclusions.
In order our results to be consistent, we also need to ensure that the effect that we are constraining is not sub-dominant with respect to the TOA induced by the purely gravitational effect of the ULDM oscillations~\cite{Khmelnitsky_2014,Smarra_2023}, which we are neglecting in our analysis.
To understand the interplay between our analysis and the analysis \textit{à la Khmelnitsky-Rubakov} performed in Ref.~\cite{Smarra_2023}, it is sufficient to notice that a non-minimally coupled ULDM candidate of mass $m$ described by the FJBD action (Eqs.~\eqref{eq:action}--\eqref{eq:fjbd}) will in general produce both an $\alpha$-\textit{dependent} residual and a \textit{propagation} residual, described respectively by Eqs.~\eqref{eq:alphadep}--\eqref{eq:st}. While the former has a dependence on the scalar coupling $\alpha$ and has a characteristic frequency $2\pi f = m$, the latter only depends on the density of the scalar field and has a characteristic frequency $2\pi f = 2m$~\cite{Khmelnitsky_2014, Smarra_2023}. Therefore, we only need to be careful selecting a value of $\rho$ which is not already excluded by the analysis \textit{à la Khmelnitsky-Rubakov} when extracting a bound on $\alpha$. Fig.~\ref{fig:comp} shows the effect of this remark on our bounds when we include the constraints on $\rho$ found in Ref.~\cite{Smarra_2023}.

Additionally, we notice in passing that the form of the signal in Eq.~\eqref{eq:dt} is similar to Eq.~(12) in Ref.~\cite{Kaplan_2022}, where direct couplings between ULDM and ordinary matter were also studied with PTA data, apart from $\mathcal{O}(1)$ numerical factors.
Mapping our signal to Eq.~(12) in Ref.~\cite{Kaplan_2022}, we find that the limits obtained here are in general agreement with their analysis.
Although the two models induce a similar TOA change, it is important to point out that they are fundamentally different. Indeed, the model presented in Ref.~\cite{Kaplan_2022} introduces a \textit{direct} and particle-dependent coupling of the scalar to matter, which implies a violation (albeit small) of the weak equivalence principle. Instead, our model relies on a universal conformal coupling of the scalar to gravity, and the weak equivalence principle is satisfied (although the strong equivalence principle is violated).

\subsection{DEF theory}
To study the  DEF theory (quadratic conformal coupling), recall that it is characterized by  $\alpha(\phi) = \beta \phi$, cf. \eqref{eq:def}. Therefore, the angular momentum sensitivity defined in Eq.~\eqref{eq:angularmomsensitivity} becomes
\begin{equation} \label{eq:betasens}
    s_I = \frac{1}{2 \beta \phi} \frac{\mathrm{d} \ln \Omega_\mathrm{obs}}{\mathrm{d} \phi} \bigg |_{N,J}.
\end{equation}
Here, we cannot neglect the dependence of $s_I$ on $\beta$; therefore we write explicitly $s_I = s_I (\beta, M)$ or $s_I = s_I (\beta)$, depending on the context. 
The induced timing residual then reads:
\begin{align}
    \Delta t(t) =& -\int\frac{\delta \Omega_\text{obs}}{\bar\Omega} dt = \nonumber\\
    =&-2\beta s_I(\beta) \int \frac{\rho}{m^2 M_\text{P}^2 }\hat{\phi}^2(\boldsymbol{x}) \cos^2(mt + \theta(\boldsymbol{x}))\nonumber \\
 = &\left.\frac{\Psi}{2m} \beta s_I (\beta) \hat{\phi}^2(\boldsymbol{x}) \sin(2mt + \theta(\boldsymbol{x}))\right|_{t_{\text {start }}} ^{t_{\text {end }}},
    \label{eq:dtbeta}
    \end{align}
where we used   $\Psi = \rho/(m^2 M_\text{P}^2)$ and  again neglect the dependence on the retarded time as well as the constant term in the cosine expansion, as it yields a linear contribution which is absorbed by the pulsar timing model~\cite{HazbounRomano2019, BlandfordNarayan1984, Ramani_2020}.
As $s_I = s_I(\beta)$, Eq.~\eqref{eq:dtbeta} depends \textit{separately} on the scalar field density, parametrized in terms of $\Psi$, and on the DEF scalar coupling $\beta$. Therefore, unlike the FJBD case, $\rho$ and $\beta$ (or, equivalently, $\Psi$ and $\beta$) must be two independent parameters of the search.

\begin{figure*}[!htb]
    \centering
    \begin{subfigure}[b]{0.49\textwidth}
        \includegraphics[width=\textwidth]{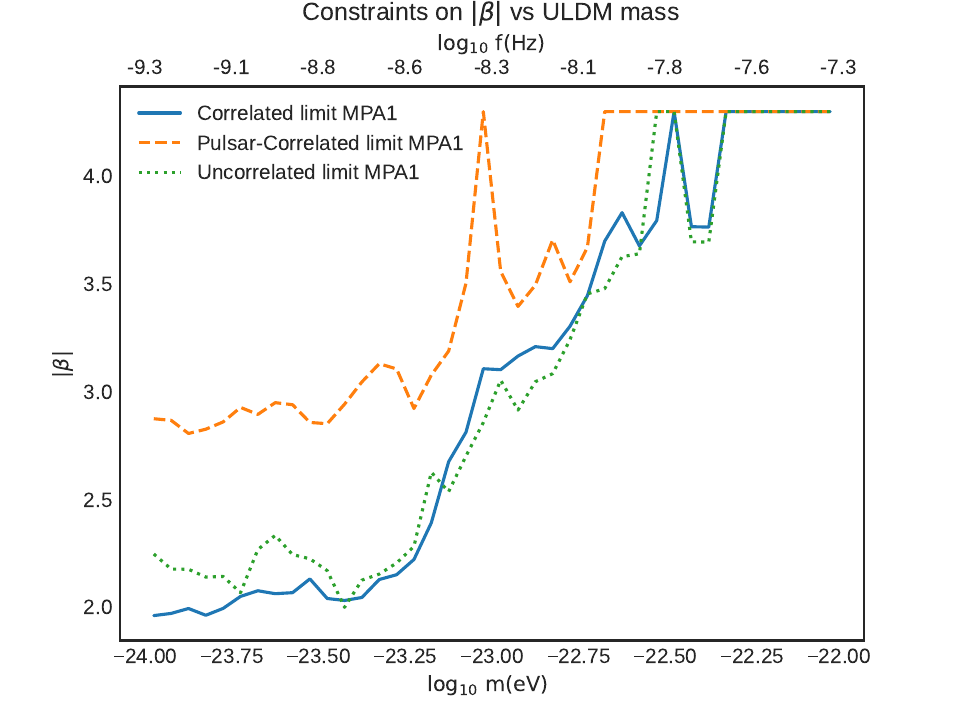}
        \label{fig:beta_neg_100}
    \end{subfigure}
    \begin{subfigure}[b]{0.49\textwidth}
        \includegraphics[width=\textwidth]{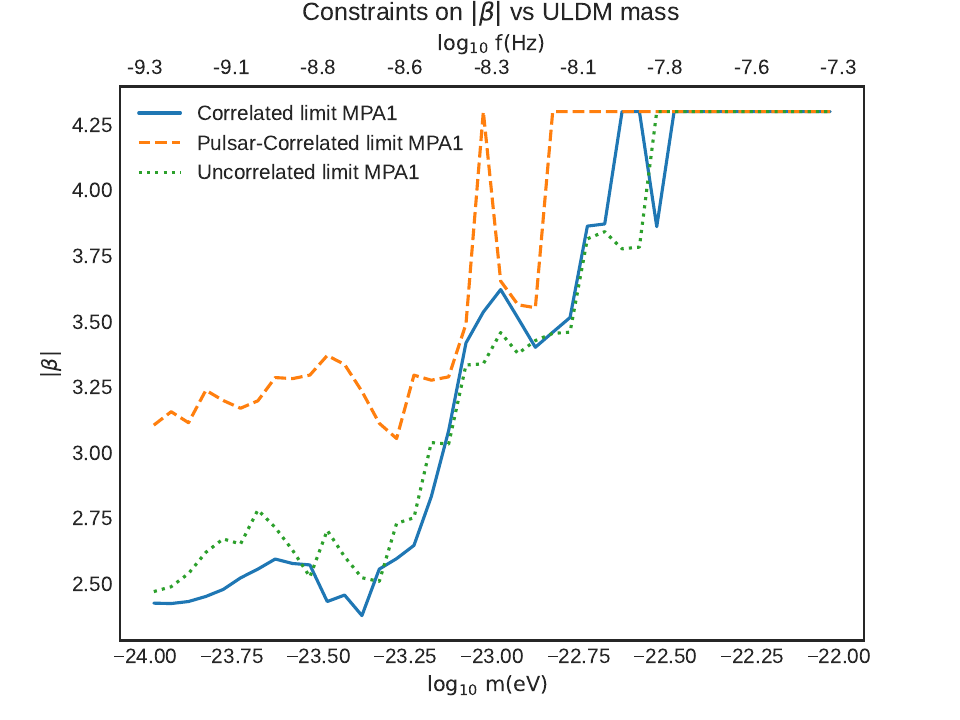}
        \label{fig:beta_neg_50}
    \end{subfigure}
    \begin{subfigure}[b]{0.49\textwidth}
        \includegraphics[width=\textwidth]{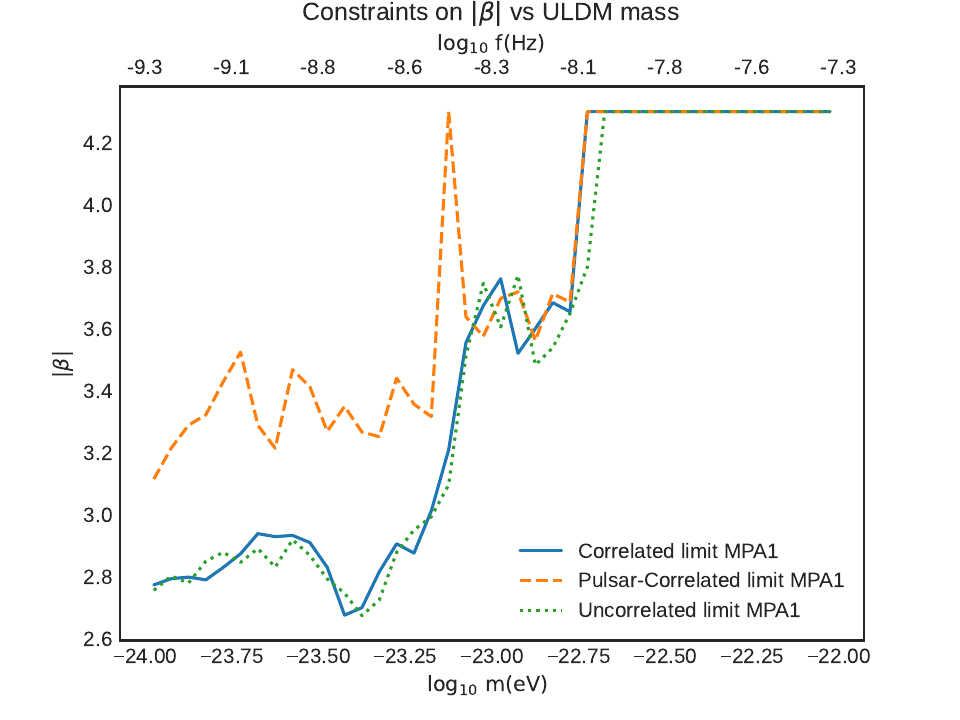}
        \label{fig:beta_neg_30}
    \end{subfigure}    
    \vspace{-0.5\baselineskip} 
    \caption{Upper limits on $\lvert\beta\rvert$  ($\beta < 0$) at 95\% credibility versus the ULDM mass. The top left panel shows the results for $\rho = \rho_\text{DM}$, the top right panel assumes $\rho = 0.5~\rho_\text{DM}$ and the bottom panel displays the results for $\rho = 0.3~\rho_\text{DM}$. We compare results for the \textit{correlated, pulsar-correlated} and \textit{uncorrelated} scenarios in solid, dashed and dotted lines, respectively. Whenever the bound is prior dominated, the upper limits represent the upper end (in absolute value) of our prior. Priors on the parameters relevant for the search are chosen according to the scheme presented in Table~\ref{tab:priors}.} 
    \label{fig:beta_neg}
\end{figure*}

\begin{figure*}[!htb]
    \centering
    \begin{subfigure}[b]{0.49\textwidth}
        \includegraphics[width=\textwidth]{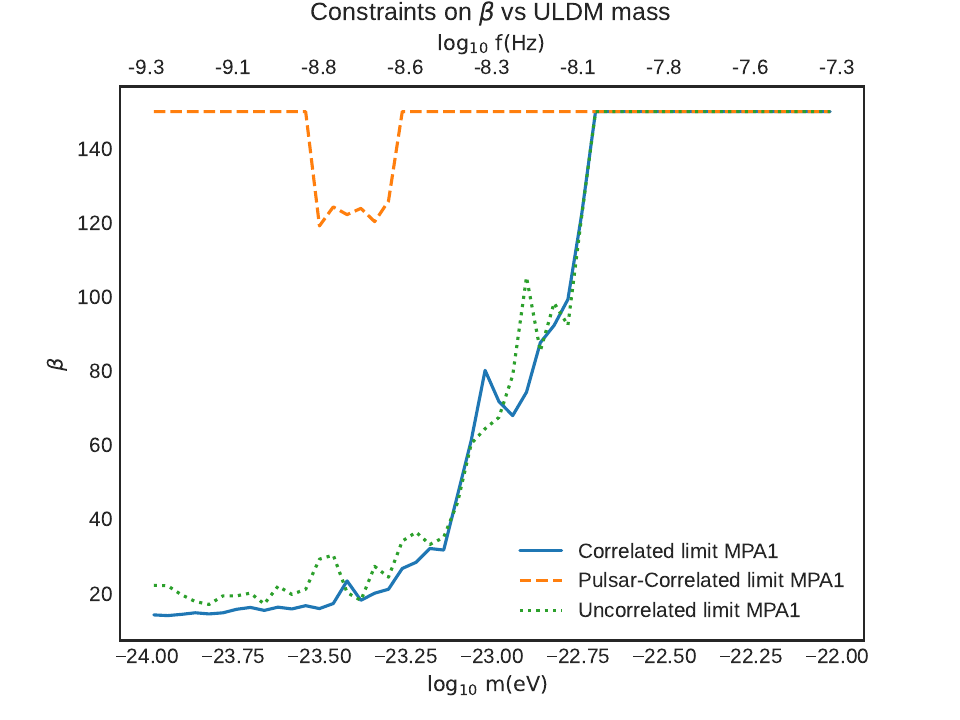}
        \label{fig:beta_pos_100}
    \end{subfigure}
    \begin{subfigure}[b]{0.49\textwidth}
        \includegraphics[width=\textwidth]{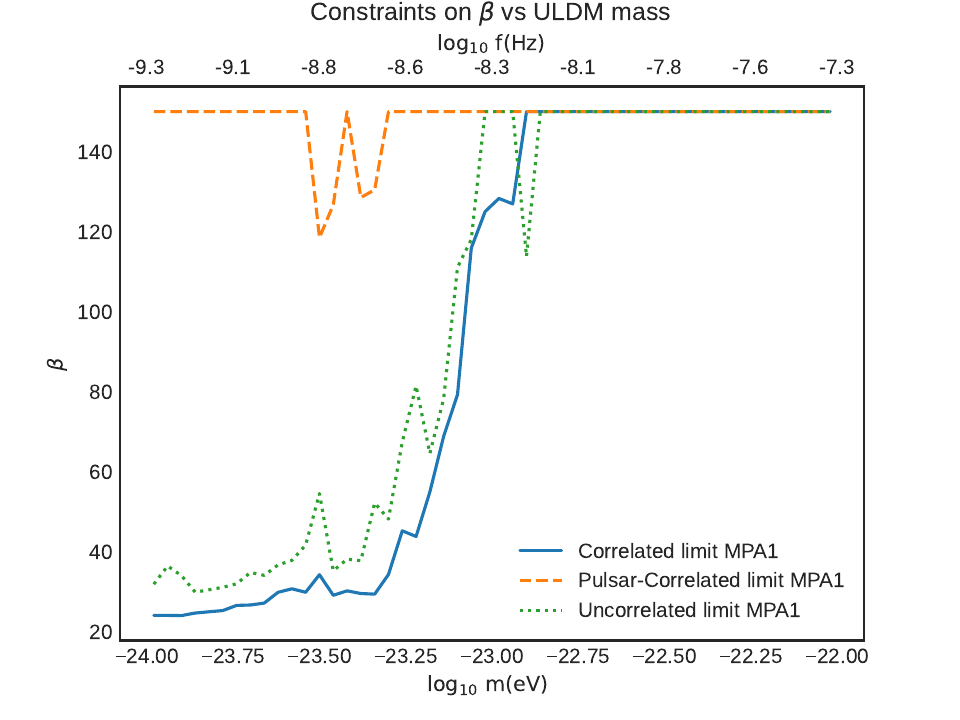}
        \label{fig:beta_pos_50}
    \end{subfigure}
    \begin{subfigure}[b]{0.49\textwidth}
        \includegraphics[width=\textwidth]{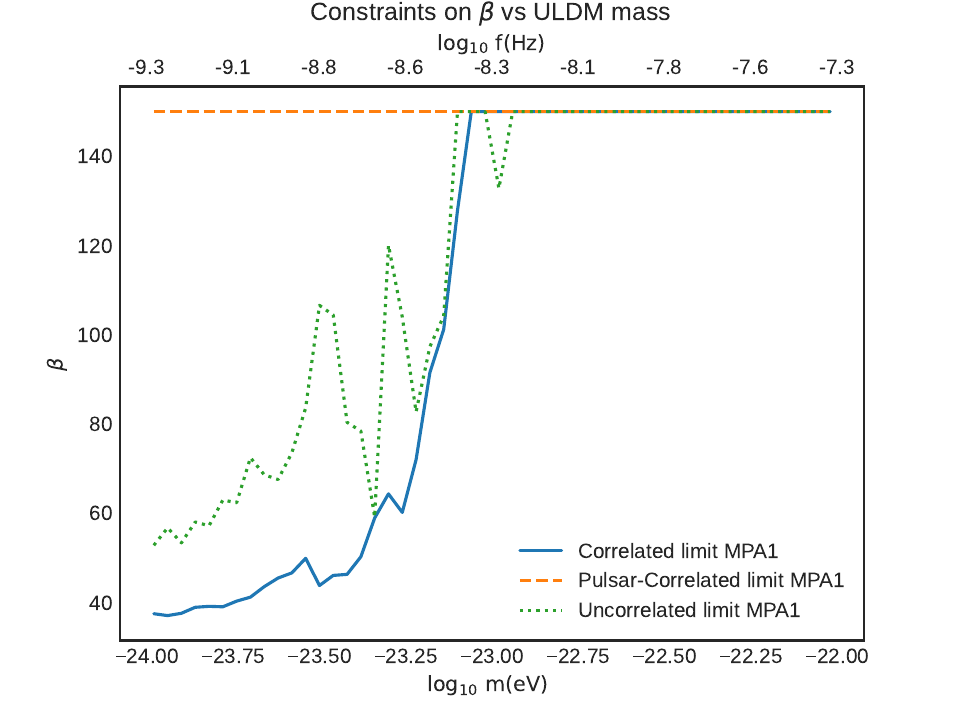}
        \label{fig:beta_pos_30}
    \end{subfigure}    
    \vspace{-0.5\baselineskip} 
    \caption{ Upper limits on $\beta$  ($\beta > 0$) at 95\% credibility versus the ULDM mass. The top left panel shows the results for $\rho = \rho_\text{DM}$, the top right panel assumes $\rho = 0.5~\rho_\text{DM}$ and the bottom panel displays the results for $\rho = 0.3~\rho_\text{DM}$. We compare results for the \textit{correlated, pulsar-correlated} and \textit{uncorrelated} scenarios in solid, dashed and dotted lines, respectively. Whenever the bound is prior dominated, the upper limits represent the upper end of our prior. We notice that the \textit{pulsar-correlated} case yields valid bounds almost only when the scalar field density is $\rho = \rho_\text{DM}$, while it is completely unconstraining when $\rho = 0.3~\rho_\text{DM}$.  Priors on the parameters relevant for the search are chosen according to the scheme presented in Table~\ref{tab:priors}.} 
    \label{fig:beta_pos}
\end{figure*}

This is a crucial observation: in the FJBD case, we constrained a combination of $\rho$ and $\alpha$, namely $\Psi$, and we rephrased the results into bounds on $\alpha$ \textit{a-posteriori}, choosing a reference value of $\rho$ only in post-processing. Here, instead, we are forced to impose an explicit prior on $\rho$. \\
Therefore, we focus on two reference values for the scalar field density; namely, $\rho = 0.5~\rho_\text{DM}$ and $\rho = \rho_\text{DM}$.
Moreover, the deterministic residual in Eq.~\eqref{eq:dtbeta} depends on the sign of $\beta$, and not only on its absolute value. While one might na\"{i}vely think that a sign flip $\beta \rightarrow - \beta$ could be absorbed by a redefinition of the random phase $\theta(\boldsymbol{x})$, the sensitivity $s_I$ does actually depend on the sign of $\beta$, as shown in Ref.~\cite{Kuntz_2024} (see \textit{e.g.} Figs.~5--6 of that work). Hence, in the following, we present results for both positive and negative values of $\beta$. As for negative $\beta$, values $\beta \lesssim -4.3$ would generate non-perturbative strong-field effect inducing $\mathcal{O}(1)$ variations from GR~\cite{Damour_1993}, and are therefore not considered in the present work
(being ruled out by binary pulsars~\cite{Shao:2017gwu}). 
Fig.~\ref{fig:beta_neg} displays the upper bounds on $\lvert\beta\rvert$ for the \textit{correlated, pulsar-correlated} and \textit{uncorrelated} limits and for the selected MPA1 EoS, showing that our analysis implies $\lvert\beta\rvert \lesssim 2.2 $  in the range $10^{-24}~\text{eV} \lesssim m \lesssim 10^{-23.5}~\text{eV}$ for $\rho = \rho_{\text{DM}}$. We also plot the constraints obtained for a scalar field constituting 50\% or 30\% of DM in Fig.~\ref{fig:beta_neg}. Whenever our analysis is prior dominated, the upper limits represent the upper end (in absolute value) of our prior, namely $\lvert\beta\rvert = 4.3$. As expected from the form of the signal in Eq.~\eqref{eq:dtbeta}, larger values of the scalar field density yield stronger (and wider) constraints. For positive $\beta$, instead, we focus on $\beta <150 $, as the code provided in Ref.~\cite{Kuntz_2024} to compute the sensitivities is unstable for higher values of $\beta$. 
The results are presented in Fig.~\ref{fig:beta_pos} for the same choices of scalar field density. Even in this case, larger values of the scalar field density translate into more constraining upper bounds, which can be as low as $\beta\lesssim 20$ for $\rho = \rho_\text{DM}$. Again, the upper bounds are chosen to coincide with the upper end of our prior, namely $\beta = 150$, whenever our analysis is prior dominated.\\
Finally, let us remind that the analysis in Ref.~\cite{Smarra_2023} excludes ultralight scalar field densities $\rho\gtrsim 0.3~\rho_\text{DM}$ for masses $10^{-24}~\text{eV} \lesssim m \lesssim 10^{-23.7}~\text{eV}$ and $\rho\gtrsim 0.7~\rho_\text{DM}$ in the mass range $10^{-23.7}~\text{eV} \lesssim m \lesssim 10^{-23.4}~\text{eV}$. This remark should be taken into account when interpreting the results in Figs.~\ref{fig:beta_neg}-\ref{fig:beta_pos}.

\newpage
\section{Conclusions}\label{sec:conclusions}

Conformally coupled ULDM induces periodic variations in the gravitational mass and in the radius of pulsars~\cite{Kuntz_2024}, with a timescale given by the mass of the particle. By conservation of angular momentum, this translates into an oscillating behavior of the pulsar spin frequency, which is accurately measured by the PTA collaborations. This effect can be used to set constraints on the coupling of ULDM to matter, characterized by the conformal factor $\mathcal{A}(\phi)$ linking the Einstein and Jordan frame metrics.
In this work, we have analyzed the FJBD and the DEF scalar-tensor theories with an ultralight scalar mass, under the assumption that the scalar field constitutes (part of) the DM, thus exploiting different functional forms of the conformal factor $\mathcal{A}(\phi)$. 

In the FJBD theory, where $\mathcal{A}(\phi) \sim 1 + \alpha \phi$, we find $\log_{10}\alpha \lesssim -4.5$ across the entire frequency range considered, vastly overperforming both Cassini bounds~\cite{Bertotti_2003} and the constraints from the pulsar in a triple stellar system system~\cite{Ransom_2014, Archibald_2018, Voisin_2020}. Moreover, for masses $m\sim 10^{-23}~\text{eV}$, our analysis yields the even tighter bound
$\log_{10}\alpha \lesssim -7$.
Let us  recall, however, that the previously mentioned bounds have a wider range of applicability than ours, since they also constrain massless scalar-tensor theories.

In the DEF theory, where $\mathcal{A} \simeq 1 + \beta \phi^2$, we distinguish between positive and negative values of $\beta$, which yield a different expression for the sensitivity $s_I$. In the low mass range, we find $-2 \lesssim \beta \lesssim 20$ for a scalar field density $\rho = \rho_\text{DM}$. We also explore how this bound relaxes when the scalar field density constitutes 50\% or 30\% of the DM density $\rho_\text{DM}$. Once again, this is competitive with respect to existing bounds that can be found in the literature~\cite{Shao:2017gwu,Mendes:2014ufa,Anderson:2017phb}. 

In summary, we have shown that PTA data alone can constrain  conformal ULDM couplings for masses below $\sim 10^{-21}$\,eV at levels not yet explored by other observations. These models include scenarios where the ULDM constitutes all the dark matter in the Universe, and scenarios where the ULDM is a fraction of the total dark matter content.  All the future  improvements of PTA searches (e.g. those related to SKA \cite{Janssen:2014dka}) will  impact  the searches we performed in this work. Furthermore, it was recently emphasized in \cite{Kim:2023pkx} that the  effects coming from the interference of the different modes comprising the ULDM field (recall \eqref{eq:scal} and the velocity dispersion $\sigma_\phi\sim 10^{-3}$) will generate a signal at frequencies below $m\sigma_\phi$ that may allow our current analysis to access ULDM models of higher masses. 

\acknowledgments
C. Smarra wishes to thank Cecilia Sgalletta for useful discussions about data processing.
E. Barausse, A. Kuntz and C. Smarra acknowledge support from the European Union’s H2020 ERC Consolidator Grant ``GRavity from Astrophysical to Microscopic Scales'' (Grant No. GRAMS-815673), the PRIN 2022 grant ``GUVIRP - Gravity tests in the UltraViolet and InfraRed with Pulsar timing'', and the EU Horizon 2020 Research and Innovation Programme under the Marie Sklodowska-Curie Grant Agreement No. 101007855.
The research leading to these results has received funding from the Spanish Ministry of Science and Innovation (PID2020-115845GB-I00/AEI/10.13039/501100011033).
IFAE is partially funded by the CERCA program of the Generalitat de Catalunya.  D. Blas acknowledges the support from the Departament de Recerca i Universitats de la Generalitat de Catalunya al Grup de Recerca i Universitats from Generalitat de Catalunya to the Grup de Recerca 00649 (Codi: 2021 SGR 00649). D. L\'opez Nacir acknowledges support from   University of Buenos Aires and CONICET. LS was supported by the National SKA Program of China (2020SKA0120300), the Beijing Natural Science Foundation (1242018), and the Max Planck Partner Group Program funded by the Max Planck Society. J.Antoniadis acknowledges support from the European Commission (ARGOS-CDS; Grant Agreement number: 101094354)
The Nan\c{c}ay radio Observatory is operated by the Paris Observatory, associated with the French Centre National de la Recherche Scientifique (CNRS), and partially supported by the Region Centre in France. I. Cognard, L. Guillemot and G. Theureau acknowledge financial support from ``Programme National de Cosmologie and Galaxies'' (PNCG), and ``Programme National Hautes Energies'' (PNHE) funded by CNRS/INSU-IN2P3-INP, CEA and CNES, France. I. Cognard, L. Guillemot and G. Theureau acknowledge financial support from Agence Nationale de la Recherche (ANR-18-CE31-0015), France.
Pulsar research at Jodrell Bank Centre for Astrophysics is supported by an STFC Consolidated Grant (ST/T000414/1; ST/X001229/1).
This work is also supported as part of the “LEGACY” MPG-CAS collaboration on 1052 low-frequency gravitational wave astronomy. 
D. Perrodin acknowledges support from the PRIN 2022 grant “GUVIRP - Gravity tests in the UltraViolet and InfraRed with Pulsar timing” and the INAF 2023 Large Grant “Gravitational Wave Detection using Pulsar Timing Arrays”.
\clearpage
\onecolumngrid
\appendix
\section{Parameters of the search}\label{sec:app_par}
Table \ref{tab:priors} summarizes the parameters used in the search along with their priors. We will add a label \textit{F} if the parameter is used only in the FJBD analysis, while a label \textit{D} will signal parameters used only in the DEF analysis. As stated in the main text, if a pulsar mass is measured from other experiments to be $M \pm \delta M$, we  sample the mass parameter from a normal prior distribution centered on $M$, with uncertainty $\delta M$ and truncated below $M_\text{min} = 1.1\,M_\odot$ and above $M_\text{max} = (2.2\,M_\odot, 2.4\,M_\odot, 2.2\,M_\odot)$ for the (AP4~\cite{Akmal_1998}, MPA1~\cite{Muther_1987}, SLy~\cite{Douchin_2001}) EoS, respectively.
We will label this as TruncNorm($\mu, \sigma$), where $\mu = M$ and $\sigma = \delta M$. Otherwise, we will just assume a uniform prior between $M_\text{min}$ and  $M_\text{max}$ for the three EoS choices.
We also plot the constraints on the FJBD scalar coupling $\alpha$ for the different EoS of the pulsar interior presented in Ref.~\cite{Kuntz_2024}.

\begin{table}[H]
\renewcommand{\arraystretch}{1.2}
\centering
\caption{Parameters employed for the search along with their respective priors. In the correlated limit, the parameter $\hat\phi^2_\text{P}$ is accounted for by a redefinition of $\Psi$, while in the pulsar-correlated regime $\hat\phi^2_\text{P} = \hat\phi^2$ is a common free parameter. Only the pulsars whose masses have been measured from other experiments are presented, along with the relevant reference. For the other pulsars, we choose uniform priors with support $[M_\text{min}, M_\text{max}]$ (see main text for more details). We display the priors on the DEF scalar coupling $\beta$ both for $\beta < 0$ and $\beta > 0$. $\mathcal{U}$ stands for the uniform distribution and $\mathcal{N}$ stands for the TruncNorm distribution (see main text). The white noise parameters EFAC (TOA error Excess FACtor) and EQUAD (TOA Error excess in QUADrature) are introduced for every receiver-backend system in every pulsar.}
\begin{tabular}{|c|c|c|c|}
\hline  \textbf{Parameter} & \textbf{Description} & \textbf{Prior} & \textbf{Occurrence} \\ \hline
\hline \multicolumn{4}{|c|}{ White Noise $\left(\sigma = E_\text{f}^2 \sigma^2_\text{TOA} + E_\text{q}^2\right)$} \\ \hline
\hline$E_\text{f}$ & EFAC & $\mathcal{U}(0,10)$ & 1 per pulsar \\
\hline$E_\text{q}$ & EQUAD & Log$_{10}$-$\mathcal{U}(-10,-5)$ & 1 per pulsar \\  \hline
\hline \multicolumn{4}{|c|}{ Red Noise (RN) } \\  \hline
\hline$A_{\text {red }}$ & RN power-law amplitude & Log$_{10}$-$\mathcal{U}(-20,-11)$ & 1 per pulsar \\
\hline$\gamma_{\text {red }}$ & RN power-law spectral index & $\mathcal{U}(0,10)$ & 1 per pulsar \\  \hline
\hline \multicolumn{4}{|c|}{ ULDM (\textit{F}) } \\  \hline
\hline$\Psi$ & ULDM signal amplitude & $\text{Log}_{10}$-$\mathcal{U}(-20,-12)$ & 1 per PTA \\
\hline$m~[\mathrm{eV}]$ & ULDM mass & Log$_{10}$-$\mathcal{U}(-24,-21)$ & 1 per PTA \\
\hline$\hat{\phi}^2$ & Pulsar factor & $e^{-x}$ & 1 per pulsar \\
\hline$\theta$ & Pulsar signal phase &$\mathcal{U}(0,2 \pi)$ & 1 per pulsar \\  \hline
\hline \multicolumn{4}{|c|}{ ULDM (\textit{D}) } \\  \hline
\hline$ f_\text{DM} $ & ULDM fraction & $\mathcal{U}(0.01,0.30)$ & 1 per PTA \\
\hline $\beta$ & DEF scalar coupling & \begin{tabular}{@{}p{4cm}@{}}$\mathcal{U}(-4.3,0)$ \\ or \\ $\mathcal{U}(0,150)$\end{tabular}  & 1 per PTA \\
\hline$m~[\mathrm{eV}]$ & ULDM mass & Log$_{10}$-$\mathcal{U}(-24,-21)$ & 1 per PTA \\
\hline$\hat{\phi}^2$ & Pulsar factor & $e^{-x}$ & 1 per pulsar \\
\hline$\theta$ & Pulsar signal phase & $\mathcal{U}(0,2 \pi)$ & 1 per pulsar \\  \hline
\hline \multicolumn{4}{|c|}{ Common spatially Uncorrelated Red Noise (CURN) } \\  \hline
\hline$A_{\mathrm{GWB}}$ & CURN strain amplitude & Log$_{10}$-$\mathcal{U}(-18,-11)$ & 1 per PTA \\
\hline$\gamma_{\mathrm{GWB}}$ & CURN spectral index & $\mathcal{U}(0,7)$ & 1 per PTA \\
\hline \multicolumn{4}{|c|}{ Pulsar Masses } \\  \hline
\hline$\text{M}_\text{J0030}$ \cite{Miller_2019} &  PSR~\text{J0030+0451} mass & $\mathcal{N}(1.44, 0.15)$ & 1 per PTA \\
\hline$\text{M}_\text{J1012} $ \cite{Sanchez_2020} & PSR~J1012+5307 mass & $\mathcal{N}(1.72, 0.16)$ & 1 per PTA \\
\hline$\text{M}_\text{J1713} $ \cite{Chen_2011} & PSR~J1713+0747 mass & $\mathcal{N}(1.3, 0.2)$ &  1 per PTA \\
\hline$\text{M}_\text{J1738} $ \cite{Antoniadis_2012} & PSR~J1738+0333 mass & $\mathcal{N}(1.47, 0.07)$ & 1 per PTA  \\
\hline$\text{M}_\text{J1909} $ \cite{Jacoby_2005} & PSR~J1909-3744 mass & $\mathcal{N}(1.438, 0.024)$ & 1 per PTA \\
\hline
\end{tabular}
\label{tab:priors}
\end{table}

\begin{figure}
    \centering
    \begin{subfigure}{0.49\textwidth}
        \includegraphics[width=\linewidth]{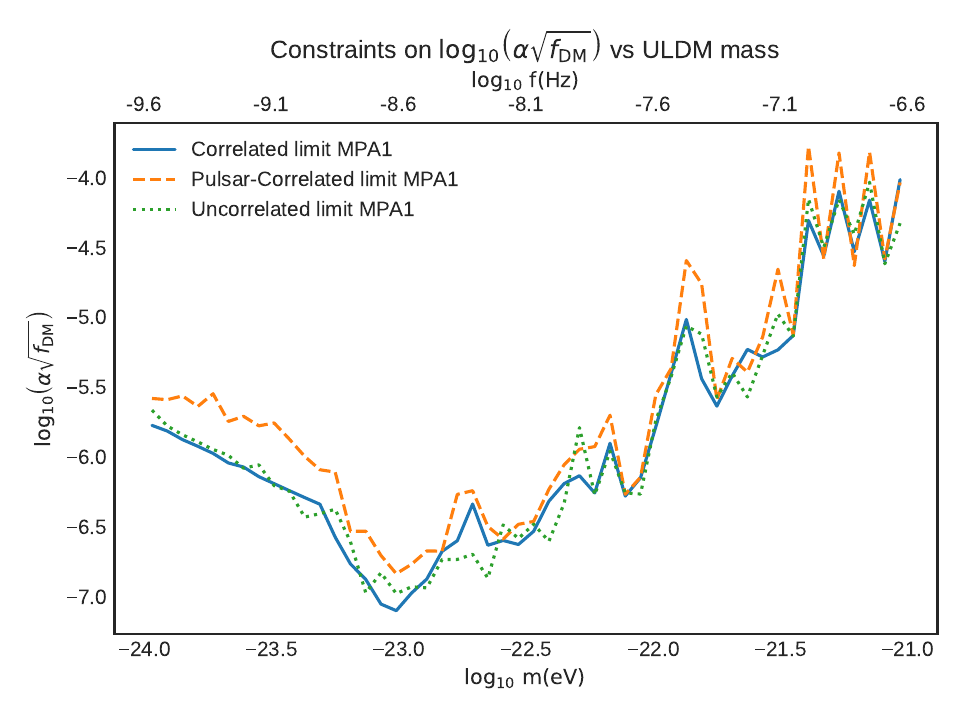}
        \label{fig:MPA1}
    \end{subfigure}
    \hfill
    \begin{subfigure}{0.49\textwidth}
        \includegraphics[width=\linewidth]{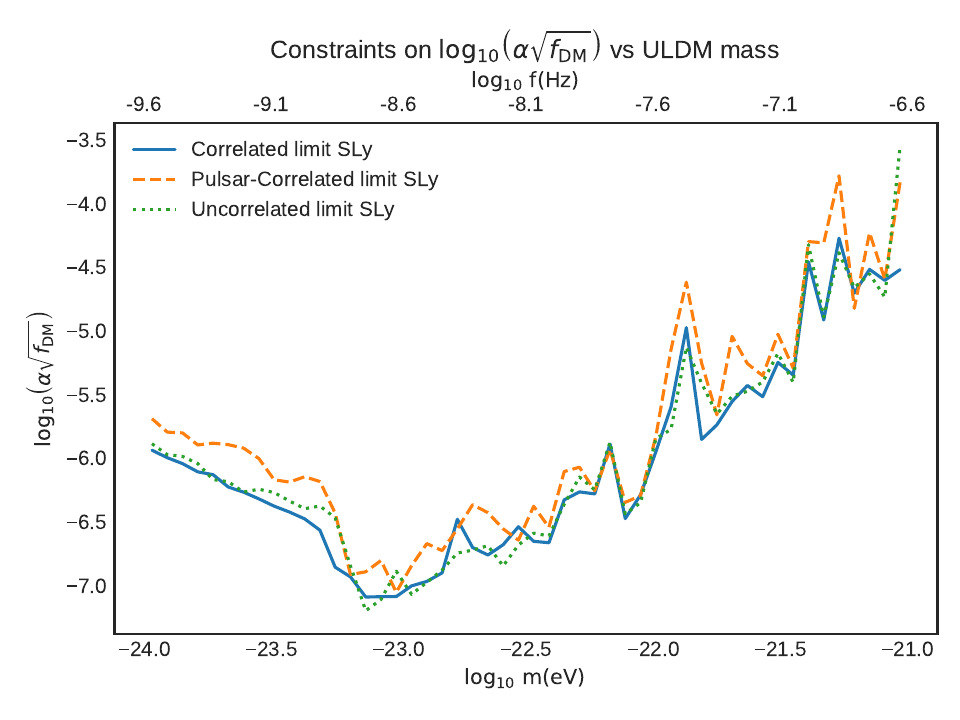}
        \label{fig:SLy}       
    \end{subfigure}
    
    \caption{Upper limits on $\log_{10}\alpha$ at 95\% credibility versus the ULDM mass, using the prescription for $\rho_\text{DM}$ detailed in the main text. We compare results for the \textit{correlated, pulsar-correlated} and \textit{uncorrelated} scenarios in solid, dashed and dotted lines, respectively. The results are obtained using the MPA1 (left panel) and the SLy (right panel) EoS, while the priors on the parameters of the search are chosen according to Table \ref{tab:priors}.}
    \label{fig:main}
\end{figure}
\newpage
\bibliography{biblio.bib}

\begin{thebibliography}{117}%
\makeatletter
\providecommand \@ifxundefined [1]{%
 \@ifx{#1\undefined}
}%
\providecommand \@ifnum [1]{%
 \ifnum #1\expandafter \@firstoftwo
 \else \expandafter \@secondoftwo
 \fi
}%
\providecommand \@ifx [1]{%
 \ifx #1\expandafter \@firstoftwo
 \else \expandafter \@secondoftwo
 \fi
}%
\providecommand \natexlab [1]{#1}%
\providecommand \enquote  [1]{``#1''}%
\providecommand \bibnamefont  [1]{#1}%
\providecommand \bibfnamefont [1]{#1}%
\providecommand \citenamefont [1]{#1}%
\providecommand \href@noop [0]{\@secondoftwo}%
\providecommand \href [0]{\begingroup \@sanitize@url \@href}%
\providecommand \@href[1]{\@@startlink{#1}\@@href}%
\providecommand \@@href[1]{\endgroup#1\@@endlink}%
\providecommand \@sanitize@url [0]{\catcode `\\12\catcode `\$12\catcode
  `\&12\catcode `\#12\catcode `\^12\catcode `\_12\catcode `\%12\relax}%
\providecommand \@@startlink[1]{}%
\providecommand \@@endlink[0]{}%
\providecommand \url  [0]{\begingroup\@sanitize@url \@url }%
\providecommand \@url [1]{\endgroup\@href {#1}{\urlprefix }}%
\providecommand \urlprefix  [0]{URL }%
\providecommand \Eprint [0]{\href }%
\providecommand \doibase [0]{http://dx.doi.org/}%
\providecommand \selectlanguage [0]{\@gobble}%
\providecommand \bibinfo  [0]{\@secondoftwo}%
\providecommand \bibfield  [0]{\@secondoftwo}%
\providecommand \translation [1]{[#1]}%
\providecommand \BibitemOpen [0]{}%
\providecommand \bibitemStop [0]{}%
\providecommand \bibitemNoStop [0]{.\EOS\space}%
\providecommand \EOS [0]{\spacefactor3000\relax}%
\providecommand \BibitemShut  [1]{\csname bibitem#1\endcsname}%
\let\auto@bib@innerbib\@empty
\bibitem [{\citenamefont {{Flores}}\ and\ \citenamefont
  {{Primack}}(1994)}]{Flores_1994}%
  \BibitemOpen
  \bibfield  {author} {\bibinfo {author} {\bibfnamefont {Ricardo~A.}\
  \bibnamefont {{Flores}}}\ and\ \bibinfo {author} {\bibfnamefont {Joel~R.}\
  \bibnamefont {{Primack}}},\ }\bibfield  {title} {\enquote {\bibinfo {title}
  {{Observational and Theoretical Constraints on Singular Dark Matter
  Halos}},}\ }\href {\doibase 10.1086/187350} {\bibfield  {journal} {\bibinfo
  {journal} {\apjl}\ }\textbf {\bibinfo {volume} {427}},\ \bibinfo {pages} {L1}
  (\bibinfo {year} {1994})},\ \Eprint {http://arxiv.org/abs/astro-ph/9402004}
  {arXiv:astro-ph/9402004 [astro-ph]} \BibitemShut {NoStop}%
\bibitem [{\citenamefont {Moore}(1994)}]{Moore_1994}%
  \BibitemOpen
  \bibfield  {author} {\bibinfo {author} {\bibfnamefont {Ben}\ \bibnamefont
  {Moore}},\ }\bibfield  {title} {\enquote {\bibinfo {title} {Evidence against
  dissipation-less dark matter from observations of galaxy haloes},}\ }\href
  {\doibase 10.1038/370629a0} {\bibfield  {journal} {\bibinfo  {journal}
  {Nature}\ }\textbf {\bibinfo {volume} {370}},\ \bibinfo {pages} {629--631}
  (\bibinfo {year} {1994})}\BibitemShut {NoStop}%
\bibitem [{\citenamefont {{Karukes, E. V.}}\ \emph {et~al.}(2015)\citenamefont
  {{Karukes, E. V.}}, \citenamefont {{Salucci, P.}},\ and\ \citenamefont
  {{Gentile, G.}}}]{Karukes_2015}%
  \BibitemOpen
  \bibfield  {author} {\bibinfo {author} {\bibnamefont {{Karukes, E. V.}}},
  \bibinfo {author} {\bibnamefont {{Salucci, P.}}}, \ and\ \bibinfo {author}
  {\bibnamefont {{Gentile, G.}}},\ }\bibfield  {title} {\enquote {\bibinfo
  {title} {The dark matter distribution in the spiral ngc 3198 out to 0.22
  rvir},}\ }\href {\doibase 10.1051/0004-6361/201425339} {\bibfield  {journal}
  {\bibinfo  {journal} {A\&A}\ }\textbf {\bibinfo {volume} {578}},\ \bibinfo
  {pages} {A13} (\bibinfo {year} {2015})}\BibitemShut {NoStop}%
\bibitem [{\citenamefont {Klypin}\ \emph {et~al.}(1999)\citenamefont {Klypin},
  \citenamefont {Kravtsov}, \citenamefont {Valenzuela},\ and\ \citenamefont
  {Prada}}]{Klypin_1999}%
  \BibitemOpen
  \bibfield  {author} {\bibinfo {author} {\bibfnamefont {Anatoly}\ \bibnamefont
  {Klypin}}, \bibinfo {author} {\bibfnamefont {Andrey~V.}\ \bibnamefont
  {Kravtsov}}, \bibinfo {author} {\bibfnamefont {Octavio}\ \bibnamefont
  {Valenzuela}}, \ and\ \bibinfo {author} {\bibfnamefont {Francisco}\
  \bibnamefont {Prada}},\ }\bibfield  {title} {\enquote {\bibinfo {title}
  {Where are the missing galactic satellites?}}\ }\href {\doibase
  10.1086/307643} {\bibfield  {journal} {\bibinfo  {journal} {The Astrophysical
  Journal}\ }\textbf {\bibinfo {volume} {522}},\ \bibinfo {pages} {82--92}
  (\bibinfo {year} {1999})}\BibitemShut {NoStop}%
\bibitem [{\citenamefont {{Moore}}\ \emph {et~al.}(1999)\citenamefont
  {{Moore}}, \citenamefont {{Ghigna}}, \citenamefont {{Governato}},
  \citenamefont {{Lake}}, \citenamefont {{Quinn}}, \citenamefont {{Stadel}},\
  and\ \citenamefont {{Tozzi}}}]{Moore_1999}%
  \BibitemOpen
  \bibfield  {author} {\bibinfo {author} {\bibfnamefont {Ben}\ \bibnamefont
  {{Moore}}}, \bibinfo {author} {\bibfnamefont {Sebastiano}\ \bibnamefont
  {{Ghigna}}}, \bibinfo {author} {\bibfnamefont {Fabio}\ \bibnamefont
  {{Governato}}}, \bibinfo {author} {\bibfnamefont {George}\ \bibnamefont
  {{Lake}}}, \bibinfo {author} {\bibfnamefont {Thomas}\ \bibnamefont
  {{Quinn}}}, \bibinfo {author} {\bibfnamefont {Joachim}\ \bibnamefont
  {{Stadel}}}, \ and\ \bibinfo {author} {\bibfnamefont {Paolo}\ \bibnamefont
  {{Tozzi}}},\ }\bibfield  {title} {\enquote {\bibinfo {title} {{Dark Matter
  Substructure within Galactic Halos}},}\ }\href {\doibase 10.1086/312287}
  {\bibfield  {journal} {\bibinfo  {journal} {\apjl}\ }\textbf {\bibinfo
  {volume} {524}},\ \bibinfo {pages} {L19--L22} (\bibinfo {year} {1999})},\
  \Eprint {http://arxiv.org/abs/astro-ph/9907411} {arXiv:astro-ph/9907411
  [astro-ph]} \BibitemShut {NoStop}%
\bibitem [{\citenamefont {{Boylan-Kolchin}}\ \emph {et~al.}(2011)\citenamefont
  {{Boylan-Kolchin}}, \citenamefont {{Bullock}},\ and\ \citenamefont
  {{Kaplinghat}}}]{Boylan-Kolchin_2011}%
  \BibitemOpen
  \bibfield  {author} {\bibinfo {author} {\bibfnamefont {Michael}\ \bibnamefont
  {{Boylan-Kolchin}}}, \bibinfo {author} {\bibfnamefont {James~S.}\
  \bibnamefont {{Bullock}}}, \ and\ \bibinfo {author} {\bibfnamefont {Manoj}\
  \bibnamefont {{Kaplinghat}}},\ }\bibfield  {title} {\enquote {\bibinfo
  {title} {{Too big to fail? The puzzling darkness of massive Milky Way
  subhaloes}},}\ }\href {\doibase 10.1111/j.1745-3933.2011.01074.x} {\bibfield
  {journal} {\bibinfo  {journal} {\mnras}\ }\textbf {\bibinfo {volume} {415}},\
  \bibinfo {pages} {L40--L44} (\bibinfo {year} {2011})},\ \Eprint
  {http://arxiv.org/abs/1103.0007} {arXiv:1103.0007 [astro-ph.CO]} \BibitemShut
  {NoStop}%
\bibitem [{\citenamefont {Navarro}\ \emph {et~al.}(1996)\citenamefont
  {Navarro}, \citenamefont {Eke},\ and\ \citenamefont {Frenk}}]{Navarro_1996}%
  \BibitemOpen
  \bibfield  {author} {\bibinfo {author} {\bibfnamefont {J.~F.}\ \bibnamefont
  {Navarro}}, \bibinfo {author} {\bibfnamefont {V.~R.}\ \bibnamefont {Eke}}, \
  and\ \bibinfo {author} {\bibfnamefont {C.~S.}\ \bibnamefont {Frenk}},\
  }\bibfield  {title} {\enquote {\bibinfo {title} {The cores of dwarf galaxy
  haloes},}\ }\href {\doibase 10.1093/mnras/283.3.l72} {\bibfield  {journal}
  {\bibinfo  {journal} {Monthly Notices of the Royal Astronomical Society}\
  }\textbf {\bibinfo {volume} {283}},\ \bibinfo {pages} {L72--L78} (\bibinfo
  {year} {1996})}\BibitemShut {NoStop}%
\bibitem [{\citenamefont {Governato}\ \emph {et~al.}(2012)\citenamefont
  {Governato}, \citenamefont {Zolotov}, \citenamefont {Pontzen}, \citenamefont
  {Christensen}, \citenamefont {Oh}, \citenamefont {Brooks}, \citenamefont
  {Quinn}, \citenamefont {Shen},\ and\ \citenamefont
  {Wadsley}}]{Governato_2012}%
  \BibitemOpen
  \bibfield  {author} {\bibinfo {author} {\bibfnamefont {F.}~\bibnamefont
  {Governato}}, \bibinfo {author} {\bibfnamefont {A.}~\bibnamefont {Zolotov}},
  \bibinfo {author} {\bibfnamefont {A.}~\bibnamefont {Pontzen}}, \bibinfo
  {author} {\bibfnamefont {C.}~\bibnamefont {Christensen}}, \bibinfo {author}
  {\bibfnamefont {S.~H.}\ \bibnamefont {Oh}}, \bibinfo {author} {\bibfnamefont
  {A.~M.}\ \bibnamefont {Brooks}}, \bibinfo {author} {\bibfnamefont
  {T.}~\bibnamefont {Quinn}}, \bibinfo {author} {\bibfnamefont
  {S.}~\bibnamefont {Shen}}, \ and\ \bibinfo {author} {\bibfnamefont
  {J.}~\bibnamefont {Wadsley}},\ }\bibfield  {title} {\enquote {\bibinfo
  {title} {{Cuspy no more: how outflows affect the central dark matter and
  baryon distribution in $\Lambda$ cold dark matter galaxies}},}\ }\href
  {\doibase 10.1111/j.1365-2966.2012.20696.x} {\bibfield  {journal} {\bibinfo
  {journal} {Monthly Notices of the Royal Astronomical Society}\ }\textbf
  {\bibinfo {volume} {422}},\ \bibinfo {pages} {1231--1240} (\bibinfo {year}
  {2012})},\ \Eprint
  {http://arxiv.org/abs/https://academic.oup.com/mnras/article-pdf/422/2/1231/3466602/mnras0422-1231.pdf}
  {https://academic.oup.com/mnras/article-pdf/422/2/1231/3466602/mnras0422-1231.pdf}
  \BibitemShut {NoStop}%
\bibitem [{\citenamefont {Brooks}\ \emph {et~al.}(2013)\citenamefont {Brooks},
  \citenamefont {Kuhlen}, \citenamefont {Zolotov},\ and\ \citenamefont
  {Hooper}}]{Brooks_2013}%
  \BibitemOpen
  \bibfield  {author} {\bibinfo {author} {\bibfnamefont {Alyson~M.}\
  \bibnamefont {Brooks}}, \bibinfo {author} {\bibfnamefont {Michael}\
  \bibnamefont {Kuhlen}}, \bibinfo {author} {\bibfnamefont {Adi}\ \bibnamefont
  {Zolotov}}, \ and\ \bibinfo {author} {\bibfnamefont {Dan}\ \bibnamefont
  {Hooper}},\ }\bibfield  {title} {\enquote {\bibinfo {title} {A baryonic
  solution to the missing satellites problem},}\ }\href {\doibase
  10.1088/0004-637X/765/1/22} {\bibfield  {journal} {\bibinfo  {journal} {The
  Astrophysical Journal}\ }\textbf {\bibinfo {volume} {765}},\ \bibinfo {pages}
  {22} (\bibinfo {year} {2013})}\BibitemShut {NoStop}%
\bibitem [{\citenamefont {Chan}\ \emph {et~al.}(2015)\citenamefont {Chan},
  \citenamefont {Kereš}, \citenamefont {Oñorbe}, \citenamefont {Hopkins},
  \citenamefont {Muratov}, \citenamefont {Faucher-Giguère},\ and\
  \citenamefont {Quataert}}]{Chan_2015}%
  \BibitemOpen
  \bibfield  {author} {\bibinfo {author} {\bibfnamefont {T.~K.}\ \bibnamefont
  {Chan}}, \bibinfo {author} {\bibfnamefont {D.}~\bibnamefont {Kereš}},
  \bibinfo {author} {\bibfnamefont {J.}~\bibnamefont {Oñorbe}}, \bibinfo
  {author} {\bibfnamefont {P.~F.}\ \bibnamefont {Hopkins}}, \bibinfo {author}
  {\bibfnamefont {A.~L.}\ \bibnamefont {Muratov}}, \bibinfo {author}
  {\bibfnamefont {C.-A.}\ \bibnamefont {Faucher-Giguère}}, \ and\ \bibinfo
  {author} {\bibfnamefont {E.}~\bibnamefont {Quataert}},\ }\bibfield  {title}
  {\enquote {\bibinfo {title} {{The impact of baryonic physics on the structure
  of dark matter haloes: the view from the FIRE cosmological simulations}},}\
  }\href {\doibase 10.1093/mnras/stv2165} {\bibfield  {journal} {\bibinfo
  {journal} {Monthly Notices of the Royal Astronomical Society}\ }\textbf
  {\bibinfo {volume} {454}},\ \bibinfo {pages} {2981--3001} (\bibinfo {year}
  {2015})},\ \Eprint
  {http://arxiv.org/abs/https://academic.oup.com/mnras/article-pdf/454/3/2981/4038253/stv2165.pdf}
  {https://academic.oup.com/mnras/article-pdf/454/3/2981/4038253/stv2165.pdf}
  \BibitemShut {NoStop}%
\bibitem [{\citenamefont {Oñorbe}\ \emph {et~al.}(2015)\citenamefont
  {Oñorbe}, \citenamefont {Boylan-Kolchin}, \citenamefont {Bullock},
  \citenamefont {Hopkins}, \citenamefont {Kereš}, \citenamefont
  {Faucher-Giguère}, \citenamefont {Quataert},\ and\ \citenamefont
  {Murray}}]{Onorbe_2015}%
  \BibitemOpen
  \bibfield  {author} {\bibinfo {author} {\bibfnamefont {Jose}\ \bibnamefont
  {Oñorbe}}, \bibinfo {author} {\bibfnamefont {Michael}\ \bibnamefont
  {Boylan-Kolchin}}, \bibinfo {author} {\bibfnamefont {James~S.}\ \bibnamefont
  {Bullock}}, \bibinfo {author} {\bibfnamefont {Philip~F.}\ \bibnamefont
  {Hopkins}}, \bibinfo {author} {\bibfnamefont {Dušan}\ \bibnamefont
  {Kereš}}, \bibinfo {author} {\bibfnamefont {Claude-André}\ \bibnamefont
  {Faucher-Giguère}}, \bibinfo {author} {\bibfnamefont {Eliot}\ \bibnamefont
  {Quataert}}, \ and\ \bibinfo {author} {\bibfnamefont {Norman}\ \bibnamefont
  {Murray}},\ }\bibfield  {title} {\enquote {\bibinfo {title} {{Forged in fire:
  cusps, cores and baryons in low-mass dwarf galaxies}},}\ }\href {\doibase
  10.1093/mnras/stv2072} {\bibfield  {journal} {\bibinfo  {journal} {Monthly
  Notices of the Royal Astronomical Society}\ }\textbf {\bibinfo {volume}
  {454}},\ \bibinfo {pages} {2092--2106} (\bibinfo {year} {2015})},\ \Eprint
  {http://arxiv.org/abs/https://academic.oup.com/mnras/article-pdf/454/2/2092/9503920/stv2072.pdf}
  {https://academic.oup.com/mnras/article-pdf/454/2/2092/9503920/stv2072.pdf}
  \BibitemShut {NoStop}%
\bibitem [{\citenamefont {Read}\ \emph {et~al.}(2016)\citenamefont {Read},
  \citenamefont {Agertz},\ and\ \citenamefont {Collins}}]{Read_2016}%
  \BibitemOpen
  \bibfield  {author} {\bibinfo {author} {\bibfnamefont {J.~I.}\ \bibnamefont
  {Read}}, \bibinfo {author} {\bibfnamefont {O.}~\bibnamefont {Agertz}}, \ and\
  \bibinfo {author} {\bibfnamefont {M.~L.~M.}\ \bibnamefont {Collins}},\
  }\bibfield  {title} {\enquote {\bibinfo {title} {{Dark matter cores all the
  way down}},}\ }\href {\doibase 10.1093/mnras/stw713} {\bibfield  {journal}
  {\bibinfo  {journal} {Monthly Notices of the Royal Astronomical Society}\
  }\textbf {\bibinfo {volume} {459}},\ \bibinfo {pages} {2573--2590} (\bibinfo
  {year} {2016})},\ \Eprint
  {http://arxiv.org/abs/https://academic.oup.com/mnras/article-pdf/459/3/2573/8105757/stw713.pdf}
  {https://academic.oup.com/mnras/article-pdf/459/3/2573/8105757/stw713.pdf}
  \BibitemShut {NoStop}%
\bibitem [{\citenamefont {Hu}\ \emph {et~al.}(2000)\citenamefont {Hu},
  \citenamefont {Barkana},\ and\ \citenamefont {Gruzinov}}]{Hu:2000ke}%
  \BibitemOpen
  \bibfield  {author} {\bibinfo {author} {\bibfnamefont {Wayne}\ \bibnamefont
  {Hu}}, \bibinfo {author} {\bibfnamefont {Rennan}\ \bibnamefont {Barkana}}, \
  and\ \bibinfo {author} {\bibfnamefont {Andrei}\ \bibnamefont {Gruzinov}},\
  }\bibfield  {title} {\enquote {\bibinfo {title} {{Cold and fuzzy dark
  matter}},}\ }\href {\doibase 10.1103/PhysRevLett.85.1158} {\bibfield
  {journal} {\bibinfo  {journal} {Phys. Rev. Lett.}\ }\textbf {\bibinfo
  {volume} {85}},\ \bibinfo {pages} {1158--1161} (\bibinfo {year} {2000})},\
  \Eprint {http://arxiv.org/abs/astro-ph/0003365} {arXiv:astro-ph/0003365}
  \BibitemShut {NoStop}%
\bibitem [{\citenamefont {Hui}\ \emph {et~al.}(2017)\citenamefont {Hui},
  \citenamefont {Ostriker}, \citenamefont {Tremaine},\ and\ \citenamefont
  {Witten}}]{Hui_2017}%
  \BibitemOpen
  \bibfield  {author} {\bibinfo {author} {\bibfnamefont {Lam}\ \bibnamefont
  {Hui}}, \bibinfo {author} {\bibfnamefont {Jeremiah~P.}\ \bibnamefont
  {Ostriker}}, \bibinfo {author} {\bibfnamefont {Scott}\ \bibnamefont
  {Tremaine}}, \ and\ \bibinfo {author} {\bibfnamefont {Edward}\ \bibnamefont
  {Witten}},\ }\bibfield  {title} {\enquote {\bibinfo {title} {Ultralight
  scalars as cosmological dark matter},}\ }\href {\doibase
  10.1103/physrevd.95.043541} {\bibfield  {journal} {\bibinfo  {journal}
  {Physical Review D}\ }\textbf {\bibinfo {volume} {95}} (\bibinfo {year}
  {2017}),\ 10.1103/physrevd.95.043541}\BibitemShut {NoStop}%
\bibitem [{\citenamefont {Green}\ \emph {et~al.}(1988)\citenamefont {Green},
  \citenamefont {Schwarz},\ and\ \citenamefont {Witten}}]{Green_1987}%
  \BibitemOpen
  \bibfield  {author} {\bibinfo {author} {\bibfnamefont {Michael~B.}\
  \bibnamefont {Green}}, \bibinfo {author} {\bibfnamefont {J.~H.}\ \bibnamefont
  {Schwarz}}, \ and\ \bibinfo {author} {\bibfnamefont {Edward}\ \bibnamefont
  {Witten}},\ }\href@noop {} {\emph {\bibinfo {title} {{SUPERSTRING THEORY.
  VOL. 1: INTRODUCTION}}}},\ Cambridge Monographs on Mathematical Physics\
  (\bibinfo {year} {1988})\BibitemShut {NoStop}%
\bibitem [{\citenamefont {Svrcek}\ and\ \citenamefont
  {Witten}(2006)}]{Svrcek_2006}%
  \BibitemOpen
  \bibfield  {author} {\bibinfo {author} {\bibfnamefont {Peter}\ \bibnamefont
  {Svrcek}}\ and\ \bibinfo {author} {\bibfnamefont {Edward}\ \bibnamefont
  {Witten}},\ }\bibfield  {title} {\enquote {\bibinfo {title} {Axions in string
  theory},}\ }\href {\doibase 10.1088/1126-6708/2006/06/051} {\bibfield
  {journal} {\bibinfo  {journal} {Journal of High Energy Physics}\ }\textbf
  {\bibinfo {volume} {2006}},\ \bibinfo {pages} {051} (\bibinfo {year}
  {2006})}\BibitemShut {NoStop}%
\bibitem [{\citenamefont {Arvanitaki}\ \emph
  {et~al.}(2010{\natexlab{a}})\citenamefont {Arvanitaki}, \citenamefont
  {Dimopoulos}, \citenamefont {Dubovsky}, \citenamefont {Kaloper},\ and\
  \citenamefont {March-Russell}}]{Arvanitaki_2010}%
  \BibitemOpen
  \bibfield  {author} {\bibinfo {author} {\bibfnamefont {Asimina}\ \bibnamefont
  {Arvanitaki}}, \bibinfo {author} {\bibfnamefont {Savas}\ \bibnamefont
  {Dimopoulos}}, \bibinfo {author} {\bibfnamefont {Sergei}\ \bibnamefont
  {Dubovsky}}, \bibinfo {author} {\bibfnamefont {Nemanja}\ \bibnamefont
  {Kaloper}}, \ and\ \bibinfo {author} {\bibfnamefont {John}\ \bibnamefont
  {March-Russell}},\ }\bibfield  {title} {\enquote {\bibinfo {title} {String
  axiverse},}\ }\href {\doibase 10.1103/PhysRevD.81.123530} {\bibfield
  {journal} {\bibinfo  {journal} {Phys. Rev. D}\ }\textbf {\bibinfo {volume}
  {81}},\ \bibinfo {pages} {123530} (\bibinfo {year}
  {2010}{\natexlab{a}})}\BibitemShut {NoStop}%
\bibitem [{\citenamefont {Hlozek}\ \emph {et~al.}(2015)\citenamefont {Hlozek},
  \citenamefont {Grin}, \citenamefont {Marsh},\ and\ \citenamefont
  {Ferreira}}]{Hlozek_2015}%
  \BibitemOpen
  \bibfield  {author} {\bibinfo {author} {\bibfnamefont {Ren\'ee}\ \bibnamefont
  {Hlozek}}, \bibinfo {author} {\bibfnamefont {Daniel}\ \bibnamefont {Grin}},
  \bibinfo {author} {\bibfnamefont {David J.~E.}\ \bibnamefont {Marsh}}, \ and\
  \bibinfo {author} {\bibfnamefont {Pedro~G.}\ \bibnamefont {Ferreira}},\
  }\bibfield  {title} {\enquote {\bibinfo {title} {A search for ultralight
  axions using precision cosmological data},}\ }\href {\doibase
  10.1103/PhysRevD.91.103512} {\bibfield  {journal} {\bibinfo  {journal} {Phys.
  Rev. D}\ }\textbf {\bibinfo {volume} {91}},\ \bibinfo {pages} {103512}
  (\bibinfo {year} {2015})}\BibitemShut {NoStop}%
\bibitem [{\citenamefont {Ir\ifmmode \check{s}\else
  \v{s}\fi{}i\ifmmode~\check{c}\else \v{c}\fi{}}\ \emph
  {et~al.}(2017)\citenamefont {Ir\ifmmode \check{s}\else
  \v{s}\fi{}i\ifmmode~\check{c}\else \v{c}\fi{}}, \citenamefont {Viel},
  \citenamefont {Haehnelt}, \citenamefont {Bolton},\ and\ \citenamefont
  {Becker}}]{Irsic_2017}%
  \BibitemOpen
  \bibfield  {author} {\bibinfo {author} {\bibfnamefont {Vid}\ \bibnamefont
  {Ir\ifmmode \check{s}\else \v{s}\fi{}i\ifmmode~\check{c}\else \v{c}\fi{}}},
  \bibinfo {author} {\bibfnamefont {Matteo}\ \bibnamefont {Viel}}, \bibinfo
  {author} {\bibfnamefont {Martin~G.}\ \bibnamefont {Haehnelt}}, \bibinfo
  {author} {\bibfnamefont {James~S.}\ \bibnamefont {Bolton}}, \ and\ \bibinfo
  {author} {\bibfnamefont {George~D.}\ \bibnamefont {Becker}},\ }\bibfield
  {title} {\enquote {\bibinfo {title} {First constraints on fuzzy dark matter
  from lyman-alpha forest data and hydrodynamical simulations},}\ }\href
  {\doibase 10.1103/PhysRevLett.119.031302} {\bibfield  {journal} {\bibinfo
  {journal} {Phys. Rev. Lett.}\ }\textbf {\bibinfo {volume} {119}},\ \bibinfo
  {pages} {031302} (\bibinfo {year} {2017})}\BibitemShut {NoStop}%
\bibitem [{\citenamefont {Armengaud}\ \emph {et~al.}(2017)\citenamefont
  {Armengaud}, \citenamefont {Palanque-Delabrouille}, \citenamefont {Yèche},
  \citenamefont {Marsh},\ and\ \citenamefont {Baur}}]{Armengaud_2017}%
  \BibitemOpen
  \bibfield  {author} {\bibinfo {author} {\bibfnamefont {Eric}\ \bibnamefont
  {Armengaud}}, \bibinfo {author} {\bibfnamefont {Nathalie}\ \bibnamefont
  {Palanque-Delabrouille}}, \bibinfo {author} {\bibfnamefont {Christophe}\
  \bibnamefont {Yèche}}, \bibinfo {author} {\bibfnamefont {David J.~E.}\
  \bibnamefont {Marsh}}, \ and\ \bibinfo {author} {\bibfnamefont {Julien}\
  \bibnamefont {Baur}},\ }\bibfield  {title} {\enquote {\bibinfo {title}
  {{Constraining the mass of light bosonic dark matter using SDSS Lyman-alpha
  forest}},}\ }\href {\doibase 10.1093/mnras/stx1870} {\bibfield  {journal}
  {\bibinfo  {journal} {Monthly Notices of the Royal Astronomical Society}\
  }\textbf {\bibinfo {volume} {471}},\ \bibinfo {pages} {4606--4614} (\bibinfo
  {year} {2017})},\ \Eprint
  {http://arxiv.org/abs/https://academic.oup.com/mnras/article-pdf/471/4/4606/19635244/stx1870.pdf}
  {https://academic.oup.com/mnras/article-pdf/471/4/4606/19635244/stx1870.pdf}
  \BibitemShut {NoStop}%
\bibitem [{\citenamefont {Kobayashi}\ \emph {et~al.}(2017)\citenamefont
  {Kobayashi}, \citenamefont {Murgia}, \citenamefont {Simone}, \citenamefont
  {Ir{\v{s}}i{\v{c}}},\ and\ \citenamefont {Viel}}]{Kobayashi_2017}%
  \BibitemOpen
  \bibfield  {author} {\bibinfo {author} {\bibfnamefont {Takeshi}\ \bibnamefont
  {Kobayashi}}, \bibinfo {author} {\bibfnamefont {Riccardo}\ \bibnamefont
  {Murgia}}, \bibinfo {author} {\bibfnamefont {Andrea~De}\ \bibnamefont
  {Simone}}, \bibinfo {author} {\bibfnamefont {Vid}\ \bibnamefont
  {Ir{\v{s}}i{\v{c}}}}, \ and\ \bibinfo {author} {\bibfnamefont {Matteo}\
  \bibnamefont {Viel}},\ }\bibfield  {title} {\enquote {\bibinfo {title}
  {Lyman-alpha constraints on ultralight scalar dark matter: Implications for
  the early and late universe},}\ }\href {\doibase 10.1103/physrevd.96.123514}
  {\bibfield  {journal} {\bibinfo  {journal} {Physical Review D}\ }\textbf
  {\bibinfo {volume} {96}} (\bibinfo {year} {2017}),\
  10.1103/physrevd.96.123514}\BibitemShut {NoStop}%
\bibitem [{\citenamefont {Nori}\ \emph {et~al.}(2018)\citenamefont {Nori},
  \citenamefont {Murgia}, \citenamefont {Iršič}, \citenamefont {Baldi},\ and\
  \citenamefont {Viel}}]{Nori_2018}%
  \BibitemOpen
  \bibfield  {author} {\bibinfo {author} {\bibfnamefont {Matteo}\ \bibnamefont
  {Nori}}, \bibinfo {author} {\bibfnamefont {Riccardo}\ \bibnamefont {Murgia}},
  \bibinfo {author} {\bibfnamefont {Vid}\ \bibnamefont {Iršič}}, \bibinfo
  {author} {\bibfnamefont {Marco}\ \bibnamefont {Baldi}}, \ and\ \bibinfo
  {author} {\bibfnamefont {Matteo}\ \bibnamefont {Viel}},\ }\bibfield  {title}
  {\enquote {\bibinfo {title} {{Lyman-alpha forest and non-linear structure
  characterization in Fuzzy Dark Matter cosmologies}},}\ }\href {\doibase
  10.1093/mnras/sty2888} {\bibfield  {journal} {\bibinfo  {journal} {Monthly
  Notices of the Royal Astronomical Society}\ }\textbf {\bibinfo {volume}
  {482}},\ \bibinfo {pages} {3227--3243} (\bibinfo {year} {2018})},\ \Eprint
  {http://arxiv.org/abs/https://academic.oup.com/mnras/article-pdf/482/3/3227/26653692/sty2888.pdf}
  {https://academic.oup.com/mnras/article-pdf/482/3/3227/26653692/sty2888.pdf}
  \BibitemShut {NoStop}%
\bibitem [{\citenamefont {Rogers}\ and\ \citenamefont
  {Peiris}(2021)}]{Rogers_2021}%
  \BibitemOpen
  \bibfield  {author} {\bibinfo {author} {\bibfnamefont {Keir~K.}\ \bibnamefont
  {Rogers}}\ and\ \bibinfo {author} {\bibfnamefont {Hiranya~V.}\ \bibnamefont
  {Peiris}},\ }\bibfield  {title} {\enquote {\bibinfo {title} {Strong bound on
  canonical ultralight axion dark matter from the lyman-alpha forest},}\ }\href
  {\doibase 10.1103/PhysRevLett.126.071302} {\bibfield  {journal} {\bibinfo
  {journal} {Phys. Rev. Lett.}\ }\textbf {\bibinfo {volume} {126}},\ \bibinfo
  {pages} {071302} (\bibinfo {year} {2021})}\BibitemShut {NoStop}%
\bibitem [{\citenamefont {Schive}\ \emph {et~al.}(2014)\citenamefont {Schive},
  \citenamefont {Liao}, \citenamefont {Woo}, \citenamefont {Wong},
  \citenamefont {Chiueh}, \citenamefont {Broadhurst},\ and\ \citenamefont
  {Hwang}}]{Schive_2014}%
  \BibitemOpen
  \bibfield  {author} {\bibinfo {author} {\bibfnamefont {Hsi-Yu}\ \bibnamefont
  {Schive}}, \bibinfo {author} {\bibfnamefont {Ming-Hsuan}\ \bibnamefont
  {Liao}}, \bibinfo {author} {\bibfnamefont {Tak-Pong}\ \bibnamefont {Woo}},
  \bibinfo {author} {\bibfnamefont {Shing-Kwong}\ \bibnamefont {Wong}},
  \bibinfo {author} {\bibfnamefont {Tzihong}\ \bibnamefont {Chiueh}}, \bibinfo
  {author} {\bibfnamefont {Tom}\ \bibnamefont {Broadhurst}}, \ and\ \bibinfo
  {author} {\bibfnamefont {W-Y.~Pauchy}\ \bibnamefont {Hwang}},\ }\bibfield
  {title} {\enquote {\bibinfo {title} {Understanding the core-halo relation of
  quantum wave dark matter from 3d simulations},}\ }\href {\doibase
  10.1103/PhysRevLett.113.261302} {\bibfield  {journal} {\bibinfo  {journal}
  {Phys. Rev. Lett.}\ }\textbf {\bibinfo {volume} {113}},\ \bibinfo {pages}
  {261302} (\bibinfo {year} {2014})}\BibitemShut {NoStop}%
\bibitem [{\citenamefont {Zhang}\ \emph {et~al.}(2019)\citenamefont {Zhang},
  \citenamefont {Liu},\ and\ \citenamefont {Chu}}]{Zhang_2019}%
  \BibitemOpen
  \bibfield  {author} {\bibinfo {author} {\bibfnamefont {Jiajun}\ \bibnamefont
  {Zhang}}, \bibinfo {author} {\bibfnamefont {Hantao}\ \bibnamefont {Liu}}, \
  and\ \bibinfo {author} {\bibfnamefont {Ming-Chung}\ \bibnamefont {Chu}},\
  }\bibfield  {title} {\enquote {\bibinfo {title} {Cosmological simulation for
  fuzzy dark matter model},}\ }\href {\doibase 10.3389/fspas.2018.00048}
  {\bibfield  {journal} {\bibinfo  {journal} {Frontiers in Astronomy and Space
  Sciences}\ }\textbf {\bibinfo {volume} {5}} (\bibinfo {year} {2019}),\
  10.3389/fspas.2018.00048}\BibitemShut {NoStop}%
\bibitem [{\citenamefont {Bar}\ \emph {et~al.}(2018)\citenamefont {Bar},
  \citenamefont {Blas}, \citenamefont {Blum},\ and\ \citenamefont
  {Sibiryakov}}]{Bar:2018acw}%
  \BibitemOpen
  \bibfield  {author} {\bibinfo {author} {\bibfnamefont {Nitsan}\ \bibnamefont
  {Bar}}, \bibinfo {author} {\bibfnamefont {Diego}\ \bibnamefont {Blas}},
  \bibinfo {author} {\bibfnamefont {Kfir}\ \bibnamefont {Blum}}, \ and\
  \bibinfo {author} {\bibfnamefont {Sergey}\ \bibnamefont {Sibiryakov}},\
  }\bibfield  {title} {\enquote {\bibinfo {title} {{Galactic rotation curves
  versus ultralight dark matter: Implications of the soliton-host halo
  relation}},}\ }\href {\doibase 10.1103/PhysRevD.98.083027} {\bibfield
  {journal} {\bibinfo  {journal} {Phys. Rev. D}\ }\textbf {\bibinfo {volume}
  {98}},\ \bibinfo {pages} {083027} (\bibinfo {year} {2018})},\ \Eprint
  {http://arxiv.org/abs/1805.00122} {arXiv:1805.00122 [astro-ph.CO]}
  \BibitemShut {NoStop}%
\bibitem [{\citenamefont {Hayashi}\ \emph {et~al.}(2021)\citenamefont
  {Hayashi}, \citenamefont {Ferreira},\ and\ \citenamefont
  {Chan}}]{Hayashi_2021}%
  \BibitemOpen
  \bibfield  {author} {\bibinfo {author} {\bibfnamefont {Kohei}\ \bibnamefont
  {Hayashi}}, \bibinfo {author} {\bibfnamefont {Elisa G.~M.}\ \bibnamefont
  {Ferreira}}, \ and\ \bibinfo {author} {\bibfnamefont {Hei Yin~Jowett}\
  \bibnamefont {Chan}},\ }\bibfield  {title} {\enquote {\bibinfo {title}
  {Narrowing the mass range of fuzzy dark matter with ultrafaint dwarfs},}\
  }\href {\doibase 10.3847/2041-8213/abf501} {\bibfield  {journal} {\bibinfo
  {journal} {The Astrophysical Journal Letters}\ }\textbf {\bibinfo {volume}
  {912}},\ \bibinfo {pages} {L3} (\bibinfo {year} {2021})}\BibitemShut
  {NoStop}%
\bibitem [{\citenamefont {Dalal}\ and\ \citenamefont
  {Kravtsov}(2022)}]{Dalal_2022}%
  \BibitemOpen
  \bibfield  {author} {\bibinfo {author} {\bibfnamefont {Neal}\ \bibnamefont
  {Dalal}}\ and\ \bibinfo {author} {\bibfnamefont {Andrey}\ \bibnamefont
  {Kravtsov}},\ }\href {\doibase https://doi.org/10.48550/arXiv.2203.05750}
  {\enquote {\bibinfo {title} {Not so fuzzy: excluding fdm with sizes and
  stellar kinematics of ultra-faint dwarf galaxies},}\ } (\bibinfo {year}
  {2022}),\ \Eprint {http://arxiv.org/abs/2203.05750} {arXiv:2203.05750
  [astro-ph.CO]} \BibitemShut {NoStop}%
\bibitem [{\citenamefont {Marsh}\ and\ \citenamefont
  {Niemeyer}(2019)}]{Marsh:2018zyw}%
  \BibitemOpen
  \bibfield  {author} {\bibinfo {author} {\bibfnamefont {David J.~E.}\
  \bibnamefont {Marsh}}\ and\ \bibinfo {author} {\bibfnamefont {Jens~C.}\
  \bibnamefont {Niemeyer}},\ }\bibfield  {title} {\enquote {\bibinfo {title}
  {{Strong Constraints on Fuzzy Dark Matter from Ultrafaint Dwarf Galaxy
  Eridanus II}},}\ }\href {\doibase 10.1103/PhysRevLett.123.051103} {\bibfield
  {journal} {\bibinfo  {journal} {Phys. Rev. Lett.}\ }\textbf {\bibinfo
  {volume} {123}},\ \bibinfo {pages} {051103} (\bibinfo {year} {2019})},\
  \Eprint {http://arxiv.org/abs/1810.08543} {arXiv:1810.08543 [astro-ph.CO]}
  \BibitemShut {NoStop}%
\bibitem [{\citenamefont {Ferreira}(2021)}]{Ferreira:2020fam}%
  \BibitemOpen
  \bibfield  {author} {\bibinfo {author} {\bibfnamefont {Elisa G.~M.}\
  \bibnamefont {Ferreira}},\ }\bibfield  {title} {\enquote {\bibinfo {title}
  {{Ultra-light dark matter}},}\ }\href {\doibase 10.1007/s00159-021-00135-6}
  {\bibfield  {journal} {\bibinfo  {journal} {Astron. Astrophys. Rev.}\
  }\textbf {\bibinfo {volume} {29}},\ \bibinfo {pages} {7} (\bibinfo {year}
  {2021})},\ \Eprint {http://arxiv.org/abs/2005.03254} {arXiv:2005.03254
  [astro-ph.CO]} \BibitemShut {NoStop}%
\bibitem [{\citenamefont {Arvanitaki}\ \emph
  {et~al.}(2010{\natexlab{b}})\citenamefont {Arvanitaki}, \citenamefont
  {Dimopoulos}, \citenamefont {Dubovsky}, \citenamefont {Kaloper},\ and\
  \citenamefont {March-Russell}}]{Arvanitaki:2009fg}%
  \BibitemOpen
  \bibfield  {author} {\bibinfo {author} {\bibfnamefont {Asimina}\ \bibnamefont
  {Arvanitaki}}, \bibinfo {author} {\bibfnamefont {Savas}\ \bibnamefont
  {Dimopoulos}}, \bibinfo {author} {\bibfnamefont {Sergei}\ \bibnamefont
  {Dubovsky}}, \bibinfo {author} {\bibfnamefont {Nemanja}\ \bibnamefont
  {Kaloper}}, \ and\ \bibinfo {author} {\bibfnamefont {John}\ \bibnamefont
  {March-Russell}},\ }\bibfield  {title} {\enquote {\bibinfo {title} {{String
  Axiverse}},}\ }\href {\doibase 10.1103/PhysRevD.81.123530} {\bibfield
  {journal} {\bibinfo  {journal} {Phys. Rev. D}\ }\textbf {\bibinfo {volume}
  {81}},\ \bibinfo {pages} {123530} (\bibinfo {year} {2010}{\natexlab{b}})},\
  \Eprint {http://arxiv.org/abs/0905.4720} {arXiv:0905.4720 [hep-th]}
  \BibitemShut {NoStop}%
\bibitem [{\citenamefont {Hamaide}\ \emph {et~al.}(2022)\citenamefont
  {Hamaide}, \citenamefont {M\"uller},\ and\ \citenamefont
  {Marsh}}]{Hamaide:2022rwi}%
  \BibitemOpen
  \bibfield  {author} {\bibinfo {author} {\bibfnamefont {Louis}\ \bibnamefont
  {Hamaide}}, \bibinfo {author} {\bibfnamefont {Hendrik}\ \bibnamefont
  {M\"uller}}, \ and\ \bibinfo {author} {\bibfnamefont {David J.~E.}\
  \bibnamefont {Marsh}},\ }\bibfield  {title} {\enquote {\bibinfo {title}
  {{Searching for dilaton fields in the Lyman-\ensuremath{\alpha} forest}},}\
  }\href {\doibase 10.1103/PhysRevD.106.123509} {\bibfield  {journal} {\bibinfo
   {journal} {Phys. Rev. D}\ }\textbf {\bibinfo {volume} {106}},\ \bibinfo
  {pages} {123509} (\bibinfo {year} {2022})},\ \Eprint
  {http://arxiv.org/abs/2210.03705} {arXiv:2210.03705 [astro-ph.CO]}
  \BibitemShut {NoStop}%
\bibitem [{\citenamefont {Khmelnitsky}\ and\ \citenamefont
  {Rubakov}(2014)}]{Khmelnitsky_2014}%
  \BibitemOpen
  \bibfield  {author} {\bibinfo {author} {\bibfnamefont {Andrei}\ \bibnamefont
  {Khmelnitsky}}\ and\ \bibinfo {author} {\bibfnamefont {Valery}\ \bibnamefont
  {Rubakov}},\ }\bibfield  {title} {\enquote {\bibinfo {title} {Pulsar timing
  signal from ultralight scalar dark matter},}\ }\href {\doibase
  10.1088/1475-7516/2014/02/019} {\bibfield  {journal} {\bibinfo  {journal}
  {Journal of Cosmology and Astroparticle Physics}\ }\textbf {\bibinfo {volume}
  {2014}},\ \bibinfo {pages} {019--019} (\bibinfo {year} {2014})}\BibitemShut
  {NoStop}%
\bibitem [{\citenamefont {{Foster}}\ and\ \citenamefont
  {{Backer}}(1990)}]{FosterBacker1990}%
  \BibitemOpen
  \bibfield  {author} {\bibinfo {author} {\bibfnamefont {R.~S.}\ \bibnamefont
  {{Foster}}}\ and\ \bibinfo {author} {\bibfnamefont {D.~C.}\ \bibnamefont
  {{Backer}}},\ }\bibfield  {title} {\enquote {\bibinfo {title} {{Constructing
  a Pulsar Timing Array}},}\ }\href {\doibase 10.1086/169195} {\bibfield
  {journal} {\bibinfo  {journal} {\apj}\ }\textbf {\bibinfo {volume} {361}},\
  \bibinfo {pages} {300} (\bibinfo {year} {1990})}\BibitemShut {NoStop}%
\bibitem [{\citenamefont {{Manchester}}\ \emph {et~al.}(2013)\citenamefont
  {{Manchester}}, \citenamefont {{Hobbs}}, \citenamefont {{Bailes}},
  \citenamefont {{Coles}}, \citenamefont {{van Straten}}, \citenamefont
  {{Keith}}, \citenamefont {{Shannon}}, \citenamefont {{Bhat}}, \citenamefont
  {{Brown}}, \citenamefont {{Burke-Spolaor}}, \citenamefont {{Champion}},
  \citenamefont {{Chaudhary}}, \citenamefont {{Edwards}}, \citenamefont
  {{Hampson}}, \citenamefont {{Hotan}}, \citenamefont {{Jameson}},
  \citenamefont {{Jenet}}, \citenamefont {{Kesteven}}, \citenamefont {{Khoo}},
  \citenamefont {{Kocz}}, \citenamefont {{Maciesiak}}, \citenamefont
  {{Oslowski}}, \citenamefont {{Ravi}}, \citenamefont {{Reynolds}},
  \citenamefont {{Sarkissian}}, \citenamefont {{Verbiest}}, \citenamefont
  {{Wen}}, \citenamefont {{Wilson}}, \citenamefont {{Yardley}}, \citenamefont
  {{Yan}},\ and\ \citenamefont {{You}}}]{ManchesterHobbs2013}%
  \BibitemOpen
  \bibfield  {author} {\bibinfo {author} {\bibfnamefont {R.~N.}\ \bibnamefont
  {{Manchester}}}, \bibinfo {author} {\bibfnamefont {G.}~\bibnamefont
  {{Hobbs}}}, \bibinfo {author} {\bibfnamefont {M.}~\bibnamefont {{Bailes}}},
  \bibinfo {author} {\bibfnamefont {W.~A.}\ \bibnamefont {{Coles}}}, \bibinfo
  {author} {\bibfnamefont {W.}~\bibnamefont {{van Straten}}}, \bibinfo {author}
  {\bibfnamefont {M.~J.}\ \bibnamefont {{Keith}}}, \bibinfo {author}
  {\bibfnamefont {R.~M.}\ \bibnamefont {{Shannon}}}, \bibinfo {author}
  {\bibfnamefont {N.~D.~R.}\ \bibnamefont {{Bhat}}}, \bibinfo {author}
  {\bibfnamefont {A.}~\bibnamefont {{Brown}}}, \bibinfo {author} {\bibfnamefont
  {S.~G.}\ \bibnamefont {{Burke-Spolaor}}}, \bibinfo {author} {\bibfnamefont
  {D.~J.}\ \bibnamefont {{Champion}}}, \bibinfo {author} {\bibfnamefont
  {A.}~\bibnamefont {{Chaudhary}}}, \bibinfo {author} {\bibfnamefont {R.~T.}\
  \bibnamefont {{Edwards}}}, \bibinfo {author} {\bibfnamefont {G.}~\bibnamefont
  {{Hampson}}}, \bibinfo {author} {\bibfnamefont {A.~W.}\ \bibnamefont
  {{Hotan}}}, \bibinfo {author} {\bibfnamefont {A.}~\bibnamefont {{Jameson}}},
  \bibinfo {author} {\bibfnamefont {F.~A.}\ \bibnamefont {{Jenet}}}, \bibinfo
  {author} {\bibfnamefont {M.~J.}\ \bibnamefont {{Kesteven}}}, \bibinfo
  {author} {\bibfnamefont {J.}~\bibnamefont {{Khoo}}}, \bibinfo {author}
  {\bibfnamefont {J.}~\bibnamefont {{Kocz}}}, \bibinfo {author} {\bibfnamefont
  {K.}~\bibnamefont {{Maciesiak}}}, \bibinfo {author} {\bibfnamefont
  {S.}~\bibnamefont {{Oslowski}}}, \bibinfo {author} {\bibfnamefont
  {V.}~\bibnamefont {{Ravi}}}, \bibinfo {author} {\bibfnamefont {J.~R.}\
  \bibnamefont {{Reynolds}}}, \bibinfo {author} {\bibfnamefont {J.~M.}\
  \bibnamefont {{Sarkissian}}}, \bibinfo {author} {\bibfnamefont {J.~P.~W.}\
  \bibnamefont {{Verbiest}}}, \bibinfo {author} {\bibfnamefont {Z.~L.}\
  \bibnamefont {{Wen}}}, \bibinfo {author} {\bibfnamefont {W.~E.}\ \bibnamefont
  {{Wilson}}}, \bibinfo {author} {\bibfnamefont {D.}~\bibnamefont {{Yardley}}},
  \bibinfo {author} {\bibfnamefont {W.~M.}\ \bibnamefont {{Yan}}}, \ and\
  \bibinfo {author} {\bibfnamefont {X.~P.}\ \bibnamefont {{You}}},\ }\bibfield
  {title} {\enquote {\bibinfo {title} {{The Parkes Pulsar Timing Array
  Project}},}\ }\href {\doibase 10.1017/pasa.2012.017} {\bibfield  {journal}
  {\bibinfo  {journal} {\pasa}\ }\textbf {\bibinfo {volume} {30}},\ \bibinfo
  {eid} {e017} (\bibinfo {year} {2013})},\ \Eprint
  {http://arxiv.org/abs/1210.6130} {arXiv:1210.6130 [astro-ph.IM]} \BibitemShut
  {NoStop}%
\bibitem [{\citenamefont {{McLaughlin}}(2013)}]{McLaughlin2013}%
  \BibitemOpen
  \bibfield  {author} {\bibinfo {author} {\bibfnamefont {M.~A.}\ \bibnamefont
  {{McLaughlin}}},\ }\bibfield  {title} {\enquote {\bibinfo {title} {{The North
  American Nanohertz Observatory for Gravitational Waves}},}\ }\href {\doibase
  10.1088/0264-9381/30/22/224008} {\bibfield  {journal} {\bibinfo  {journal}
  {Classical and Quantum Gravity}\ }\textbf {\bibinfo {volume} {30}},\ \bibinfo
  {eid} {224008} (\bibinfo {year} {2013})},\ \Eprint
  {http://arxiv.org/abs/1310.0758} {arXiv:1310.0758 [astro-ph.IM]} \BibitemShut
  {NoStop}%
\bibitem [{\citenamefont {{Kramer}}\ and\ \citenamefont
  {{Champion}}(2013)}]{KramerChampion2013}%
  \BibitemOpen
  \bibfield  {author} {\bibinfo {author} {\bibfnamefont {Michael}\ \bibnamefont
  {{Kramer}}}\ and\ \bibinfo {author} {\bibfnamefont {David~J.}\ \bibnamefont
  {{Champion}}},\ }\bibfield  {title} {\enquote {\bibinfo {title} {{The
  European Pulsar Timing Array and the Large European Array for Pulsars}},}\
  }\href {\doibase 10.1088/0264-9381/30/22/224009} {\bibfield  {journal}
  {\bibinfo  {journal} {Classical and Quantum Gravity}\ }\textbf {\bibinfo
  {volume} {30}},\ \bibinfo {eid} {224009} (\bibinfo {year}
  {2013})}\BibitemShut {NoStop}%
\bibitem [{\citenamefont {{Joshi}}\ \emph {et~al.}(2018)\citenamefont
  {{Joshi}}, \citenamefont {{Arumugasamy}}, \citenamefont {{Bagchi}},
  \citenamefont {{Bandyopadhyay}}, \citenamefont {{Basu}}, \citenamefont
  {{Dhanda Batra}}, \citenamefont {{Bethapudi}}, \citenamefont {{Choudhary}},
  \citenamefont {{De}}, \citenamefont {{Dey}}, \citenamefont {{Gopakumar}},
  \citenamefont {{Gupta}}, \citenamefont {{Krishnakumar}}, \citenamefont
  {{Maan}}, \citenamefont {{Manoharan}}, \citenamefont {{Naidu}}, \citenamefont
  {{Nandi}}, \citenamefont {{Pathak}}, \citenamefont {{Surnis}},\ and\
  \citenamefont {{Susobhanan}}}]{InPTA}%
  \BibitemOpen
  \bibfield  {author} {\bibinfo {author} {\bibfnamefont {Bhal~Chandra}\
  \bibnamefont {{Joshi}}}, \bibinfo {author} {\bibfnamefont {Prakash}\
  \bibnamefont {{Arumugasamy}}}, \bibinfo {author} {\bibfnamefont {Manjari}\
  \bibnamefont {{Bagchi}}}, \bibinfo {author} {\bibfnamefont {Debades}\
  \bibnamefont {{Bandyopadhyay}}}, \bibinfo {author} {\bibfnamefont {Avishek}\
  \bibnamefont {{Basu}}}, \bibinfo {author} {\bibfnamefont {Neelam}\
  \bibnamefont {{Dhanda Batra}}}, \bibinfo {author} {\bibfnamefont {Suryarao}\
  \bibnamefont {{Bethapudi}}}, \bibinfo {author} {\bibfnamefont {Arpita}\
  \bibnamefont {{Choudhary}}}, \bibinfo {author} {\bibfnamefont {Kishalay}\
  \bibnamefont {{De}}}, \bibinfo {author} {\bibfnamefont {L.}~\bibnamefont
  {{Dey}}}, \bibinfo {author} {\bibfnamefont {A.}~\bibnamefont {{Gopakumar}}},
  \bibinfo {author} {\bibfnamefont {Y.}~\bibnamefont {{Gupta}}}, \bibinfo
  {author} {\bibfnamefont {M.~A.}\ \bibnamefont {{Krishnakumar}}}, \bibinfo
  {author} {\bibfnamefont {Yogesh}\ \bibnamefont {{Maan}}}, \bibinfo {author}
  {\bibfnamefont {P.~K.}\ \bibnamefont {{Manoharan}}}, \bibinfo {author}
  {\bibfnamefont {Arun}\ \bibnamefont {{Naidu}}}, \bibinfo {author}
  {\bibfnamefont {Rana}\ \bibnamefont {{Nandi}}}, \bibinfo {author}
  {\bibfnamefont {Dhruv}\ \bibnamefont {{Pathak}}}, \bibinfo {author}
  {\bibfnamefont {Mayuresh}\ \bibnamefont {{Surnis}}}, \ and\ \bibinfo {author}
  {\bibfnamefont {Abhimanyu}\ \bibnamefont {{Susobhanan}}},\ }\bibfield
  {title} {\enquote {\bibinfo {title} {{Precision pulsar timing with the ORT
  and the GMRT and its applications in pulsar astrophysics}},}\ }\href
  {\doibase 10.1007/s12036-018-9549-y} {\bibfield  {journal} {\bibinfo
  {journal} {Journal of Astrophysics and Astronomy}\ }\textbf {\bibinfo
  {volume} {39}},\ \bibinfo {eid} {51} (\bibinfo {year} {2018})}\BibitemShut
  {NoStop}%
\bibitem [{\citenamefont {{Bailes}}\ \emph {et~al.}(2020)\citenamefont
  {{Bailes}}, \citenamefont {{Jameson}}, \citenamefont {{Abbate}},
  \citenamefont {{Barr}}, \citenamefont {{Bhat}}, \citenamefont {{Bondonneau}},
  \citenamefont {{Burgay}}, \citenamefont {{Buchner}}, \citenamefont
  {{Camilo}}, \citenamefont {{Champion}}, \citenamefont {{Cognard}},
  \citenamefont {{Demorest}}, \citenamefont {{Freire}}, \citenamefont
  {{Gautam}}, \citenamefont {{Geyer}}, \citenamefont {{Griessmeier}},
  \citenamefont {{Guillemot}}, \citenamefont {{Hu}}, \citenamefont
  {{Jankowski}}, \citenamefont {{Johnston}}, \citenamefont {{Karastergiou}},
  \citenamefont {{Karuppusamy}}, \citenamefont {{Kaur}}, \citenamefont
  {{Keith}}, \citenamefont {{Kramer}}, \citenamefont {{van Leeuwen}},
  \citenamefont {{Lower}}, \citenamefont {{Maan}}, \citenamefont
  {{McLaughlin}}, \citenamefont {{Meyers}}, \citenamefont {{Os{\l}owski}},
  \citenamefont {{Oswald}}, \citenamefont {{Parthasarathy}}, \citenamefont
  {{Pennucci}}, \citenamefont {{Posselt}}, \citenamefont {{Possenti}},
  \citenamefont {{Ransom}}, \citenamefont {{Reardon}}, \citenamefont
  {{Ridolfi}}, \citenamefont {{Schollar}}, \citenamefont {{Serylak}},
  \citenamefont {{Shaifullah}}, \citenamefont {{Shamohammadi}}, \citenamefont
  {{Shannon}}, \citenamefont {{Sobey}}, \citenamefont {{Song}}, \citenamefont
  {{Spiewak}}, \citenamefont {{Stairs}}, \citenamefont {{Stappers}},
  \citenamefont {{van Straten}}, \citenamefont {{Szary}}, \citenamefont
  {{Theureau}}, \citenamefont {{Venkatraman Krishnan}}, \citenamefont
  {{Weltevrede}}, \citenamefont {{Wex}}, \citenamefont {{Abbott}},
  \citenamefont {{Adams}}, \citenamefont {{Burger}}, \citenamefont
  {{Gamatham}}, \citenamefont {{Gouws}}, \citenamefont {{Horn}}, \citenamefont
  {{Hugo}}, \citenamefont {{Joubert}}, \citenamefont {{Manley}}, \citenamefont
  {{McAlpine}}, \citenamefont {{Passmoor}}, \citenamefont {{Peens-Hough}},
  \citenamefont {{Ramudzuli}}, \citenamefont {{Rust}}, \citenamefont {{Salie}},
  \citenamefont {{Schwardt}}, \citenamefont {{Siebrits}}, \citenamefont {{Van
  Tonder}}, \citenamefont {{Van Tonder}},\ and\ \citenamefont
  {{Welz}}}]{MeerTime}%
  \BibitemOpen
  \bibfield  {author} {\bibinfo {author} {\bibfnamefont {M.}~\bibnamefont
  {{Bailes}}}, \bibinfo {author} {\bibfnamefont {A.}~\bibnamefont {{Jameson}}},
  \bibinfo {author} {\bibfnamefont {F.}~\bibnamefont {{Abbate}}}, \bibinfo
  {author} {\bibfnamefont {E.~D.}\ \bibnamefont {{Barr}}}, \bibinfo {author}
  {\bibfnamefont {N.~D.~R.}\ \bibnamefont {{Bhat}}}, \bibinfo {author}
  {\bibfnamefont {L.}~\bibnamefont {{Bondonneau}}}, \bibinfo {author}
  {\bibfnamefont {M.}~\bibnamefont {{Burgay}}}, \bibinfo {author}
  {\bibfnamefont {S.~J.}\ \bibnamefont {{Buchner}}}, \bibinfo {author}
  {\bibfnamefont {F.}~\bibnamefont {{Camilo}}}, \bibinfo {author}
  {\bibfnamefont {D.~J.}\ \bibnamefont {{Champion}}}, \bibinfo {author}
  {\bibfnamefont {I.}~\bibnamefont {{Cognard}}}, \bibinfo {author}
  {\bibfnamefont {P.~B.}\ \bibnamefont {{Demorest}}}, \bibinfo {author}
  {\bibfnamefont {P.~C.~C.}\ \bibnamefont {{Freire}}}, \bibinfo {author}
  {\bibfnamefont {T.}~\bibnamefont {{Gautam}}}, \bibinfo {author}
  {\bibfnamefont {M.}~\bibnamefont {{Geyer}}}, \bibinfo {author} {\bibfnamefont
  {J.~M.}\ \bibnamefont {{Griessmeier}}}, \bibinfo {author} {\bibfnamefont
  {L.}~\bibnamefont {{Guillemot}}}, \bibinfo {author} {\bibfnamefont
  {H.}~\bibnamefont {{Hu}}}, \bibinfo {author} {\bibfnamefont {F.}~\bibnamefont
  {{Jankowski}}}, \bibinfo {author} {\bibfnamefont {S.}~\bibnamefont
  {{Johnston}}}, \bibinfo {author} {\bibfnamefont {A.}~\bibnamefont
  {{Karastergiou}}}, \bibinfo {author} {\bibfnamefont {R.}~\bibnamefont
  {{Karuppusamy}}}, \bibinfo {author} {\bibfnamefont {D.}~\bibnamefont
  {{Kaur}}}, \bibinfo {author} {\bibfnamefont {M.~J.}\ \bibnamefont {{Keith}}},
  \bibinfo {author} {\bibfnamefont {M.}~\bibnamefont {{Kramer}}}, \bibinfo
  {author} {\bibfnamefont {J.}~\bibnamefont {{van Leeuwen}}}, \bibinfo {author}
  {\bibfnamefont {M.~E.}\ \bibnamefont {{Lower}}}, \bibinfo {author}
  {\bibfnamefont {Y.}~\bibnamefont {{Maan}}}, \bibinfo {author} {\bibfnamefont
  {M.~A.}\ \bibnamefont {{McLaughlin}}}, \bibinfo {author} {\bibfnamefont
  {B.~W.}\ \bibnamefont {{Meyers}}}, \bibinfo {author} {\bibfnamefont
  {S.}~\bibnamefont {{Os{\l}owski}}}, \bibinfo {author} {\bibfnamefont {L.~S.}\
  \bibnamefont {{Oswald}}}, \bibinfo {author} {\bibfnamefont {A.}~\bibnamefont
  {{Parthasarathy}}}, \bibinfo {author} {\bibfnamefont {T.}~\bibnamefont
  {{Pennucci}}}, \bibinfo {author} {\bibfnamefont {B.}~\bibnamefont
  {{Posselt}}}, \bibinfo {author} {\bibfnamefont {A.}~\bibnamefont
  {{Possenti}}}, \bibinfo {author} {\bibfnamefont {S.~M.}\ \bibnamefont
  {{Ransom}}}, \bibinfo {author} {\bibfnamefont {D.~J.}\ \bibnamefont
  {{Reardon}}}, \bibinfo {author} {\bibfnamefont {A.}~\bibnamefont
  {{Ridolfi}}}, \bibinfo {author} {\bibfnamefont {C.~T.~G.}\ \bibnamefont
  {{Schollar}}}, \bibinfo {author} {\bibfnamefont {M.}~\bibnamefont
  {{Serylak}}}, \bibinfo {author} {\bibfnamefont {G.}~\bibnamefont
  {{Shaifullah}}}, \bibinfo {author} {\bibfnamefont {M.}~\bibnamefont
  {{Shamohammadi}}}, \bibinfo {author} {\bibfnamefont {R.~M.}\ \bibnamefont
  {{Shannon}}}, \bibinfo {author} {\bibfnamefont {C.}~\bibnamefont {{Sobey}}},
  \bibinfo {author} {\bibfnamefont {X.}~\bibnamefont {{Song}}}, \bibinfo
  {author} {\bibfnamefont {R.}~\bibnamefont {{Spiewak}}}, \bibinfo {author}
  {\bibfnamefont {I.~H.}\ \bibnamefont {{Stairs}}}, \bibinfo {author}
  {\bibfnamefont {B.~W.}\ \bibnamefont {{Stappers}}}, \bibinfo {author}
  {\bibfnamefont {W.}~\bibnamefont {{van Straten}}}, \bibinfo {author}
  {\bibfnamefont {A.}~\bibnamefont {{Szary}}}, \bibinfo {author} {\bibfnamefont
  {G.}~\bibnamefont {{Theureau}}}, \bibinfo {author} {\bibfnamefont
  {V.}~\bibnamefont {{Venkatraman Krishnan}}}, \bibinfo {author} {\bibfnamefont
  {P.}~\bibnamefont {{Weltevrede}}}, \bibinfo {author} {\bibfnamefont
  {N.}~\bibnamefont {{Wex}}}, \bibinfo {author} {\bibfnamefont {T.~D.}\
  \bibnamefont {{Abbott}}}, \bibinfo {author} {\bibfnamefont {G.~B.}\
  \bibnamefont {{Adams}}}, \bibinfo {author} {\bibfnamefont {J.~P.}\
  \bibnamefont {{Burger}}}, \bibinfo {author} {\bibfnamefont {R.~R.~G.}\
  \bibnamefont {{Gamatham}}}, \bibinfo {author} {\bibfnamefont
  {M.}~\bibnamefont {{Gouws}}}, \bibinfo {author} {\bibfnamefont {D.~M.}\
  \bibnamefont {{Horn}}}, \bibinfo {author} {\bibfnamefont {B.}~\bibnamefont
  {{Hugo}}}, \bibinfo {author} {\bibfnamefont {A.~F.}\ \bibnamefont
  {{Joubert}}}, \bibinfo {author} {\bibfnamefont {J.~R.}\ \bibnamefont
  {{Manley}}}, \bibinfo {author} {\bibfnamefont {K.}~\bibnamefont
  {{McAlpine}}}, \bibinfo {author} {\bibfnamefont {S.~S.}\ \bibnamefont
  {{Passmoor}}}, \bibinfo {author} {\bibfnamefont {A.}~\bibnamefont
  {{Peens-Hough}}}, \bibinfo {author} {\bibfnamefont {Z.~R.}\ \bibnamefont
  {{Ramudzuli}}}, \bibinfo {author} {\bibfnamefont {A.}~\bibnamefont {{Rust}}},
  \bibinfo {author} {\bibfnamefont {S.}~\bibnamefont {{Salie}}}, \bibinfo
  {author} {\bibfnamefont {L.~C.}\ \bibnamefont {{Schwardt}}}, \bibinfo
  {author} {\bibfnamefont {R.}~\bibnamefont {{Siebrits}}}, \bibinfo {author}
  {\bibfnamefont {G.}~\bibnamefont {{Van Tonder}}}, \bibinfo {author}
  {\bibfnamefont {V.}~\bibnamefont {{Van Tonder}}}, \ and\ \bibinfo {author}
  {\bibfnamefont {M.~G.}\ \bibnamefont {{Welz}}},\ }\bibfield  {title}
  {\enquote {\bibinfo {title} {{The MeerKAT telescope as a pulsar facility:
  System verification and early science results from MeerTime}},}\ }\href
  {\doibase 10.1017/pasa.2020.19} {\bibfield  {journal} {\bibinfo  {journal}
  {\pasa}\ }\textbf {\bibinfo {volume} {37}},\ \bibinfo {eid} {e028} (\bibinfo
  {year} {2020})},\ \Eprint {http://arxiv.org/abs/2005.14366} {arXiv:2005.14366
  [astro-ph.IM]} \BibitemShut {NoStop}%
\bibitem [{\citenamefont {{EPTA Collaboration}}\ \emph
  {et~al.}(2023{\natexlab{a}})\citenamefont {{EPTA Collaboration}},
  \citenamefont {Smarra}, \citenamefont {Goncharov}, \citenamefont {Barausse},
  \citenamefont {Antoniadis}, \citenamefont {Babak}, \citenamefont {Nielsen},
  \citenamefont {Bassa}, \citenamefont {Berthereau}, \citenamefont {Bonetti},
  \citenamefont {Bortolas}, \citenamefont {Brook}, \citenamefont {Burgay},
  \citenamefont {Caballero}, \citenamefont {Chalumeau}, \citenamefont
  {Champion}, \citenamefont {Chanlaridis}, \citenamefont {Chen}, \citenamefont
  {Cognard}, \citenamefont {Desvignes}, \citenamefont {Falxa}, \citenamefont
  {Ferdman}, \citenamefont {Franchini}, \citenamefont {Gair}, \citenamefont
  {Graikou}, \citenamefont {Grie\ss{}meier}, \citenamefont {Guillemot},
  \citenamefont {Guo}, \citenamefont {Hu}, \citenamefont {Iraci}, \citenamefont
  {Izquierdo-Villalba}, \citenamefont {Jang}, \citenamefont {Jawor},
  \citenamefont {Janssen}, \citenamefont {Jessner}, \citenamefont
  {Karuppusamy}, \citenamefont {Keane}, \citenamefont {Keith}, \citenamefont
  {Kramer}, \citenamefont {Krishnakumar}, \citenamefont {Lackeos},
  \citenamefont {Lee}, \citenamefont {Liu}, \citenamefont {Liu}, \citenamefont
  {Lyne}, \citenamefont {McKee}, \citenamefont {Main}, \citenamefont
  {Mickaliger}, \citenamefont {Ni\ifmmode~\mbox{\c{t}}\else \c{t}\fi{}u},
  \citenamefont {Parthasarathy}, \citenamefont {Perera}, \citenamefont
  {Perrodin}, \citenamefont {Petiteau}, \citenamefont {Porayko}, \citenamefont
  {Possenti}, \citenamefont {Leclere}, \citenamefont {Samajdar}, \citenamefont
  {Sanidas}, \citenamefont {Sesana}, \citenamefont {Shaifullah}, \citenamefont
  {Speri}, \citenamefont {Spiewak}, \citenamefont {Stappers}, \citenamefont
  {Susarla}, \citenamefont {Theureau}, \citenamefont {Tiburzi}, \citenamefont
  {van~der Wateren}, \citenamefont {Vecchio}, \citenamefont {Krishnan},
  \citenamefont {Wang}, \citenamefont {Wang},\ and\ \citenamefont
  {Wu}}]{Smarra_2023}%
  \BibitemOpen
  \bibfield  {author} {\bibinfo {author} {\bibnamefont {{EPTA Collaboration}}},
  \bibinfo {author} {\bibfnamefont {Clemente}\ \bibnamefont {Smarra}}, \bibinfo
  {author} {\bibfnamefont {Boris}\ \bibnamefont {Goncharov}}, \bibinfo {author}
  {\bibfnamefont {Enrico}\ \bibnamefont {Barausse}}, \bibinfo {author}
  {\bibfnamefont {J.}~\bibnamefont {Antoniadis}}, \bibinfo {author}
  {\bibfnamefont {S.}~\bibnamefont {Babak}}, \bibinfo {author} {\bibfnamefont
  {A.-S.~Bak}\ \bibnamefont {Nielsen}}, \bibinfo {author} {\bibfnamefont
  {C.~G.}\ \bibnamefont {Bassa}}, \bibinfo {author} {\bibfnamefont
  {A.}~\bibnamefont {Berthereau}}, \bibinfo {author} {\bibfnamefont
  {M.}~\bibnamefont {Bonetti}}, \bibinfo {author} {\bibfnamefont
  {E.}~\bibnamefont {Bortolas}}, \bibinfo {author} {\bibfnamefont {P.~R.}\
  \bibnamefont {Brook}}, \bibinfo {author} {\bibfnamefont {M.}~\bibnamefont
  {Burgay}}, \bibinfo {author} {\bibfnamefont {R.~N.}\ \bibnamefont
  {Caballero}}, \bibinfo {author} {\bibfnamefont {A.}~\bibnamefont
  {Chalumeau}}, \bibinfo {author} {\bibfnamefont {D.~J.}\ \bibnamefont
  {Champion}}, \bibinfo {author} {\bibfnamefont {S.}~\bibnamefont
  {Chanlaridis}}, \bibinfo {author} {\bibfnamefont {S.}~\bibnamefont {Chen}},
  \bibinfo {author} {\bibfnamefont {I.}~\bibnamefont {Cognard}}, \bibinfo
  {author} {\bibfnamefont {G.}~\bibnamefont {Desvignes}}, \bibinfo {author}
  {\bibfnamefont {M.}~\bibnamefont {Falxa}}, \bibinfo {author} {\bibfnamefont
  {R.~D.}\ \bibnamefont {Ferdman}}, \bibinfo {author} {\bibfnamefont
  {A.}~\bibnamefont {Franchini}}, \bibinfo {author} {\bibfnamefont {J.~R.}\
  \bibnamefont {Gair}}, \bibinfo {author} {\bibfnamefont {E.}~\bibnamefont
  {Graikou}}, \bibinfo {author} {\bibfnamefont {J.-M.}\ \bibnamefont
  {Grie\ss{}meier}}, \bibinfo {author} {\bibfnamefont {L.}~\bibnamefont
  {Guillemot}}, \bibinfo {author} {\bibfnamefont {Y.~J.}\ \bibnamefont {Guo}},
  \bibinfo {author} {\bibfnamefont {H.}~\bibnamefont {Hu}}, \bibinfo {author}
  {\bibfnamefont {F.}~\bibnamefont {Iraci}}, \bibinfo {author} {\bibfnamefont
  {D.}~\bibnamefont {Izquierdo-Villalba}}, \bibinfo {author} {\bibfnamefont
  {J.}~\bibnamefont {Jang}}, \bibinfo {author} {\bibfnamefont {J.}~\bibnamefont
  {Jawor}}, \bibinfo {author} {\bibfnamefont {G.~H.}\ \bibnamefont {Janssen}},
  \bibinfo {author} {\bibfnamefont {A.}~\bibnamefont {Jessner}}, \bibinfo
  {author} {\bibfnamefont {R.}~\bibnamefont {Karuppusamy}}, \bibinfo {author}
  {\bibfnamefont {E.~F.}\ \bibnamefont {Keane}}, \bibinfo {author}
  {\bibfnamefont {M.~J.}\ \bibnamefont {Keith}}, \bibinfo {author}
  {\bibfnamefont {M.}~\bibnamefont {Kramer}}, \bibinfo {author} {\bibfnamefont
  {M.~A.}\ \bibnamefont {Krishnakumar}}, \bibinfo {author} {\bibfnamefont
  {K.}~\bibnamefont {Lackeos}}, \bibinfo {author} {\bibfnamefont {K.~J.}\
  \bibnamefont {Lee}}, \bibinfo {author} {\bibfnamefont {K.}~\bibnamefont
  {Liu}}, \bibinfo {author} {\bibfnamefont {Y.}~\bibnamefont {Liu}}, \bibinfo
  {author} {\bibfnamefont {A.~G.}\ \bibnamefont {Lyne}}, \bibinfo {author}
  {\bibfnamefont {J.~W.}\ \bibnamefont {McKee}}, \bibinfo {author}
  {\bibfnamefont {R.~A.}\ \bibnamefont {Main}}, \bibinfo {author}
  {\bibfnamefont {M.~B.}\ \bibnamefont {Mickaliger}}, \bibinfo {author}
  {\bibfnamefont {I.~C.}\ \bibnamefont {Ni\ifmmode~\mbox{\c{t}}\else
  \c{t}\fi{}u}}, \bibinfo {author} {\bibfnamefont {A.}~\bibnamefont
  {Parthasarathy}}, \bibinfo {author} {\bibfnamefont {B.~B.~P.}\ \bibnamefont
  {Perera}}, \bibinfo {author} {\bibfnamefont {D.}~\bibnamefont {Perrodin}},
  \bibinfo {author} {\bibfnamefont {A.}~\bibnamefont {Petiteau}}, \bibinfo
  {author} {\bibfnamefont {N.~K.}\ \bibnamefont {Porayko}}, \bibinfo {author}
  {\bibfnamefont {A.}~\bibnamefont {Possenti}}, \bibinfo {author}
  {\bibfnamefont {H.~Quelquejay}\ \bibnamefont {Leclere}}, \bibinfo {author}
  {\bibfnamefont {A.}~\bibnamefont {Samajdar}}, \bibinfo {author}
  {\bibfnamefont {S.~A.}\ \bibnamefont {Sanidas}}, \bibinfo {author}
  {\bibfnamefont {A.}~\bibnamefont {Sesana}}, \bibinfo {author} {\bibfnamefont
  {G.}~\bibnamefont {Shaifullah}}, \bibinfo {author} {\bibfnamefont
  {L.}~\bibnamefont {Speri}}, \bibinfo {author} {\bibfnamefont
  {R.}~\bibnamefont {Spiewak}}, \bibinfo {author} {\bibfnamefont {B.~W.}\
  \bibnamefont {Stappers}}, \bibinfo {author} {\bibfnamefont {S.~C.}\
  \bibnamefont {Susarla}}, \bibinfo {author} {\bibfnamefont {G.}~\bibnamefont
  {Theureau}}, \bibinfo {author} {\bibfnamefont {C.}~\bibnamefont {Tiburzi}},
  \bibinfo {author} {\bibfnamefont {E.}~\bibnamefont {van~der Wateren}},
  \bibinfo {author} {\bibfnamefont {A.}~\bibnamefont {Vecchio}}, \bibinfo
  {author} {\bibfnamefont {V.~Venkatraman}\ \bibnamefont {Krishnan}}, \bibinfo
  {author} {\bibfnamefont {J.}~\bibnamefont {Wang}}, \bibinfo {author}
  {\bibfnamefont {L.}~\bibnamefont {Wang}}, \ and\ \bibinfo {author}
  {\bibfnamefont {Z.}~\bibnamefont {Wu}},\ }\bibfield  {title} {\enquote
  {\bibinfo {title} {Second data release from the european pulsar timing array:
  Challenging the ultralight dark matter paradigm},}\ }\href {\doibase
  10.1103/PhysRevLett.131.171001} {\bibfield  {journal} {\bibinfo  {journal}
  {Phys. Rev. Lett.}\ }\textbf {\bibinfo {volume} {131}},\ \bibinfo {pages}
  {171001} (\bibinfo {year} {2023}{\natexlab{a}})}\BibitemShut {NoStop}%
\bibitem [{\citenamefont {Porayko}\ \emph {et~al.}(2018)\citenamefont
  {Porayko}, \citenamefont {Zhu}, \citenamefont {Levin}, \citenamefont {Hui},
  \citenamefont {Hobbs}, \citenamefont {Grudskaya}, \citenamefont {Postnov},
  \citenamefont {Bailes}, \citenamefont {Bhat}, \citenamefont {Coles},
  \citenamefont {Dai}, \citenamefont {Dempsey}, \citenamefont {Keith},
  \citenamefont {Kerr}, \citenamefont {Kramer}, \citenamefont {Lasky},
  \citenamefont {Manchester}, \citenamefont {Osłowski}, \citenamefont
  {Parthasarathy}, \citenamefont {Ravi}, \citenamefont {Reardon}, \citenamefont
  {Rosado}, \citenamefont {Russell}, \citenamefont {Shannon}, \citenamefont
  {Spiewak}, \citenamefont {van Straten}, \citenamefont {Toomey}, \citenamefont
  {Wang}, \citenamefont {Wen},\ and\ \citenamefont {You}}]{Porayko_2018}%
  \BibitemOpen
  \bibfield  {author} {\bibinfo {author} {\bibfnamefont {Nataliya~K.}\
  \bibnamefont {Porayko}}, \bibinfo {author} {\bibfnamefont {Xingjiang}\
  \bibnamefont {Zhu}}, \bibinfo {author} {\bibfnamefont {Yuri}\ \bibnamefont
  {Levin}}, \bibinfo {author} {\bibfnamefont {Lam}\ \bibnamefont {Hui}},
  \bibinfo {author} {\bibfnamefont {George}\ \bibnamefont {Hobbs}}, \bibinfo
  {author} {\bibfnamefont {Aleksandra}\ \bibnamefont {Grudskaya}}, \bibinfo
  {author} {\bibfnamefont {Konstantin}\ \bibnamefont {Postnov}}, \bibinfo
  {author} {\bibfnamefont {Matthew}\ \bibnamefont {Bailes}}, \bibinfo {author}
  {\bibfnamefont {N.D.~Ramesh}\ \bibnamefont {Bhat}}, \bibinfo {author}
  {\bibfnamefont {William}\ \bibnamefont {Coles}}, \bibinfo {author}
  {\bibfnamefont {Shi}\ \bibnamefont {Dai}}, \bibinfo {author} {\bibfnamefont
  {James}\ \bibnamefont {Dempsey}}, \bibinfo {author} {\bibfnamefont
  {Michael~J.}\ \bibnamefont {Keith}}, \bibinfo {author} {\bibfnamefont
  {Matthew}\ \bibnamefont {Kerr}}, \bibinfo {author} {\bibfnamefont {Michael}\
  \bibnamefont {Kramer}}, \bibinfo {author} {\bibfnamefont {Paul~D.}\
  \bibnamefont {Lasky}}, \bibinfo {author} {\bibfnamefont {Richard~N.}\
  \bibnamefont {Manchester}}, \bibinfo {author} {\bibfnamefont {Stefan}\
  \bibnamefont {Osłowski}}, \bibinfo {author} {\bibfnamefont {Aditya}\
  \bibnamefont {Parthasarathy}}, \bibinfo {author} {\bibfnamefont {Vikram}\
  \bibnamefont {Ravi}}, \bibinfo {author} {\bibfnamefont {Daniel~J.}\
  \bibnamefont {Reardon}}, \bibinfo {author} {\bibfnamefont {Pablo~A.}\
  \bibnamefont {Rosado}}, \bibinfo {author} {\bibfnamefont {Christopher~J.}\
  \bibnamefont {Russell}}, \bibinfo {author} {\bibfnamefont {Ryan~M.}\
  \bibnamefont {Shannon}}, \bibinfo {author} {\bibfnamefont {Renée}\
  \bibnamefont {Spiewak}}, \bibinfo {author} {\bibfnamefont {Willem}\
  \bibnamefont {van Straten}}, \bibinfo {author} {\bibfnamefont {Lawrence}\
  \bibnamefont {Toomey}}, \bibinfo {author} {\bibfnamefont {Jingbo}\
  \bibnamefont {Wang}}, \bibinfo {author} {\bibfnamefont {Linqing}\
  \bibnamefont {Wen}}, \ and\ \bibinfo {author} {\bibfnamefont {Xiaopeng}\
  \bibnamefont {You}},\ }\bibfield  {title} {\enquote {\bibinfo {title} {Parkes
  pulsar timing array constraints on ultralight scalar-field dark matter},}\
  }\href {\doibase 10.1103/physrevd.98.102002} {\bibfield  {journal} {\bibinfo
  {journal} {Physical Review D}\ }\textbf {\bibinfo {volume} {98}} (\bibinfo
  {year} {2018}),\ 10.1103/physrevd.98.102002}\BibitemShut {NoStop}%
\bibitem [{\citenamefont {Kaplan}\ \emph {et~al.}(2022)\citenamefont {Kaplan},
  \citenamefont {Mitridate},\ and\ \citenamefont {Trickle}}]{Kaplan_2022}%
  \BibitemOpen
  \bibfield  {author} {\bibinfo {author} {\bibfnamefont {David~E.}\
  \bibnamefont {Kaplan}}, \bibinfo {author} {\bibfnamefont {Andrea}\
  \bibnamefont {Mitridate}}, \ and\ \bibinfo {author} {\bibfnamefont {Tanner}\
  \bibnamefont {Trickle}},\ }\bibfield  {title} {\enquote {\bibinfo {title}
  {Constraining fundamental constant variations from ultralight dark matter
  with pulsar timing arrays},}\ }\href {\doibase 10.1103/PhysRevD.106.035032}
  {\bibfield  {journal} {\bibinfo  {journal} {Phys. Rev. D}\ }\textbf {\bibinfo
  {volume} {106}},\ \bibinfo {pages} {035032} (\bibinfo {year}
  {2022})}\BibitemShut {NoStop}%
\bibitem [{\citenamefont {Afzal}\ \emph {et~al.}(2023)\citenamefont {Afzal},
  \citenamefont {Agazie}, \citenamefont {Anumarlapudi}, \citenamefont
  {Archibald}, \citenamefont {Arzoumanian}, \citenamefont {Baker},
  \citenamefont {Bécsy}, \citenamefont {Blanco-Pillado}, \citenamefont
  {Blecha}, \citenamefont {Boddy}, \citenamefont {Brazier}, \citenamefont
  {Brook}, \citenamefont {Burke-Spolaor}, \citenamefont {Burnette},
  \citenamefont {Case}, \citenamefont {Charisi}, \citenamefont {Chatterjee},
  \citenamefont {Chatziioannou}, \citenamefont {Cheeseboro}, \citenamefont
  {Chen}, \citenamefont {Cohen}, \citenamefont {Cordes}, \citenamefont
  {Cornish}, \citenamefont {Crawford}, \citenamefont {Cromartie}, \citenamefont
  {Crowter}, \citenamefont {Cutler}, \citenamefont {DeCesar}, \citenamefont
  {DeGan}, \citenamefont {Demorest}, \citenamefont {Deng}, \citenamefont
  {Dolch}, \citenamefont {Drachler}, \citenamefont {von Eckardstein},
  \citenamefont {Ferrara}, \citenamefont {Fiore}, \citenamefont {Fonseca},
  \citenamefont {Freedman}, \citenamefont {Garver-Daniels}, \citenamefont
  {Gentile}, \citenamefont {Gersbach}, \citenamefont {Glaser}, \citenamefont
  {Good}, \citenamefont {Guertin}, \citenamefont {Gültekin}, \citenamefont
  {Hazboun}, \citenamefont {Hourihane}, \citenamefont {Islo}, \citenamefont
  {Jennings}, \citenamefont {Johnson}, \citenamefont {Jones}, \citenamefont
  {Kaiser}, \citenamefont {Kaplan}, \citenamefont {Kelley}, \citenamefont
  {Kerr}, \citenamefont {Key}, \citenamefont {Laal}, \citenamefont {Lam},
  \citenamefont {Lamb}, \citenamefont {Lazio}, \citenamefont {Lee},
  \citenamefont {Lewandowska}, \citenamefont {dos Santos}, \citenamefont
  {Littenberg}, \citenamefont {Liu}, \citenamefont {Lorimer}, \citenamefont
  {Luo}, \citenamefont {Lynch}, \citenamefont {Ma}, \citenamefont {Madison},
  \citenamefont {McEwen}, \citenamefont {McKee}, \citenamefont {McLaughlin},
  \citenamefont {McMann}, \citenamefont {Meyers}, \citenamefont {Meyers},
  \citenamefont {Mingarelli}, \citenamefont {Mitridate}, \citenamefont {Nay},
  \citenamefont {Natarajan}, \citenamefont {Ng}, \citenamefont {Nice},
  \citenamefont {Ocker}, \citenamefont {Olum}, \citenamefont {Pennucci},
  \citenamefont {Perera}, \citenamefont {Petrov}, \citenamefont {Pol},
  \citenamefont {Radovan}, \citenamefont {Ransom}, \citenamefont {Ray},
  \citenamefont {Romano}, \citenamefont {Sardesai}, \citenamefont
  {Schmiedekamp}, \citenamefont {Schmiedekamp}, \citenamefont {Schmitz},
  \citenamefont {Schröder}, \citenamefont {Schult}, \citenamefont
  {Shapiro-Albert}, \citenamefont {Siemens}, \citenamefont {Simon},
  \citenamefont {Siwek}, \citenamefont {Stairs}, \citenamefont {Stinebring},
  \citenamefont {Stovall}, \citenamefont {Stratmann}, \citenamefont {Sun},
  \citenamefont {Susobhanan}, \citenamefont {Swiggum}, \citenamefont {Taylor},
  \citenamefont {Taylor}, \citenamefont {Trickle}, \citenamefont {Turner},
  \citenamefont {Unal}, \citenamefont {Vallisneri}, \citenamefont {Verma},
  \citenamefont {Vigeland}, \citenamefont {Wahl}, \citenamefont {Wang},
  \citenamefont {Witt}, \citenamefont {Wright}, \citenamefont {Young},
  \citenamefont {Zurek},\ and\ \citenamefont {Collaboration}}]{Afzal_2023}%
  \BibitemOpen
  \bibfield  {author} {\bibinfo {author} {\bibfnamefont {Adeela}\ \bibnamefont
  {Afzal}}, \bibinfo {author} {\bibfnamefont {Gabriella}\ \bibnamefont
  {Agazie}}, \bibinfo {author} {\bibfnamefont {Akash}\ \bibnamefont
  {Anumarlapudi}}, \bibinfo {author} {\bibfnamefont {Anne~M.}\ \bibnamefont
  {Archibald}}, \bibinfo {author} {\bibfnamefont {Zaven}\ \bibnamefont
  {Arzoumanian}}, \bibinfo {author} {\bibfnamefont {Paul~T.}\ \bibnamefont
  {Baker}}, \bibinfo {author} {\bibfnamefont {Bence}\ \bibnamefont {Bécsy}},
  \bibinfo {author} {\bibfnamefont {Jose~Juan}\ \bibnamefont {Blanco-Pillado}},
  \bibinfo {author} {\bibfnamefont {Laura}\ \bibnamefont {Blecha}}, \bibinfo
  {author} {\bibfnamefont {Kimberly~K.}\ \bibnamefont {Boddy}}, \bibinfo
  {author} {\bibfnamefont {Adam}\ \bibnamefont {Brazier}}, \bibinfo {author}
  {\bibfnamefont {Paul~R.}\ \bibnamefont {Brook}}, \bibinfo {author}
  {\bibfnamefont {Sarah}\ \bibnamefont {Burke-Spolaor}}, \bibinfo {author}
  {\bibfnamefont {Rand}\ \bibnamefont {Burnette}}, \bibinfo {author}
  {\bibfnamefont {Robin}\ \bibnamefont {Case}}, \bibinfo {author}
  {\bibfnamefont {Maria}\ \bibnamefont {Charisi}}, \bibinfo {author}
  {\bibfnamefont {Shami}\ \bibnamefont {Chatterjee}}, \bibinfo {author}
  {\bibfnamefont {Katerina}\ \bibnamefont {Chatziioannou}}, \bibinfo {author}
  {\bibfnamefont {Belinda~D.}\ \bibnamefont {Cheeseboro}}, \bibinfo {author}
  {\bibfnamefont {Siyuan}\ \bibnamefont {Chen}}, \bibinfo {author}
  {\bibfnamefont {Tyler}\ \bibnamefont {Cohen}}, \bibinfo {author}
  {\bibfnamefont {James~M.}\ \bibnamefont {Cordes}}, \bibinfo {author}
  {\bibfnamefont {Neil~J.}\ \bibnamefont {Cornish}}, \bibinfo {author}
  {\bibfnamefont {Fronefield}\ \bibnamefont {Crawford}}, \bibinfo {author}
  {\bibfnamefont {H.~Thankful}\ \bibnamefont {Cromartie}}, \bibinfo {author}
  {\bibfnamefont {Kathryn}\ \bibnamefont {Crowter}}, \bibinfo {author}
  {\bibfnamefont {Curt~J.}\ \bibnamefont {Cutler}}, \bibinfo {author}
  {\bibfnamefont {Megan~E.}\ \bibnamefont {DeCesar}}, \bibinfo {author}
  {\bibfnamefont {Dallas}\ \bibnamefont {DeGan}}, \bibinfo {author}
  {\bibfnamefont {Paul~B.}\ \bibnamefont {Demorest}}, \bibinfo {author}
  {\bibfnamefont {Heling}\ \bibnamefont {Deng}}, \bibinfo {author}
  {\bibfnamefont {Timothy}\ \bibnamefont {Dolch}}, \bibinfo {author}
  {\bibfnamefont {Brendan}\ \bibnamefont {Drachler}}, \bibinfo {author}
  {\bibfnamefont {Richard}\ \bibnamefont {von Eckardstein}}, \bibinfo {author}
  {\bibfnamefont {Elizabeth~C.}\ \bibnamefont {Ferrara}}, \bibinfo {author}
  {\bibfnamefont {William}\ \bibnamefont {Fiore}}, \bibinfo {author}
  {\bibfnamefont {Emmanuel}\ \bibnamefont {Fonseca}}, \bibinfo {author}
  {\bibfnamefont {Gabriel~E.}\ \bibnamefont {Freedman}}, \bibinfo {author}
  {\bibfnamefont {Nate}\ \bibnamefont {Garver-Daniels}}, \bibinfo {author}
  {\bibfnamefont {Peter~A.}\ \bibnamefont {Gentile}}, \bibinfo {author}
  {\bibfnamefont {Kyle~A.}\ \bibnamefont {Gersbach}}, \bibinfo {author}
  {\bibfnamefont {Joseph}\ \bibnamefont {Glaser}}, \bibinfo {author}
  {\bibfnamefont {Deborah~C.}\ \bibnamefont {Good}}, \bibinfo {author}
  {\bibfnamefont {Lydia}\ \bibnamefont {Guertin}}, \bibinfo {author}
  {\bibfnamefont {Kayhan}\ \bibnamefont {Gültekin}}, \bibinfo {author}
  {\bibfnamefont {Jeffrey~S.}\ \bibnamefont {Hazboun}}, \bibinfo {author}
  {\bibfnamefont {Sophie}\ \bibnamefont {Hourihane}}, \bibinfo {author}
  {\bibfnamefont {Kristina}\ \bibnamefont {Islo}}, \bibinfo {author}
  {\bibfnamefont {Ross~J.}\ \bibnamefont {Jennings}}, \bibinfo {author}
  {\bibfnamefont {Aaron~D.}\ \bibnamefont {Johnson}}, \bibinfo {author}
  {\bibfnamefont {Megan~L.}\ \bibnamefont {Jones}}, \bibinfo {author}
  {\bibfnamefont {Andrew~R.}\ \bibnamefont {Kaiser}}, \bibinfo {author}
  {\bibfnamefont {David~L.}\ \bibnamefont {Kaplan}}, \bibinfo {author}
  {\bibfnamefont {Luke~Zoltan}\ \bibnamefont {Kelley}}, \bibinfo {author}
  {\bibfnamefont {Matthew}\ \bibnamefont {Kerr}}, \bibinfo {author}
  {\bibfnamefont {Joey~S.}\ \bibnamefont {Key}}, \bibinfo {author}
  {\bibfnamefont {Nima}\ \bibnamefont {Laal}}, \bibinfo {author} {\bibfnamefont
  {Michael~T.}\ \bibnamefont {Lam}}, \bibinfo {author} {\bibfnamefont
  {William~G.}\ \bibnamefont {Lamb}}, \bibinfo {author} {\bibfnamefont
  {T.~Joseph~W.}\ \bibnamefont {Lazio}}, \bibinfo {author} {\bibfnamefont
  {Vincent S.~H.}\ \bibnamefont {Lee}}, \bibinfo {author} {\bibfnamefont
  {Natalia}\ \bibnamefont {Lewandowska}}, \bibinfo {author} {\bibfnamefont
  {Rafael R.~Lino}\ \bibnamefont {dos Santos}}, \bibinfo {author}
  {\bibfnamefont {Tyson~B.}\ \bibnamefont {Littenberg}}, \bibinfo {author}
  {\bibfnamefont {Tingting}\ \bibnamefont {Liu}}, \bibinfo {author}
  {\bibfnamefont {Duncan~R.}\ \bibnamefont {Lorimer}}, \bibinfo {author}
  {\bibfnamefont {Jing}\ \bibnamefont {Luo}}, \bibinfo {author} {\bibfnamefont
  {Ryan~S.}\ \bibnamefont {Lynch}}, \bibinfo {author} {\bibfnamefont
  {Chung-Pei}\ \bibnamefont {Ma}}, \bibinfo {author} {\bibfnamefont
  {Dustin~R.}\ \bibnamefont {Madison}}, \bibinfo {author} {\bibfnamefont
  {Alexander}\ \bibnamefont {McEwen}}, \bibinfo {author} {\bibfnamefont
  {James~W.}\ \bibnamefont {McKee}}, \bibinfo {author} {\bibfnamefont
  {Maura~A.}\ \bibnamefont {McLaughlin}}, \bibinfo {author} {\bibfnamefont
  {Natasha}\ \bibnamefont {McMann}}, \bibinfo {author} {\bibfnamefont
  {Bradley~W.}\ \bibnamefont {Meyers}}, \bibinfo {author} {\bibfnamefont
  {Patrick~M.}\ \bibnamefont {Meyers}}, \bibinfo {author} {\bibfnamefont
  {Chiara M.~F.}\ \bibnamefont {Mingarelli}}, \bibinfo {author} {\bibfnamefont
  {Andrea}\ \bibnamefont {Mitridate}}, \bibinfo {author} {\bibfnamefont
  {Jonathan}\ \bibnamefont {Nay}}, \bibinfo {author} {\bibfnamefont
  {Priyamvada}\ \bibnamefont {Natarajan}}, \bibinfo {author} {\bibfnamefont
  {Cherry}\ \bibnamefont {Ng}}, \bibinfo {author} {\bibfnamefont {David~J.}\
  \bibnamefont {Nice}}, \bibinfo {author} {\bibfnamefont {Stella~Koch}\
  \bibnamefont {Ocker}}, \bibinfo {author} {\bibfnamefont {Ken~D.}\
  \bibnamefont {Olum}}, \bibinfo {author} {\bibfnamefont {Timothy~T.}\
  \bibnamefont {Pennucci}}, \bibinfo {author} {\bibfnamefont {Benetge B.~P.}\
  \bibnamefont {Perera}}, \bibinfo {author} {\bibfnamefont {Polina}\
  \bibnamefont {Petrov}}, \bibinfo {author} {\bibfnamefont {Nihan~S.}\
  \bibnamefont {Pol}}, \bibinfo {author} {\bibfnamefont {Henri~A.}\
  \bibnamefont {Radovan}}, \bibinfo {author} {\bibfnamefont {Scott~M.}\
  \bibnamefont {Ransom}}, \bibinfo {author} {\bibfnamefont {Paul~S.}\
  \bibnamefont {Ray}}, \bibinfo {author} {\bibfnamefont {Joseph~D.}\
  \bibnamefont {Romano}}, \bibinfo {author} {\bibfnamefont {Shashwat~C.}\
  \bibnamefont {Sardesai}}, \bibinfo {author} {\bibfnamefont {Ann}\
  \bibnamefont {Schmiedekamp}}, \bibinfo {author} {\bibfnamefont {Carl}\
  \bibnamefont {Schmiedekamp}}, \bibinfo {author} {\bibfnamefont {Kai}\
  \bibnamefont {Schmitz}}, \bibinfo {author} {\bibfnamefont {Tobias}\
  \bibnamefont {Schröder}}, \bibinfo {author} {\bibfnamefont {Levi}\
  \bibnamefont {Schult}}, \bibinfo {author} {\bibfnamefont {Brent~J.}\
  \bibnamefont {Shapiro-Albert}}, \bibinfo {author} {\bibfnamefont {Xavier}\
  \bibnamefont {Siemens}}, \bibinfo {author} {\bibfnamefont {Joseph}\
  \bibnamefont {Simon}}, \bibinfo {author} {\bibfnamefont {Magdalena~S.}\
  \bibnamefont {Siwek}}, \bibinfo {author} {\bibfnamefont {Ingrid~H.}\
  \bibnamefont {Stairs}}, \bibinfo {author} {\bibfnamefont {Daniel~R.}\
  \bibnamefont {Stinebring}}, \bibinfo {author} {\bibfnamefont {Kevin}\
  \bibnamefont {Stovall}}, \bibinfo {author} {\bibfnamefont {Peter}\
  \bibnamefont {Stratmann}}, \bibinfo {author} {\bibfnamefont {Jerry~P.}\
  \bibnamefont {Sun}}, \bibinfo {author} {\bibfnamefont {Abhimanyu}\
  \bibnamefont {Susobhanan}}, \bibinfo {author} {\bibfnamefont {Joseph~K.}\
  \bibnamefont {Swiggum}}, \bibinfo {author} {\bibfnamefont {Jacob}\
  \bibnamefont {Taylor}}, \bibinfo {author} {\bibfnamefont {Stephen~R.}\
  \bibnamefont {Taylor}}, \bibinfo {author} {\bibfnamefont {Tanner}\
  \bibnamefont {Trickle}}, \bibinfo {author} {\bibfnamefont {Jacob~E.}\
  \bibnamefont {Turner}}, \bibinfo {author} {\bibfnamefont {Caner}\
  \bibnamefont {Unal}}, \bibinfo {author} {\bibfnamefont {Michele}\
  \bibnamefont {Vallisneri}}, \bibinfo {author} {\bibfnamefont {Sonali}\
  \bibnamefont {Verma}}, \bibinfo {author} {\bibfnamefont {Sarah~J.}\
  \bibnamefont {Vigeland}}, \bibinfo {author} {\bibfnamefont {Haley~M.}\
  \bibnamefont {Wahl}}, \bibinfo {author} {\bibfnamefont {Qiaohong}\
  \bibnamefont {Wang}}, \bibinfo {author} {\bibfnamefont {Caitlin~A.}\
  \bibnamefont {Witt}}, \bibinfo {author} {\bibfnamefont {David}\ \bibnamefont
  {Wright}}, \bibinfo {author} {\bibfnamefont {Olivia}\ \bibnamefont {Young}},
  \bibinfo {author} {\bibfnamefont {Kathryn~M.}\ \bibnamefont {Zurek}}, \ and\
  \bibinfo {author} {\bibfnamefont {The~NANOGrav}\ \bibnamefont
  {Collaboration}},\ }\bibfield  {title} {\enquote {\bibinfo {title} {The
  nanograv 15 yr data set: Search for signals from new physics},}\ }\href
  {\doibase 10.3847/2041-8213/acdc91} {\bibfield  {journal} {\bibinfo
  {journal} {The Astrophysical Journal Letters}\ }\textbf {\bibinfo {volume}
  {951}},\ \bibinfo {pages} {L11} (\bibinfo {year} {2023})}\BibitemShut
  {NoStop}%
\bibitem [{\citenamefont {Blas}\ \emph {et~al.}(2017)\citenamefont {Blas},
  \citenamefont {Nacir},\ and\ \citenamefont {Sibiryakov}}]{Blas_2017}%
  \BibitemOpen
  \bibfield  {author} {\bibinfo {author} {\bibfnamefont {Diego}\ \bibnamefont
  {Blas}}, \bibinfo {author} {\bibfnamefont {Diana~L{\'{o} }pez}\ \bibnamefont
  {Nacir}}, \ and\ \bibinfo {author} {\bibfnamefont {Sergey}\ \bibnamefont
  {Sibiryakov}},\ }\bibfield  {title} {\enquote {\bibinfo {title} {Ultralight
  dark matter resonates with binary pulsars},}\ }\href {\doibase
  10.1103/physrevlett.118.261102} {\bibfield  {journal} {\bibinfo  {journal}
  {Physical Review Letters}\ }\textbf {\bibinfo {volume} {118}} (\bibinfo
  {year} {2017}),\ 10.1103/physrevlett.118.261102}\BibitemShut {NoStop}%
\bibitem [{\citenamefont {Blas}\ \emph {et~al.}(2020)\citenamefont {Blas},
  \citenamefont {Nacir},\ and\ \citenamefont {Sibiryakov}}]{Blas_2020}%
  \BibitemOpen
  \bibfield  {author} {\bibinfo {author} {\bibfnamefont {Diego}\ \bibnamefont
  {Blas}}, \bibinfo {author} {\bibfnamefont {Diana~L{\'{o} }pez}\ \bibnamefont
  {Nacir}}, \ and\ \bibinfo {author} {\bibfnamefont {Sergey}\ \bibnamefont
  {Sibiryakov}},\ }\bibfield  {title} {\enquote {\bibinfo {title} {Secular
  effects of ultralight dark matter on binary pulsars},}\ }\href {\doibase
  10.1103/physrevd.101.063016} {\bibfield  {journal} {\bibinfo  {journal}
  {Physical Review D}\ }\textbf {\bibinfo {volume} {101}} (\bibinfo {year}
  {2020}),\ 10.1103/physrevd.101.063016}\BibitemShut {NoStop}%
\bibitem [{\citenamefont {K\r{u}s}\ \emph {et~al.}(2024)\citenamefont
  {K\r{u}s}, \citenamefont {L\'opez~Nacir},\ and\ \citenamefont
  {Urban}}]{Kus:2024vpa}%
  \BibitemOpen
  \bibfield  {author} {\bibinfo {author} {\bibfnamefont {Pavel}\ \bibnamefont
  {K\r{u}s}}, \bibinfo {author} {\bibfnamefont {Diana}\ \bibnamefont
  {L\'opez~Nacir}}, \ and\ \bibinfo {author} {\bibfnamefont {Federico~R.}\
  \bibnamefont {Urban}},\ }\bibfield  {title} {\enquote {\bibinfo {title}
  {{Bayesian sensitivity of binary pulsars to ultra-light dark matter}},}\
  }\href@noop {} {\  (\bibinfo {year} {2024})},\ \Eprint
  {http://arxiv.org/abs/2402.04099} {arXiv:2402.04099 [astro-ph.HE]}
  \BibitemShut {NoStop}%
\bibitem [{\citenamefont {Fierz}(1956)}]{Fierz:1956zz}%
  \BibitemOpen
  \bibfield  {author} {\bibinfo {author} {\bibfnamefont {M.}~\bibnamefont
  {Fierz}},\ }\bibfield  {title} {\enquote {\bibinfo {title} {{On the physical
  interpretation of P.Jordan's extended theory of gravitation}},}\ }\href@noop
  {} {\bibfield  {journal} {\bibinfo  {journal} {Helv.\ Phys.\ Acta}\ }\textbf
  {\bibinfo {volume} {29}},\ \bibinfo {pages} {128--134} (\bibinfo {year}
  {1956})}\BibitemShut {NoStop}%
\bibitem [{\citenamefont {Jordan}(1959)}]{Jordan:1959eg}%
  \BibitemOpen
  \bibfield  {author} {\bibinfo {author} {\bibfnamefont {Pascual}\ \bibnamefont
  {Jordan}},\ }\bibfield  {title} {\enquote {\bibinfo {title} {{The present
  state of Dirac's cosmological hypothesis}},}\ }\href {\doibase
  10.1007/BF01375155} {\bibfield  {journal} {\bibinfo  {journal} {Z.\ Phys.}\
  }\textbf {\bibinfo {volume} {157}},\ \bibinfo {pages} {112--121} (\bibinfo
  {year} {1959})}\BibitemShut {NoStop}%
\bibitem [{\citenamefont {Brans}\ and\ \citenamefont
  {Dicke}(1961)}]{Brans_1961}%
  \BibitemOpen
  \bibfield  {author} {\bibinfo {author} {\bibfnamefont {C.}~\bibnamefont
  {Brans}}\ and\ \bibinfo {author} {\bibfnamefont {R.~H.}\ \bibnamefont
  {Dicke}},\ }\bibfield  {title} {\enquote {\bibinfo {title} {Mach's principle
  and a relativistic theory of gravitation},}\ }\href {\doibase
  10.1103/PhysRev.124.925} {\bibfield  {journal} {\bibinfo  {journal} {Phys.
  Rev.}\ }\textbf {\bibinfo {volume} {124}},\ \bibinfo {pages} {925--935}
  (\bibinfo {year} {1961})}\BibitemShut {NoStop}%
\bibitem [{\citenamefont {Dicke}(1962)}]{Dicke62}%
  \BibitemOpen
  \bibfield  {author} {\bibinfo {author} {\bibfnamefont {R.~H.}\ \bibnamefont
  {Dicke}},\ }\bibfield  {title} {\enquote {\bibinfo {title} {Mach's principle
  and invariance under transformation of units},}\ }\href {\doibase
  10.1103/PhysRev.125.2163} {\bibfield  {journal} {\bibinfo  {journal} {Phys.
  Rev.}\ }\textbf {\bibinfo {volume} {125}},\ \bibinfo {pages} {2163--2167}
  (\bibinfo {year} {1962})}\BibitemShut {NoStop}%
\bibitem [{\citenamefont {Damour}\ and\ \citenamefont
  {Esposito-Farese}(1992)}]{Damour_1992}%
  \BibitemOpen
  \bibfield  {author} {\bibinfo {author} {\bibfnamefont {Thibault}\
  \bibnamefont {Damour}}\ and\ \bibinfo {author} {\bibfnamefont {Gilles}\
  \bibnamefont {Esposito-Farese}},\ }\bibfield  {title} {\enquote {\bibinfo
  {title} {{Tensor multiscalar theories of gravitation}},}\ }\href {\doibase
  10.1088/0264-9381/9/9/015} {\bibfield  {journal} {\bibinfo  {journal} {Class.
  Quant. Grav.}\ }\textbf {\bibinfo {volume} {9}},\ \bibinfo {pages}
  {2093--2176} (\bibinfo {year} {1992})}\BibitemShut {NoStop}%
\bibitem [{\citenamefont {Damour}\ and\ \citenamefont
  {Esposito-Far\`ese}(1993)}]{Damour_1993}%
  \BibitemOpen
  \bibfield  {author} {\bibinfo {author} {\bibfnamefont {Thibault}\
  \bibnamefont {Damour}}\ and\ \bibinfo {author} {\bibfnamefont {Gilles}\
  \bibnamefont {Esposito-Far\`ese}},\ }\bibfield  {title} {\enquote {\bibinfo
  {title} {Nonperturbative strong-field effects in tensor-scalar theories of
  gravitation},}\ }\href {\doibase 10.1103/PhysRevLett.70.2220} {\bibfield
  {journal} {\bibinfo  {journal} {Phys. Rev. Lett.}\ }\textbf {\bibinfo
  {volume} {70}},\ \bibinfo {pages} {2220--2223} (\bibinfo {year}
  {1993})}\BibitemShut {NoStop}%
\bibitem [{\citenamefont {Alsing}\ \emph {et~al.}(2012)\citenamefont {Alsing},
  \citenamefont {Berti}, \citenamefont {Will},\ and\ \citenamefont
  {Zaglauer}}]{Alsing:2011er}%
  \BibitemOpen
  \bibfield  {author} {\bibinfo {author} {\bibfnamefont {Justin}\ \bibnamefont
  {Alsing}}, \bibinfo {author} {\bibfnamefont {Emanuele}\ \bibnamefont
  {Berti}}, \bibinfo {author} {\bibfnamefont {Clifford~M.}\ \bibnamefont
  {Will}}, \ and\ \bibinfo {author} {\bibfnamefont {Helmut}\ \bibnamefont
  {Zaglauer}},\ }\bibfield  {title} {\enquote {\bibinfo {title} {{Gravitational
  radiation from compact binary systems in the massive Brans-Dicke theory of
  gravity}},}\ }\href {\doibase 10.1103/PhysRevD.85.064041} {\bibfield
  {journal} {\bibinfo  {journal} {Phys. Rev. D}\ }\textbf {\bibinfo {volume}
  {85}},\ \bibinfo {pages} {064041} (\bibinfo {year} {2012})},\ \Eprint
  {http://arxiv.org/abs/1112.4903} {arXiv:1112.4903 [gr-qc]} \BibitemShut
  {NoStop}%
\bibitem [{\citenamefont {Nordtvedt}(1968)}]{Nordtvedt68}%
  \BibitemOpen
  \bibfield  {author} {\bibinfo {author} {\bibfnamefont {Kenneth}\ \bibnamefont
  {Nordtvedt}},\ }\bibfield  {title} {\enquote {\bibinfo {title} {Equivalence
  principle for massive bodies. ii. theory},}\ }\href {\doibase
  10.1103/PhysRev.169.1017} {\bibfield  {journal} {\bibinfo  {journal} {Phys.
  Rev.}\ }\textbf {\bibinfo {volume} {169}},\ \bibinfo {pages} {1017--1025}
  (\bibinfo {year} {1968})}\BibitemShut {NoStop}%
\bibitem [{\citenamefont {{Eardley}}(1975)}]{Eardley1975ApJ}%
  \BibitemOpen
  \bibfield  {author} {\bibinfo {author} {\bibfnamefont {D.~M.}\ \bibnamefont
  {{Eardley}}},\ }\bibfield  {title} {\enquote {\bibinfo {title} {{Observable
  effects of a scalar gravitational field in a binary pulsar}},}\ }\href
  {\doibase 10.1086/181744} {\bibfield  {journal} {\bibinfo  {journal}
  {Astrophysical Journal}\ }\textbf {\bibinfo {volume} {196}},\ \bibinfo
  {pages} {L59--L62} (\bibinfo {year} {1975})}\BibitemShut {NoStop}%
\bibitem [{\citenamefont {{Will}}\ and\ \citenamefont
  {{Zaglauer}}(1989)}]{1989ApJ...346..366W}%
  \BibitemOpen
  \bibfield  {author} {\bibinfo {author} {\bibfnamefont {Clifford~M.}\
  \bibnamefont {{Will}}}\ and\ \bibinfo {author} {\bibfnamefont {Helmut~W.}\
  \bibnamefont {{Zaglauer}}},\ }\bibfield  {title} {\enquote {\bibinfo {title}
  {{Gravitational Radiation, Close Binary Systems, and the Brans-Dicke Theory
  of Gravity}},}\ }\href {\doibase 10.1086/168016} {\bibfield  {journal}
  {\bibinfo  {journal} {\apj}\ }\textbf {\bibinfo {volume} {346}},\ \bibinfo
  {pages} {366} (\bibinfo {year} {1989})}\BibitemShut {NoStop}%
\bibitem [{\citenamefont {{Will}}(1977)}]{1977ApJ...214..826W}%
  \BibitemOpen
  \bibfield  {author} {\bibinfo {author} {\bibfnamefont {C.~M.}\ \bibnamefont
  {{Will}}},\ }\bibfield  {title} {\enquote {\bibinfo {title} {{Gravitational
  radiation from binary systems in alternative metric theories of gravity:
  dipole radiation and the binary pulsar.}}}\ }\href {\doibase 10.1086/155313}
  {\bibfield  {journal} {\bibinfo  {journal} {\apj}\ }\textbf {\bibinfo
  {volume} {214}},\ \bibinfo {pages} {826--839} (\bibinfo {year}
  {1977})}\BibitemShut {NoStop}%
\bibitem [{\citenamefont {Will}(1993{\natexlab{a}})}]{Will:1993ns}%
  \BibitemOpen
  \bibfield  {author} {\bibinfo {author} {\bibfnamefont {C.~M.}\ \bibnamefont
  {Will}},\ }\href@noop {} {\emph {\bibinfo {title} {{Theory and experiment in
  gravitational physics}}}}\ (\bibinfo {year} {1993})\BibitemShut {NoStop}%
\bibitem [{\citenamefont {Damour}\ and\ \citenamefont
  {Taylor}(1992)}]{Damour:1991rd}%
  \BibitemOpen
  \bibfield  {author} {\bibinfo {author} {\bibfnamefont {Thibault}\
  \bibnamefont {Damour}}\ and\ \bibinfo {author} {\bibfnamefont {Joseph~H.}\
  \bibnamefont {Taylor}},\ }\bibfield  {title} {\enquote {\bibinfo {title}
  {{Strong field tests of relativistic gravity and binary pulsars}},}\ }\href
  {\doibase 10.1103/PhysRevD.45.1840} {\bibfield  {journal} {\bibinfo
  {journal} {Phys. Rev. D}\ }\textbf {\bibinfo {volume} {45}},\ \bibinfo
  {pages} {1840--1868} (\bibinfo {year} {1992})}\BibitemShut {NoStop}%
\bibitem [{\citenamefont {Weisberg}\ and\ \citenamefont
  {Taylor}(2004)}]{weisberg2004relativistic}%
  \BibitemOpen
  \bibfield  {author} {\bibinfo {author} {\bibfnamefont {J.~M.}\ \bibnamefont
  {Weisberg}}\ and\ \bibinfo {author} {\bibfnamefont {J.~H.}\ \bibnamefont
  {Taylor}},\ }\href@noop {} {\enquote {\bibinfo {title} {Relativistic binary
  pulsar b1913+16: Thirty years of observations and analysis},}\ } (\bibinfo
  {year} {2004}),\ \Eprint {http://arxiv.org/abs/astro-ph/0407149}
  {arXiv:astro-ph/0407149 [astro-ph]} \BibitemShut {NoStop}%
\bibitem [{\citenamefont {Kramer}\ \emph {et~al.}(2006)\citenamefont {Kramer},
  \citenamefont {Stairs}, \citenamefont {Manchester}, \citenamefont
  {McLaughlin}, \citenamefont {Lyne} \emph {et~al.}}]{Kramer:2006nb}%
  \BibitemOpen
  \bibfield  {author} {\bibinfo {author} {\bibfnamefont {M.}~\bibnamefont
  {Kramer}}, \bibinfo {author} {\bibfnamefont {Ingrid~H.}\ \bibnamefont
  {Stairs}}, \bibinfo {author} {\bibfnamefont {R.N.}\ \bibnamefont
  {Manchester}}, \bibinfo {author} {\bibfnamefont {M.A.}\ \bibnamefont
  {McLaughlin}}, \bibinfo {author} {\bibfnamefont {A.G.}\ \bibnamefont {Lyne}},
   \emph {et~al.},\ }\bibfield  {title} {\enquote {\bibinfo {title} {{Tests of
  general relativity from timing the double pulsar}},}\ }\href {\doibase
  10.1126/science.1132305} {\bibfield  {journal} {\bibinfo  {journal}
  {Science}\ }\textbf {\bibinfo {volume} {314}},\ \bibinfo {pages} {97--102}
  (\bibinfo {year} {2006})},\ \Eprint {http://arxiv.org/abs/astro-ph/0609417}
  {arXiv:astro-ph/0609417 [astro-ph]} \BibitemShut {NoStop}%
\bibitem [{\citenamefont {Ransom}\ \emph {et~al.}(2014)\citenamefont {Ransom},
  \citenamefont {Stairs}, \citenamefont {Archibald}, \citenamefont {Hessels},
  \citenamefont {Kaplan}, \citenamefont {van Kerkwijk}, \citenamefont {Boyles},
  \citenamefont {Deller}, \citenamefont {Chatterjee}, \citenamefont
  {Schechtman-Rook}, \citenamefont {Berndsen}, \citenamefont {Lynch},
  \citenamefont {Lorimer}, \citenamefont {Karako-Argaman}, \citenamefont
  {Kaspi}, \citenamefont {Kondratiev}, \citenamefont {McLaughlin},
  \citenamefont {van Leeuwen}, \citenamefont {Rosen}, \citenamefont {Roberts},\
  and\ \citenamefont {Stovall}}]{Ransom_2014}%
  \BibitemOpen
  \bibfield  {author} {\bibinfo {author} {\bibfnamefont {S.~M.}\ \bibnamefont
  {Ransom}}, \bibinfo {author} {\bibfnamefont {I.~H.}\ \bibnamefont {Stairs}},
  \bibinfo {author} {\bibfnamefont {A.~M.}\ \bibnamefont {Archibald}}, \bibinfo
  {author} {\bibfnamefont {J.~W.~T.}\ \bibnamefont {Hessels}}, \bibinfo
  {author} {\bibfnamefont {D.~L.}\ \bibnamefont {Kaplan}}, \bibinfo {author}
  {\bibfnamefont {M.~H.}\ \bibnamefont {van Kerkwijk}}, \bibinfo {author}
  {\bibfnamefont {J.}~\bibnamefont {Boyles}}, \bibinfo {author} {\bibfnamefont
  {A.~T.}\ \bibnamefont {Deller}}, \bibinfo {author} {\bibfnamefont
  {S.}~\bibnamefont {Chatterjee}}, \bibinfo {author} {\bibfnamefont
  {A.}~\bibnamefont {Schechtman-Rook}}, \bibinfo {author} {\bibfnamefont
  {A.}~\bibnamefont {Berndsen}}, \bibinfo {author} {\bibfnamefont {R.~S.}\
  \bibnamefont {Lynch}}, \bibinfo {author} {\bibfnamefont {D.~R.}\ \bibnamefont
  {Lorimer}}, \bibinfo {author} {\bibfnamefont {C.}~\bibnamefont
  {Karako-Argaman}}, \bibinfo {author} {\bibfnamefont {V.~M.}\ \bibnamefont
  {Kaspi}}, \bibinfo {author} {\bibfnamefont {V.~I.}\ \bibnamefont
  {Kondratiev}}, \bibinfo {author} {\bibfnamefont {M.~A.}\ \bibnamefont
  {McLaughlin}}, \bibinfo {author} {\bibfnamefont {J.}~\bibnamefont {van
  Leeuwen}}, \bibinfo {author} {\bibfnamefont {R.}~\bibnamefont {Rosen}},
  \bibinfo {author} {\bibfnamefont {M.~S.~E.}\ \bibnamefont {Roberts}}, \ and\
  \bibinfo {author} {\bibfnamefont {K.}~\bibnamefont {Stovall}},\ }\bibfield
  {title} {\enquote {\bibinfo {title} {A millisecond pulsar in a stellar triple
  system},}\ }\href {\doibase 10.1038/nature12917} {\bibfield  {journal}
  {\bibinfo  {journal} {Nature}\ }\textbf {\bibinfo {volume} {505}},\ \bibinfo
  {pages} {520--524} (\bibinfo {year} {2014})}\BibitemShut {NoStop}%
\bibitem [{\citenamefont {Archibald}\ \emph {et~al.}(2018)\citenamefont
  {Archibald}, \citenamefont {Gusinskaia}, \citenamefont {Hessels},
  \citenamefont {Deller}, \citenamefont {Kaplan}, \citenamefont {Lorimer},
  \citenamefont {Lynch}, \citenamefont {Ransom},\ and\ \citenamefont
  {Stairs}}]{Archibald_2018}%
  \BibitemOpen
  \bibfield  {author} {\bibinfo {author} {\bibfnamefont {Anne~M.}\ \bibnamefont
  {Archibald}}, \bibinfo {author} {\bibfnamefont {Nina~V.}\ \bibnamefont
  {Gusinskaia}}, \bibinfo {author} {\bibfnamefont {Jason W.~T.}\ \bibnamefont
  {Hessels}}, \bibinfo {author} {\bibfnamefont {Adam~T.}\ \bibnamefont
  {Deller}}, \bibinfo {author} {\bibfnamefont {David~L.}\ \bibnamefont
  {Kaplan}}, \bibinfo {author} {\bibfnamefont {Duncan~R.}\ \bibnamefont
  {Lorimer}}, \bibinfo {author} {\bibfnamefont {Ryan~S.}\ \bibnamefont
  {Lynch}}, \bibinfo {author} {\bibfnamefont {Scott~M.}\ \bibnamefont
  {Ransom}}, \ and\ \bibinfo {author} {\bibfnamefont {Ingrid~H.}\ \bibnamefont
  {Stairs}},\ }\bibfield  {title} {\enquote {\bibinfo {title} {Universality of
  free fall from the orbital motion of a pulsar in a stellar triple system},}\
  }\href {\doibase 10.1038/s41586-018-0265-1} {\bibfield  {journal} {\bibinfo
  {journal} {Nature}\ }\textbf {\bibinfo {volume} {559}},\ \bibinfo {pages}
  {73--76} (\bibinfo {year} {2018})}\BibitemShut {NoStop}%
\bibitem [{\citenamefont {{Voisin, G.}}\ \emph {et~al.}(2020)\citenamefont
  {{Voisin, G.}}, \citenamefont {{Cognard, I.}}, \citenamefont {{Freire, P. C.
  C.}}, \citenamefont {{Wex, N.}}, \citenamefont {{Guillemot, L.}},
  \citenamefont {{Desvignes, G.}}, \citenamefont {{Kramer, M.}},\ and\
  \citenamefont {{Theureau, G.}}}]{Voisin_2020}%
  \BibitemOpen
  \bibfield  {author} {\bibinfo {author} {\bibnamefont {{Voisin, G.}}},
  \bibinfo {author} {\bibnamefont {{Cognard, I.}}}, \bibinfo {author}
  {\bibnamefont {{Freire, P. C. C.}}}, \bibinfo {author} {\bibnamefont {{Wex,
  N.}}}, \bibinfo {author} {\bibnamefont {{Guillemot, L.}}}, \bibinfo {author}
  {\bibnamefont {{Desvignes, G.}}}, \bibinfo {author} {\bibnamefont {{Kramer,
  M.}}}, \ and\ \bibinfo {author} {\bibnamefont {{Theureau, G.}}},\ }\bibfield
  {title} {\enquote {\bibinfo {title} {An improved test of the strong
  equivalence principle with the pulsar in a triple star system},}\ }\href
  {\doibase 10.1051/0004-6361/202038104} {\bibfield  {journal} {\bibinfo
  {journal} {A\& A}\ }\textbf {\bibinfo {volume} {638}},\ \bibinfo {pages}
  {A24} (\bibinfo {year} {2020})}\BibitemShut {NoStop}%
\bibitem [{\citenamefont {Kramer}\ \emph {et~al.}(2021)\citenamefont {Kramer},
  \citenamefont {Stairs}, \citenamefont {Manchester}, \citenamefont {Wex},
  \citenamefont {Deller}, \citenamefont {Coles}, \citenamefont {Ali},
  \citenamefont {Burgay}, \citenamefont {Camilo}, \citenamefont {Cognard},
  \citenamefont {Damour}, \citenamefont {Desvignes}, \citenamefont {Ferdman},
  \citenamefont {Freire}, \citenamefont {Grondin}, \citenamefont {Guillemot},
  \citenamefont {Hobbs}, \citenamefont {Janssen}, \citenamefont {Karuppusamy},
  \citenamefont {Lorimer}, \citenamefont {Lyne}, \citenamefont {McKee},
  \citenamefont {McLaughlin}, \citenamefont {M\"unch}, \citenamefont {Perera},
  \citenamefont {Pol}, \citenamefont {Possenti}, \citenamefont {Sarkissian},
  \citenamefont {Stappers},\ and\ \citenamefont {Theureau}}]{Kramer_2021}%
  \BibitemOpen
  \bibfield  {author} {\bibinfo {author} {\bibfnamefont {M.}~\bibnamefont
  {Kramer}}, \bibinfo {author} {\bibfnamefont {I.~H.}\ \bibnamefont {Stairs}},
  \bibinfo {author} {\bibfnamefont {R.~N.}\ \bibnamefont {Manchester}},
  \bibinfo {author} {\bibfnamefont {N.}~\bibnamefont {Wex}}, \bibinfo {author}
  {\bibfnamefont {A.~T.}\ \bibnamefont {Deller}}, \bibinfo {author}
  {\bibfnamefont {W.~A.}\ \bibnamefont {Coles}}, \bibinfo {author}
  {\bibfnamefont {M.}~\bibnamefont {Ali}}, \bibinfo {author} {\bibfnamefont
  {M.}~\bibnamefont {Burgay}}, \bibinfo {author} {\bibfnamefont
  {F.}~\bibnamefont {Camilo}}, \bibinfo {author} {\bibfnamefont
  {I.}~\bibnamefont {Cognard}}, \bibinfo {author} {\bibfnamefont
  {T.}~\bibnamefont {Damour}}, \bibinfo {author} {\bibfnamefont
  {G.}~\bibnamefont {Desvignes}}, \bibinfo {author} {\bibfnamefont {R.~D.}\
  \bibnamefont {Ferdman}}, \bibinfo {author} {\bibfnamefont {P.~C.~C.}\
  \bibnamefont {Freire}}, \bibinfo {author} {\bibfnamefont {S.}~\bibnamefont
  {Grondin}}, \bibinfo {author} {\bibfnamefont {L.}~\bibnamefont {Guillemot}},
  \bibinfo {author} {\bibfnamefont {G.~B.}\ \bibnamefont {Hobbs}}, \bibinfo
  {author} {\bibfnamefont {G.}~\bibnamefont {Janssen}}, \bibinfo {author}
  {\bibfnamefont {R.}~\bibnamefont {Karuppusamy}}, \bibinfo {author}
  {\bibfnamefont {D.~R.}\ \bibnamefont {Lorimer}}, \bibinfo {author}
  {\bibfnamefont {A.~G.}\ \bibnamefont {Lyne}}, \bibinfo {author}
  {\bibfnamefont {J.~W.}\ \bibnamefont {McKee}}, \bibinfo {author}
  {\bibfnamefont {M.}~\bibnamefont {McLaughlin}}, \bibinfo {author}
  {\bibfnamefont {L.~E.}\ \bibnamefont {M\"unch}}, \bibinfo {author}
  {\bibfnamefont {B.~B.~P.}\ \bibnamefont {Perera}}, \bibinfo {author}
  {\bibfnamefont {N.}~\bibnamefont {Pol}}, \bibinfo {author} {\bibfnamefont
  {A.}~\bibnamefont {Possenti}}, \bibinfo {author} {\bibfnamefont
  {J.}~\bibnamefont {Sarkissian}}, \bibinfo {author} {\bibfnamefont {B.~W.}\
  \bibnamefont {Stappers}}, \ and\ \bibinfo {author} {\bibfnamefont
  {G.}~\bibnamefont {Theureau}},\ }\bibfield  {title} {\enquote {\bibinfo
  {title} {Strong-field gravity tests with the double pulsar},}\ }\href
  {\doibase 10.1103/PhysRevX.11.041050} {\bibfield  {journal} {\bibinfo
  {journal} {Phys. Rev. X}\ }\textbf {\bibinfo {volume} {11}},\ \bibinfo
  {pages} {041050} (\bibinfo {year} {2021})}\BibitemShut {NoStop}%
\bibitem [{\citenamefont {Zhao}\ \emph {et~al.}(2022)\citenamefont {Zhao},
  \citenamefont {Freire}, \citenamefont {Kramer}, \citenamefont {Shao},\ and\
  \citenamefont {Wex}}]{Zhao:2022vig}%
  \BibitemOpen
  \bibfield  {author} {\bibinfo {author} {\bibfnamefont {Junjie}\ \bibnamefont
  {Zhao}}, \bibinfo {author} {\bibfnamefont {Paulo C.~C.}\ \bibnamefont
  {Freire}}, \bibinfo {author} {\bibfnamefont {Michael}\ \bibnamefont
  {Kramer}}, \bibinfo {author} {\bibfnamefont {Lijing}\ \bibnamefont {Shao}}, \
  and\ \bibinfo {author} {\bibfnamefont {Norbert}\ \bibnamefont {Wex}},\
  }\bibfield  {title} {\enquote {\bibinfo {title} {{Closing a
  spontaneous-scalarization window with binary pulsars}},}\ }\href {\doibase
  10.1088/1361-6382/ac69a3} {\bibfield  {journal} {\bibinfo  {journal} {Class.
  Quant. Grav.}\ }\textbf {\bibinfo {volume} {39}},\ \bibinfo {pages} {11LT01}
  (\bibinfo {year} {2022})},\ \Eprint {http://arxiv.org/abs/2201.03771}
  {arXiv:2201.03771 [astro-ph.HE]} \BibitemShut {NoStop}%
\bibitem [{\citenamefont {Kuntz}\ and\ \citenamefont
  {Barausse}(2024)}]{Kuntz_2024}%
  \BibitemOpen
  \bibfield  {author} {\bibinfo {author} {\bibfnamefont {Adrien}\ \bibnamefont
  {Kuntz}}\ and\ \bibinfo {author} {\bibfnamefont {Enrico}\ \bibnamefont
  {Barausse}},\ }\href@noop {} {\enquote {\bibinfo {title} {Angular momentum
  sensitivities in scalar-tensor theories},}\ } (\bibinfo {year} {2024}),\
  \Eprint {http://arxiv.org/abs/2403.07980} {arXiv:2403.07980 [gr-qc]}
  \BibitemShut {NoStop}%
\bibitem [{\citenamefont {Damour}\ and\ \citenamefont
  {Esposito-Far\`ese}(1996)}]{Damour_1994}%
  \BibitemOpen
  \bibfield  {author} {\bibinfo {author} {\bibfnamefont {Thibault}\
  \bibnamefont {Damour}}\ and\ \bibinfo {author} {\bibfnamefont {Gilles}\
  \bibnamefont {Esposito-Far\`ese}},\ }\bibfield  {title} {\enquote {\bibinfo
  {title} {Tensor-scalar gravity and binary-pulsar experiments},}\ }\href
  {\doibase 10.1103/PhysRevD.54.1474} {\bibfield  {journal} {\bibinfo
  {journal} {Phys. Rev. D}\ }\textbf {\bibinfo {volume} {54}},\ \bibinfo
  {pages} {1474--1491} (\bibinfo {year} {1996})}\BibitemShut {NoStop}%
\bibitem [{\citenamefont {Bertotti}\ \emph {et~al.}(2003)\citenamefont
  {Bertotti}, \citenamefont {Iess},\ and\ \citenamefont
  {Tortora}}]{Bertotti_2003}%
  \BibitemOpen
  \bibfield  {author} {\bibinfo {author} {\bibfnamefont {B.}~\bibnamefont
  {Bertotti}}, \bibinfo {author} {\bibfnamefont {L.}~\bibnamefont {Iess}}, \
  and\ \bibinfo {author} {\bibfnamefont {P.}~\bibnamefont {Tortora}},\
  }\bibfield  {title} {\enquote {\bibinfo {title} {A test of general relativity
  using radio links with the cassini spacecraft},}\ }\href {\doibase
  10.1038/nature01997} {\bibfield  {journal} {\bibinfo  {journal} {Nature}\
  }\textbf {\bibinfo {volume} {425}},\ \bibinfo {pages} {374--376} (\bibinfo
  {year} {2003})}\BibitemShut {NoStop}%
\bibitem [{\citenamefont {Damour}\ and\ \citenamefont
  {Polyakov}(1994)}]{Damour:1994zq}%
  \BibitemOpen
  \bibfield  {author} {\bibinfo {author} {\bibfnamefont {T.}~\bibnamefont
  {Damour}}\ and\ \bibinfo {author} {\bibfnamefont {Alexander~M.}\ \bibnamefont
  {Polyakov}},\ }\bibfield  {title} {\enquote {\bibinfo {title} {{The String
  dilaton and a least coupling principle}},}\ }\href {\doibase
  10.1016/0550-3213(94)90143-0} {\bibfield  {journal} {\bibinfo  {journal}
  {Nucl. Phys. B}\ }\textbf {\bibinfo {volume} {423}},\ \bibinfo {pages}
  {532--558} (\bibinfo {year} {1994})},\ \Eprint
  {http://arxiv.org/abs/hep-th/9401069} {arXiv:hep-th/9401069} \BibitemShut
  {NoStop}%
\bibitem [{\citenamefont {Kim}(1987)}]{Kim:1986ax}%
  \BibitemOpen
  \bibfield  {author} {\bibinfo {author} {\bibfnamefont {Jihn~E.}\ \bibnamefont
  {Kim}},\ }\bibfield  {title} {\enquote {\bibinfo {title} {{Light
  Pseudoscalars, Particle Physics and Cosmology}},}\ }\href {\doibase
  10.1016/0370-1573(87)90017-2} {\bibfield  {journal} {\bibinfo  {journal}
  {Phys. Rept.}\ }\textbf {\bibinfo {volume} {150}},\ \bibinfo {pages} {1--177}
  (\bibinfo {year} {1987})}\BibitemShut {NoStop}%
\bibitem [{\citenamefont {Will}(1993{\natexlab{b}})}]{Will_1993}%
  \BibitemOpen
  \bibfield  {author} {\bibinfo {author} {\bibfnamefont {Clifford~M.}\
  \bibnamefont {Will}},\ }\href {\doibase 10.1017/CBO9780511564246} {\emph
  {\bibinfo {title} {Theory and Experiment in Gravitational Physics}}}\
  (\bibinfo  {publisher} {Cambridge University Press},\ \bibinfo {year}
  {1993})\BibitemShut {NoStop}%
\bibitem [{\citenamefont {Shao}\ \emph {et~al.}(2017)\citenamefont {Shao},
  \citenamefont {Sennett}, \citenamefont {Buonanno}, \citenamefont {Kramer},\
  and\ \citenamefont {Wex}}]{Shao:2017gwu}%
  \BibitemOpen
  \bibfield  {author} {\bibinfo {author} {\bibfnamefont {Lijing}\ \bibnamefont
  {Shao}}, \bibinfo {author} {\bibfnamefont {Noah}\ \bibnamefont {Sennett}},
  \bibinfo {author} {\bibfnamefont {Alessandra}\ \bibnamefont {Buonanno}},
  \bibinfo {author} {\bibfnamefont {Michael}\ \bibnamefont {Kramer}}, \ and\
  \bibinfo {author} {\bibfnamefont {Norbert}\ \bibnamefont {Wex}},\ }\bibfield
  {title} {\enquote {\bibinfo {title} {{Constraining nonperturbative
  strong-field effects in scalar-tensor gravity by combining pulsar timing and
  laser-interferometer gravitational-wave detectors}},}\ }\href {\doibase
  10.1103/PhysRevX.7.041025} {\bibfield  {journal} {\bibinfo  {journal} {Phys.
  Rev. X}\ }\textbf {\bibinfo {volume} {7}},\ \bibinfo {pages} {041025}
  (\bibinfo {year} {2017})},\ \Eprint {http://arxiv.org/abs/1704.07561}
  {arXiv:1704.07561 [gr-qc]} \BibitemShut {NoStop}%
\bibitem [{\citenamefont {Castillo}\ \emph {et~al.}(2022)\citenamefont
  {Castillo}, \citenamefont {Martin-Camalich}, \citenamefont {Terol-Calvo},
  \citenamefont {Blas}, \citenamefont {Caputo}, \citenamefont {Santos},
  \citenamefont {Sberna}, \citenamefont {Peel},\ and\ \citenamefont {Rubi\~no
  Mart\'\i{}n}}]{Castillo_2022}%
  \BibitemOpen
  \bibfield  {author} {\bibinfo {author} {\bibfnamefont {Andr\'es}\
  \bibnamefont {Castillo}}, \bibinfo {author} {\bibfnamefont {Jorge}\
  \bibnamefont {Martin-Camalich}}, \bibinfo {author} {\bibfnamefont {Jorge}\
  \bibnamefont {Terol-Calvo}}, \bibinfo {author} {\bibfnamefont {Diego}\
  \bibnamefont {Blas}}, \bibinfo {author} {\bibfnamefont {Andrea}\ \bibnamefont
  {Caputo}}, \bibinfo {author} {\bibfnamefont {Ricardo Tanaus\'u~G\'enova}\
  \bibnamefont {Santos}}, \bibinfo {author} {\bibfnamefont {Laura}\
  \bibnamefont {Sberna}}, \bibinfo {author} {\bibfnamefont {Michael}\
  \bibnamefont {Peel}}, \ and\ \bibinfo {author} {\bibfnamefont {Jose~Alberto}\
  \bibnamefont {Rubi\~no Mart\'\i{}n}},\ }\bibfield  {title} {\enquote
  {\bibinfo {title} {{Searching for dark-matter waves with PPTA and QUIJOTE
  pulsar polarimetry}},}\ }\href {\doibase 10.1088/1475-7516/2022/06/014}
  {\bibfield  {journal} {\bibinfo  {journal} {JCAP}\ }\textbf {\bibinfo
  {volume} {06}},\ \bibinfo {pages} {014} (\bibinfo {year} {2022})},\ \Eprint
  {http://arxiv.org/abs/2201.03422} {arXiv:2201.03422 [astro-ph.CO]}
  \BibitemShut {NoStop}%
\bibitem [{\citenamefont {Nesti}\ and\ \citenamefont
  {Salucci}(2013)}]{Nesti:2013uwa}%
  \BibitemOpen
  \bibfield  {author} {\bibinfo {author} {\bibfnamefont {Fabrizio}\
  \bibnamefont {Nesti}}\ and\ \bibinfo {author} {\bibfnamefont {Paolo}\
  \bibnamefont {Salucci}},\ }\bibfield  {title} {\enquote {\bibinfo {title}
  {{The Dark Matter halo of the Milky Way, AD 2013}},}\ }\href {\doibase
  10.1088/1475-7516/2013/07/016} {\bibfield  {journal} {\bibinfo  {journal}
  {JCAP}\ }\textbf {\bibinfo {volume} {07}},\ \bibinfo {pages} {016} (\bibinfo
  {year} {2013})},\ \Eprint {http://arxiv.org/abs/1304.5127} {arXiv:1304.5127
  [astro-ph.GA]} \BibitemShut {NoStop}%
\bibitem [{\citenamefont {{EPTA Collaboration}}\ \emph
  {et~al.}(2023{\natexlab{b}})\citenamefont {{EPTA Collaboration}},
  \citenamefont {{Antoniadis, J.}}, \citenamefont {{Babak, S.}}, \citenamefont
  {{Bak Nielsen, A.-S.}}, \citenamefont {{Bassa, C. G.}}, \citenamefont
  {{Berthereau, A.}}, \citenamefont {{Bonetti, M.}}, \citenamefont {{Bortolas,
  E.}}, \citenamefont {{Brook, P. R.}}, \citenamefont {{Burgay, M.}},
  \citenamefont {{Caballero, R. N.}}, \citenamefont {{Chalumeau, A.}},
  \citenamefont {{Champion, D. J.}}, \citenamefont {{Cłianlaridis, S.}},
  \citenamefont {{Chen, S.}}, \citenamefont {{Cognard, I.}}, \citenamefont
  {{Desvignes, G.}}, \citenamefont {{Falxa, M.}}, \citenamefont {{Ferdman, R.
  D.}}, \citenamefont {{Franchini, A.}}, \citenamefont {{Gair, J. R.}},
  \citenamefont {{Goncharov, B.}}, \citenamefont {{Graikou, E.}}, \citenamefont
  {{Grießmeier, J.-M.}}, \citenamefont {{Guillemot, L.}}, \citenamefont {{Guo,
  Y. J.}}, \citenamefont {{Hu, H.}}, \citenamefont {{Iraci, F.}}, \citenamefont
  {{Izquierdo-Villalba, D.}}, \citenamefont {{Jang, J.}}, \citenamefont
  {{Jawor, J.}}, \citenamefont {{Janssen, G. H.}}, \citenamefont {{Jessner,
  A.}}, \citenamefont {{Karuppusamy, R.}}, \citenamefont {{Keane, E. F.}},
  \citenamefont {{Keith, M. J.}}, \citenamefont {{Kramer, M.}}, \citenamefont
  {{Krishnakumar, M. A.}}, \citenamefont {{Lackeos, K.}}, \citenamefont {{Lee,
  K. J.}}, \citenamefont {{Liu, K.}}, \citenamefont {{Liu, Y.}}, \citenamefont
  {{Lyne, A. G.}}, \citenamefont {{McKee, J. W.}}, \citenamefont {{Main, R.
  A.}}, \citenamefont {{Mickaliger, M. B.}}, \citenamefont {{Niţu, I. C.}},
  \citenamefont {{Parthasarathy, A.}}, \citenamefont {{Perera, B. B. P.}},
  \citenamefont {{Perrodin, D.}}, \citenamefont {{Petiteau, A.}}, \citenamefont
  {{Porayko, N. K.}}, \citenamefont {{Possenti, A.}}, \citenamefont
  {{Quelquejay Leclere, H.}}, \citenamefont {{Samajdar, A.}}, \citenamefont
  {{Sanidas, S. A.}}, \citenamefont {{Sesana, A.}}, \citenamefont {{Shaifullah,
  G.}}, \citenamefont {{Speri, L.}}, \citenamefont {{Spiewak, R.}},
  \citenamefont {{Stappers, B. W.}}, \citenamefont {{Susarla, S. C.}},
  \citenamefont {{Theureau, G.}}, \citenamefont {{Tiburzi, C.}}, \citenamefont
  {{van der Wateren, E.}}, \citenamefont {{Vecchio, A.}}, \citenamefont
  {{Venkatraman Krishnan, V.}}, \citenamefont {{Verbiest, J. P. W.}},
  \citenamefont {{Wang, J.}}, \citenamefont {{Wang, L.}},\ and\ \citenamefont
  {{Wu, Z.}}}]{EPTA_I_2023}%
  \BibitemOpen
  \bibfield  {author} {\bibinfo {author} {\bibnamefont {{EPTA Collaboration}}},
  \bibinfo {author} {\bibnamefont {{Antoniadis, J.}}}, \bibinfo {author}
  {\bibnamefont {{Babak, S.}}}, \bibinfo {author} {\bibnamefont {{Bak Nielsen,
  A.-S.}}}, \bibinfo {author} {\bibnamefont {{Bassa, C. G.}}}, \bibinfo
  {author} {\bibnamefont {{Berthereau, A.}}}, \bibinfo {author} {\bibnamefont
  {{Bonetti, M.}}}, \bibinfo {author} {\bibnamefont {{Bortolas, E.}}}, \bibinfo
  {author} {\bibnamefont {{Brook, P. R.}}}, \bibinfo {author} {\bibnamefont
  {{Burgay, M.}}}, \bibinfo {author} {\bibnamefont {{Caballero, R. N.}}},
  \bibinfo {author} {\bibnamefont {{Chalumeau, A.}}}, \bibinfo {author}
  {\bibnamefont {{Champion, D. J.}}}, \bibinfo {author} {\bibnamefont
  {{Cłianlaridis, S.}}}, \bibinfo {author} {\bibnamefont {{Chen, S.}}},
  \bibinfo {author} {\bibnamefont {{Cognard, I.}}}, \bibinfo {author}
  {\bibnamefont {{Desvignes, G.}}}, \bibinfo {author} {\bibnamefont {{Falxa,
  M.}}}, \bibinfo {author} {\bibnamefont {{Ferdman, R. D.}}}, \bibinfo {author}
  {\bibnamefont {{Franchini, A.}}}, \bibinfo {author} {\bibnamefont {{Gair, J.
  R.}}}, \bibinfo {author} {\bibnamefont {{Goncharov, B.}}}, \bibinfo {author}
  {\bibnamefont {{Graikou, E.}}}, \bibinfo {author} {\bibnamefont
  {{Grießmeier, J.-M.}}}, \bibinfo {author} {\bibnamefont {{Guillemot, L.}}},
  \bibinfo {author} {\bibnamefont {{Guo, Y. J.}}}, \bibinfo {author}
  {\bibnamefont {{Hu, H.}}}, \bibinfo {author} {\bibnamefont {{Iraci, F.}}},
  \bibinfo {author} {\bibnamefont {{Izquierdo-Villalba, D.}}}, \bibinfo
  {author} {\bibnamefont {{Jang, J.}}}, \bibinfo {author} {\bibnamefont
  {{Jawor, J.}}}, \bibinfo {author} {\bibnamefont {{Janssen, G. H.}}}, \bibinfo
  {author} {\bibnamefont {{Jessner, A.}}}, \bibinfo {author} {\bibnamefont
  {{Karuppusamy, R.}}}, \bibinfo {author} {\bibnamefont {{Keane, E. F.}}},
  \bibinfo {author} {\bibnamefont {{Keith, M. J.}}}, \bibinfo {author}
  {\bibnamefont {{Kramer, M.}}}, \bibinfo {author} {\bibnamefont
  {{Krishnakumar, M. A.}}}, \bibinfo {author} {\bibnamefont {{Lackeos, K.}}},
  \bibinfo {author} {\bibnamefont {{Lee, K. J.}}}, \bibinfo {author}
  {\bibnamefont {{Liu, K.}}}, \bibinfo {author} {\bibnamefont {{Liu, Y.}}},
  \bibinfo {author} {\bibnamefont {{Lyne, A. G.}}}, \bibinfo {author}
  {\bibnamefont {{McKee, J. W.}}}, \bibinfo {author} {\bibnamefont {{Main, R.
  A.}}}, \bibinfo {author} {\bibnamefont {{Mickaliger, M. B.}}}, \bibinfo
  {author} {\bibnamefont {{Niţu, I. C.}}}, \bibinfo {author} {\bibnamefont
  {{Parthasarathy, A.}}}, \bibinfo {author} {\bibnamefont {{Perera, B. B.
  P.}}}, \bibinfo {author} {\bibnamefont {{Perrodin, D.}}}, \bibinfo {author}
  {\bibnamefont {{Petiteau, A.}}}, \bibinfo {author} {\bibnamefont {{Porayko,
  N. K.}}}, \bibinfo {author} {\bibnamefont {{Possenti, A.}}}, \bibinfo
  {author} {\bibnamefont {{Quelquejay Leclere, H.}}}, \bibinfo {author}
  {\bibnamefont {{Samajdar, A.}}}, \bibinfo {author} {\bibnamefont {{Sanidas,
  S. A.}}}, \bibinfo {author} {\bibnamefont {{Sesana, A.}}}, \bibinfo {author}
  {\bibnamefont {{Shaifullah, G.}}}, \bibinfo {author} {\bibnamefont {{Speri,
  L.}}}, \bibinfo {author} {\bibnamefont {{Spiewak, R.}}}, \bibinfo {author}
  {\bibnamefont {{Stappers, B. W.}}}, \bibinfo {author} {\bibnamefont
  {{Susarla, S. C.}}}, \bibinfo {author} {\bibnamefont {{Theureau, G.}}},
  \bibinfo {author} {\bibnamefont {{Tiburzi, C.}}}, \bibinfo {author}
  {\bibnamefont {{van der Wateren, E.}}}, \bibinfo {author} {\bibnamefont
  {{Vecchio, A.}}}, \bibinfo {author} {\bibnamefont {{Venkatraman Krishnan,
  V.}}}, \bibinfo {author} {\bibnamefont {{Verbiest, J. P. W.}}}, \bibinfo
  {author} {\bibnamefont {{Wang, J.}}}, \bibinfo {author} {\bibnamefont {{Wang,
  L.}}}, \ and\ \bibinfo {author} {\bibnamefont {{Wu, Z.}}},\ }\bibfield
  {title} {\enquote {\bibinfo {title} {The second data release from the
  european pulsar timing array - i. the dataset and timing analysis},}\ }\href
  {\doibase 10.1051/0004-6361/202346841} {\bibfield  {journal} {\bibinfo
  {journal} {A\& A}\ }\textbf {\bibinfo {volume} {678}},\ \bibinfo {pages}
  {A48} (\bibinfo {year} {2023}{\natexlab{b}})}\BibitemShut {NoStop}%
\bibitem [{\citenamefont {{Desvignes}}\ \emph {et~al.}(2016)\citenamefont
  {{Desvignes}}, \citenamefont {{Caballero}}, \citenamefont {{Lentati}},
  \citenamefont {{Verbiest}}, \citenamefont {{Champion}}, \citenamefont
  {{Stappers}}, \citenamefont {{Janssen}}, \citenamefont {{Lazarus}},
  \citenamefont {{Os{\l}owski}}, \citenamefont {{Babak}}, \citenamefont
  {{Bassa}}, \citenamefont {{Brem}}, \citenamefont {{Burgay}}, \citenamefont
  {{Cognard}}, \citenamefont {{Gair}}, \citenamefont {{Graikou}}, \citenamefont
  {{Guillemot}}, \citenamefont {{Hessels}}, \citenamefont {{Jessner}},
  \citenamefont {{Jordan}}, \citenamefont {{Karuppusamy}}, \citenamefont
  {{Kramer}}, \citenamefont {{Lassus}}, \citenamefont {{Lazaridis}},
  \citenamefont {{Lee}}, \citenamefont {{Liu}}, \citenamefont {{Lyne}},
  \citenamefont {{McKee}}, \citenamefont {{Mingarelli}}, \citenamefont
  {{Perrodin}}, \citenamefont {{Petiteau}}, \citenamefont {{Possenti}},
  \citenamefont {{Purver}}, \citenamefont {{Rosado}}, \citenamefont
  {{Sanidas}}, \citenamefont {{Sesana}}, \citenamefont {{Shaifullah}},
  \citenamefont {{Smits}}, \citenamefont {{Taylor}}, \citenamefont
  {{Theureau}}, \citenamefont {{Tiburzi}}, \citenamefont {{van Haasteren}},\
  and\ \citenamefont {{Vecchio}}}]{DesvignesCaballero2016}%
  \BibitemOpen
  \bibfield  {author} {\bibinfo {author} {\bibfnamefont {G.}~\bibnamefont
  {{Desvignes}}}, \bibinfo {author} {\bibfnamefont {R.~N.}\ \bibnamefont
  {{Caballero}}}, \bibinfo {author} {\bibfnamefont {L.}~\bibnamefont
  {{Lentati}}}, \bibinfo {author} {\bibfnamefont {J.~P.~W.}\ \bibnamefont
  {{Verbiest}}}, \bibinfo {author} {\bibfnamefont {D.~J.}\ \bibnamefont
  {{Champion}}}, \bibinfo {author} {\bibfnamefont {B.~W.}\ \bibnamefont
  {{Stappers}}}, \bibinfo {author} {\bibfnamefont {G.~H.}\ \bibnamefont
  {{Janssen}}}, \bibinfo {author} {\bibfnamefont {P.}~\bibnamefont
  {{Lazarus}}}, \bibinfo {author} {\bibfnamefont {S.}~\bibnamefont
  {{Os{\l}owski}}}, \bibinfo {author} {\bibfnamefont {S.}~\bibnamefont
  {{Babak}}}, \bibinfo {author} {\bibfnamefont {C.~G.}\ \bibnamefont
  {{Bassa}}}, \bibinfo {author} {\bibfnamefont {P.}~\bibnamefont {{Brem}}},
  \bibinfo {author} {\bibfnamefont {M.}~\bibnamefont {{Burgay}}}, \bibinfo
  {author} {\bibfnamefont {I.}~\bibnamefont {{Cognard}}}, \bibinfo {author}
  {\bibfnamefont {J.~R.}\ \bibnamefont {{Gair}}}, \bibinfo {author}
  {\bibfnamefont {E.}~\bibnamefont {{Graikou}}}, \bibinfo {author}
  {\bibfnamefont {L.}~\bibnamefont {{Guillemot}}}, \bibinfo {author}
  {\bibfnamefont {J.~W.~T.}\ \bibnamefont {{Hessels}}}, \bibinfo {author}
  {\bibfnamefont {A.}~\bibnamefont {{Jessner}}}, \bibinfo {author}
  {\bibfnamefont {C.}~\bibnamefont {{Jordan}}}, \bibinfo {author}
  {\bibfnamefont {R.}~\bibnamefont {{Karuppusamy}}}, \bibinfo {author}
  {\bibfnamefont {M.}~\bibnamefont {{Kramer}}}, \bibinfo {author}
  {\bibfnamefont {A.}~\bibnamefont {{Lassus}}}, \bibinfo {author}
  {\bibfnamefont {K.}~\bibnamefont {{Lazaridis}}}, \bibinfo {author}
  {\bibfnamefont {K.~J.}\ \bibnamefont {{Lee}}}, \bibinfo {author}
  {\bibfnamefont {K.}~\bibnamefont {{Liu}}}, \bibinfo {author} {\bibfnamefont
  {A.~G.}\ \bibnamefont {{Lyne}}}, \bibinfo {author} {\bibfnamefont
  {J.}~\bibnamefont {{McKee}}}, \bibinfo {author} {\bibfnamefont {C.~M.~F.}\
  \bibnamefont {{Mingarelli}}}, \bibinfo {author} {\bibfnamefont
  {D.}~\bibnamefont {{Perrodin}}}, \bibinfo {author} {\bibfnamefont
  {A.}~\bibnamefont {{Petiteau}}}, \bibinfo {author} {\bibfnamefont
  {A.}~\bibnamefont {{Possenti}}}, \bibinfo {author} {\bibfnamefont {M.~B.}\
  \bibnamefont {{Purver}}}, \bibinfo {author} {\bibfnamefont {P.~A.}\
  \bibnamefont {{Rosado}}}, \bibinfo {author} {\bibfnamefont {S.}~\bibnamefont
  {{Sanidas}}}, \bibinfo {author} {\bibfnamefont {A.}~\bibnamefont {{Sesana}}},
  \bibinfo {author} {\bibfnamefont {G.}~\bibnamefont {{Shaifullah}}}, \bibinfo
  {author} {\bibfnamefont {R.}~\bibnamefont {{Smits}}}, \bibinfo {author}
  {\bibfnamefont {S.~R.}\ \bibnamefont {{Taylor}}}, \bibinfo {author}
  {\bibfnamefont {G.}~\bibnamefont {{Theureau}}}, \bibinfo {author}
  {\bibfnamefont {C.}~\bibnamefont {{Tiburzi}}}, \bibinfo {author}
  {\bibfnamefont {R.}~\bibnamefont {{van Haasteren}}}, \ and\ \bibinfo {author}
  {\bibfnamefont {A.}~\bibnamefont {{Vecchio}}},\ }\bibfield  {title} {\enquote
  {\bibinfo {title} {{High-precision timing of 42 millisecond pulsars with the
  European Pulsar Timing Array}},}\ }\href {\doibase 10.1093/mnras/stw483}
  {\bibfield  {journal} {\bibinfo  {journal} {\mnras}\ }\textbf {\bibinfo
  {volume} {458}},\ \bibinfo {pages} {3341--3380} (\bibinfo {year} {2016})},\
  \Eprint {http://arxiv.org/abs/1602.08511} {arXiv:1602.08511 [astro-ph.HE]}
  \BibitemShut {NoStop}%
\bibitem [{\citenamefont {{Edwards}}\ \emph {et~al.}(2006)\citenamefont
  {{Edwards}}, \citenamefont {{Hobbs}},\ and\ \citenamefont
  {{Manchester}}}]{EdwardsHobbs2006}%
  \BibitemOpen
  \bibfield  {author} {\bibinfo {author} {\bibfnamefont {R.~T.}\ \bibnamefont
  {{Edwards}}}, \bibinfo {author} {\bibfnamefont {G.~B.}\ \bibnamefont
  {{Hobbs}}}, \ and\ \bibinfo {author} {\bibfnamefont {R.~N.}\ \bibnamefont
  {{Manchester}}},\ }\bibfield  {title} {\enquote {\bibinfo {title} {{TEMPO2, a
  new pulsar timing package - II. The timing model and precision estimates}},}\
  }\href {\doibase 10.1111/j.1365-2966.2006.10870.x} {\bibfield  {journal}
  {\bibinfo  {journal} {\mnras}\ }\textbf {\bibinfo {volume} {372}},\ \bibinfo
  {pages} {1549--1574} (\bibinfo {year} {2006})},\ \Eprint
  {http://arxiv.org/abs/astro-ph/0607664} {arXiv:astro-ph/0607664 [astro-ph]}
  \BibitemShut {NoStop}%
\bibitem [{\citenamefont {{Goncharov}}\ \emph
  {et~al.}(2021{\natexlab{a}})\citenamefont {{Goncharov}}, \citenamefont
  {{Reardon}}, \citenamefont {{Shannon}}, \citenamefont {{Zhu}}, \citenamefont
  {{Thrane}}, \citenamefont {{Bailes}}, \citenamefont {{Bhat}}, \citenamefont
  {{Dai}}, \citenamefont {{Hobbs}}, \citenamefont {{Kerr}}, \citenamefont
  {{Manchester}}, \citenamefont {{Os{\l}owski}}, \citenamefont
  {{Parthasarathy}}, \citenamefont {{Russell}}, \citenamefont {{Spiewak}},
  \citenamefont {{Thyagarajan}},\ and\ \citenamefont
  {{Wang}}}]{GoncharovReardon2021}%
  \BibitemOpen
  \bibfield  {author} {\bibinfo {author} {\bibfnamefont {Boris}\ \bibnamefont
  {{Goncharov}}}, \bibinfo {author} {\bibfnamefont {D.~J.}\ \bibnamefont
  {{Reardon}}}, \bibinfo {author} {\bibfnamefont {R.~M.}\ \bibnamefont
  {{Shannon}}}, \bibinfo {author} {\bibfnamefont {Xing-Jiang}\ \bibnamefont
  {{Zhu}}}, \bibinfo {author} {\bibfnamefont {Eric}\ \bibnamefont {{Thrane}}},
  \bibinfo {author} {\bibfnamefont {M.}~\bibnamefont {{Bailes}}}, \bibinfo
  {author} {\bibfnamefont {N.~D.~R.}\ \bibnamefont {{Bhat}}}, \bibinfo {author}
  {\bibfnamefont {S.}~\bibnamefont {{Dai}}}, \bibinfo {author} {\bibfnamefont
  {G.}~\bibnamefont {{Hobbs}}}, \bibinfo {author} {\bibfnamefont
  {M.}~\bibnamefont {{Kerr}}}, \bibinfo {author} {\bibfnamefont {R.~N.}\
  \bibnamefont {{Manchester}}}, \bibinfo {author} {\bibfnamefont
  {S.}~\bibnamefont {{Os{\l}owski}}}, \bibinfo {author} {\bibfnamefont
  {A.}~\bibnamefont {{Parthasarathy}}}, \bibinfo {author} {\bibfnamefont
  {C.~J.}\ \bibnamefont {{Russell}}}, \bibinfo {author} {\bibfnamefont
  {R.}~\bibnamefont {{Spiewak}}}, \bibinfo {author} {\bibfnamefont
  {N.}~\bibnamefont {{Thyagarajan}}}, \ and\ \bibinfo {author} {\bibfnamefont
  {J.~B.}\ \bibnamefont {{Wang}}},\ }\bibfield  {title} {\enquote {\bibinfo
  {title} {{Identifying and mitigating noise sources in precision pulsar timing
  data sets}},}\ }\href {\doibase 10.1093/mnras/staa3411} {\bibfield  {journal}
  {\bibinfo  {journal} {\mnras}\ }\textbf {\bibinfo {volume} {502}},\ \bibinfo
  {pages} {478--493} (\bibinfo {year} {2021}{\natexlab{a}})},\ \Eprint
  {http://arxiv.org/abs/2010.06109} {arXiv:2010.06109 [astro-ph.HE]}
  \BibitemShut {NoStop}%
\bibitem [{\citenamefont {{EPTA Collaboration and InPTA Collaboration}}\ \emph
  {et~al.}(2023{\natexlab{a}})\citenamefont {{EPTA Collaboration and InPTA
  Collaboration}}, \citenamefont {{Antoniadis, J.}}, \citenamefont {{Arumugam,
  P.}}, \citenamefont {{Arumugam, S.}}, \citenamefont {{Babak, S.}},
  \citenamefont {{Bagchi, M.}}, \citenamefont {{Nielsen, A.-S. Bak}},
  \citenamefont {{Bassa, C. G.}}, \citenamefont {{Bathula, A.}}, \citenamefont
  {{Berthereau, A.}}, \citenamefont {{Bonetti, M.}}, \citenamefont {{Bortolas,
  E.}}, \citenamefont {{Brook, P. R.}}, \citenamefont {{Burgay, M.}},
  \citenamefont {{Caballero, R. N.}}, \citenamefont {{Chalumeau, A.}},
  \citenamefont {{Champion, D. J.}}, \citenamefont {{Chanlaridis, S.}},
  \citenamefont {{Chen, S.}}, \citenamefont {{Cognard, I.}}, \citenamefont
  {{Dandapat, S.}}, \citenamefont {{Deb, D.}}, \citenamefont {{Desai, S.}},
  \citenamefont {{Desvignes, G.}}, \citenamefont {{Dhanda-Batra, N.}},
  \citenamefont {{Dwivedi, C.}}, \citenamefont {{Falxa, M.}}, \citenamefont
  {{Ferdman, R. D.}}, \citenamefont {{Franchini, A.}}, \citenamefont {{Gair, J.
  R.}}, \citenamefont {{Goncharov, B.}}, \citenamefont {{Gopakumar, A.}},
  \citenamefont {{Graikou, E.}}, \citenamefont {{Grießmeier, J.-M.}},
  \citenamefont {{Guillemot, L.}}, \citenamefont {{Guo, Y. J.}}, \citenamefont
  {{Gupta, Y.}}, \citenamefont {{Hisano, S.}}, \citenamefont {{Hu, H.}},
  \citenamefont {{Iraci, F.}}, \citenamefont {{Izquierdo-Villalba, D.}},
  \citenamefont {{Jang, J.}}, \citenamefont {{Jawor, J.}}, \citenamefont
  {{Janssen, G. H.}}, \citenamefont {{Jessner, A.}}, \citenamefont {{Joshi, B.
  C.}}, \citenamefont {{Kareem, F.}}, \citenamefont {{Karuppusamy, R.}},
  \citenamefont {{Keane, E. F.}}, \citenamefont {{Keith, M. J.}}, \citenamefont
  {{Kharbanda, D.}}, \citenamefont {{Kikunaga, T.}}, \citenamefont {{Kolhe,
  N.}}, \citenamefont {{Kramer, M.}}, \citenamefont {{Krishnakumar, M. A.}},
  \citenamefont {{Lackeos, K.}}, \citenamefont {{Lee, K. J.}}, \citenamefont
  {{Liu, K.}}, \citenamefont {{Liu, Y.}}, \citenamefont {{Lyne, A. G.}},
  \citenamefont {{McKee, J. W.}}, \citenamefont {{Maan, Y.}}, \citenamefont
  {{Main, R. A.}}, \citenamefont {{Mickaliger, M. B.}}, \citenamefont {{Niţu,
  I. C.}}, \citenamefont {{Nobleson, K.}}, \citenamefont {{Paladi, A. K.}},
  \citenamefont {{Parthasarathy, A.}}, \citenamefont {{Perera, B. B. P.}},
  \citenamefont {{Perrodin, D.}}, \citenamefont {{Petiteau, A.}}, \citenamefont
  {{Porayko, N. K.}}, \citenamefont {{Possenti, A.}}, \citenamefont {{Prabu,
  T.}}, \citenamefont {{Leclere, H. Quelquejay}}, \citenamefont {{Rana, P.}},
  \citenamefont {{Samajdar, A.}}, \citenamefont {{Sanidas, S. A.}},
  \citenamefont {{Sesana, A.}}, \citenamefont {{Shaifullah, G.}}, \citenamefont
  {{Singha, J.}}, \citenamefont {{Speri, L.}}, \citenamefont {{Spiewak, R.}},
  \citenamefont {{Srivastava, A.}}, \citenamefont {{Stappers, B. W.}},
  \citenamefont {{Surnis, M.}}, \citenamefont {{Susarla, S. C.}}, \citenamefont
  {{Susobhanan, A.}}, \citenamefont {{Takahashi, K.}}, \citenamefont
  {{Tarafdar, P.}}, \citenamefont {{Theureau, G.}}, \citenamefont {{Tiburzi,
  C.}}, \citenamefont {{van der Wateren, E.}}, \citenamefont {{Vecchio, A.}},
  \citenamefont {{Krishnan, V. Venkatraman}}, \citenamefont {{Verbiest, J. P.
  W.}}, \citenamefont {{Wang, J.}}, \citenamefont {{Wang, L.}},\ and\
  \citenamefont {{Wu, Z.}}}]{EPTA_II_2023}%
  \BibitemOpen
  \bibfield  {author} {\bibinfo {author} {\bibnamefont {{EPTA Collaboration and
  InPTA Collaboration}}}, \bibinfo {author} {\bibnamefont {{Antoniadis, J.}}},
  \bibinfo {author} {\bibnamefont {{Arumugam, P.}}}, \bibinfo {author}
  {\bibnamefont {{Arumugam, S.}}}, \bibinfo {author} {\bibnamefont {{Babak,
  S.}}}, \bibinfo {author} {\bibnamefont {{Bagchi, M.}}}, \bibinfo {author}
  {\bibnamefont {{Nielsen, A.-S. Bak}}}, \bibinfo {author} {\bibnamefont
  {{Bassa, C. G.}}}, \bibinfo {author} {\bibnamefont {{Bathula, A.}}}, \bibinfo
  {author} {\bibnamefont {{Berthereau, A.}}}, \bibinfo {author} {\bibnamefont
  {{Bonetti, M.}}}, \bibinfo {author} {\bibnamefont {{Bortolas, E.}}}, \bibinfo
  {author} {\bibnamefont {{Brook, P. R.}}}, \bibinfo {author} {\bibnamefont
  {{Burgay, M.}}}, \bibinfo {author} {\bibnamefont {{Caballero, R. N.}}},
  \bibinfo {author} {\bibnamefont {{Chalumeau, A.}}}, \bibinfo {author}
  {\bibnamefont {{Champion, D. J.}}}, \bibinfo {author} {\bibnamefont
  {{Chanlaridis, S.}}}, \bibinfo {author} {\bibnamefont {{Chen, S.}}}, \bibinfo
  {author} {\bibnamefont {{Cognard, I.}}}, \bibinfo {author} {\bibnamefont
  {{Dandapat, S.}}}, \bibinfo {author} {\bibnamefont {{Deb, D.}}}, \bibinfo
  {author} {\bibnamefont {{Desai, S.}}}, \bibinfo {author} {\bibnamefont
  {{Desvignes, G.}}}, \bibinfo {author} {\bibnamefont {{Dhanda-Batra, N.}}},
  \bibinfo {author} {\bibnamefont {{Dwivedi, C.}}}, \bibinfo {author}
  {\bibnamefont {{Falxa, M.}}}, \bibinfo {author} {\bibnamefont {{Ferdman, R.
  D.}}}, \bibinfo {author} {\bibnamefont {{Franchini, A.}}}, \bibinfo {author}
  {\bibnamefont {{Gair, J. R.}}}, \bibinfo {author} {\bibnamefont {{Goncharov,
  B.}}}, \bibinfo {author} {\bibnamefont {{Gopakumar, A.}}}, \bibinfo {author}
  {\bibnamefont {{Graikou, E.}}}, \bibinfo {author} {\bibnamefont
  {{Grießmeier, J.-M.}}}, \bibinfo {author} {\bibnamefont {{Guillemot, L.}}},
  \bibinfo {author} {\bibnamefont {{Guo, Y. J.}}}, \bibinfo {author}
  {\bibnamefont {{Gupta, Y.}}}, \bibinfo {author} {\bibnamefont {{Hisano,
  S.}}}, \bibinfo {author} {\bibnamefont {{Hu, H.}}}, \bibinfo {author}
  {\bibnamefont {{Iraci, F.}}}, \bibinfo {author} {\bibnamefont
  {{Izquierdo-Villalba, D.}}}, \bibinfo {author} {\bibnamefont {{Jang, J.}}},
  \bibinfo {author} {\bibnamefont {{Jawor, J.}}}, \bibinfo {author}
  {\bibnamefont {{Janssen, G. H.}}}, \bibinfo {author} {\bibnamefont {{Jessner,
  A.}}}, \bibinfo {author} {\bibnamefont {{Joshi, B. C.}}}, \bibinfo {author}
  {\bibnamefont {{Kareem, F.}}}, \bibinfo {author} {\bibnamefont {{Karuppusamy,
  R.}}}, \bibinfo {author} {\bibnamefont {{Keane, E. F.}}}, \bibinfo {author}
  {\bibnamefont {{Keith, M. J.}}}, \bibinfo {author} {\bibnamefont {{Kharbanda,
  D.}}}, \bibinfo {author} {\bibnamefont {{Kikunaga, T.}}}, \bibinfo {author}
  {\bibnamefont {{Kolhe, N.}}}, \bibinfo {author} {\bibnamefont {{Kramer,
  M.}}}, \bibinfo {author} {\bibnamefont {{Krishnakumar, M. A.}}}, \bibinfo
  {author} {\bibnamefont {{Lackeos, K.}}}, \bibinfo {author} {\bibnamefont
  {{Lee, K. J.}}}, \bibinfo {author} {\bibnamefont {{Liu, K.}}}, \bibinfo
  {author} {\bibnamefont {{Liu, Y.}}}, \bibinfo {author} {\bibnamefont {{Lyne,
  A. G.}}}, \bibinfo {author} {\bibnamefont {{McKee, J. W.}}}, \bibinfo
  {author} {\bibnamefont {{Maan, Y.}}}, \bibinfo {author} {\bibnamefont {{Main,
  R. A.}}}, \bibinfo {author} {\bibnamefont {{Mickaliger, M. B.}}}, \bibinfo
  {author} {\bibnamefont {{Niţu, I. C.}}}, \bibinfo {author} {\bibnamefont
  {{Nobleson, K.}}}, \bibinfo {author} {\bibnamefont {{Paladi, A. K.}}},
  \bibinfo {author} {\bibnamefont {{Parthasarathy, A.}}}, \bibinfo {author}
  {\bibnamefont {{Perera, B. B. P.}}}, \bibinfo {author} {\bibnamefont
  {{Perrodin, D.}}}, \bibinfo {author} {\bibnamefont {{Petiteau, A.}}},
  \bibinfo {author} {\bibnamefont {{Porayko, N. K.}}}, \bibinfo {author}
  {\bibnamefont {{Possenti, A.}}}, \bibinfo {author} {\bibnamefont {{Prabu,
  T.}}}, \bibinfo {author} {\bibnamefont {{Leclere, H. Quelquejay}}}, \bibinfo
  {author} {\bibnamefont {{Rana, P.}}}, \bibinfo {author} {\bibnamefont
  {{Samajdar, A.}}}, \bibinfo {author} {\bibnamefont {{Sanidas, S. A.}}},
  \bibinfo {author} {\bibnamefont {{Sesana, A.}}}, \bibinfo {author}
  {\bibnamefont {{Shaifullah, G.}}}, \bibinfo {author} {\bibnamefont {{Singha,
  J.}}}, \bibinfo {author} {\bibnamefont {{Speri, L.}}}, \bibinfo {author}
  {\bibnamefont {{Spiewak, R.}}}, \bibinfo {author} {\bibnamefont {{Srivastava,
  A.}}}, \bibinfo {author} {\bibnamefont {{Stappers, B. W.}}}, \bibinfo
  {author} {\bibnamefont {{Surnis, M.}}}, \bibinfo {author} {\bibnamefont
  {{Susarla, S. C.}}}, \bibinfo {author} {\bibnamefont {{Susobhanan, A.}}},
  \bibinfo {author} {\bibnamefont {{Takahashi, K.}}}, \bibinfo {author}
  {\bibnamefont {{Tarafdar, P.}}}, \bibinfo {author} {\bibnamefont {{Theureau,
  G.}}}, \bibinfo {author} {\bibnamefont {{Tiburzi, C.}}}, \bibinfo {author}
  {\bibnamefont {{van der Wateren, E.}}}, \bibinfo {author} {\bibnamefont
  {{Vecchio, A.}}}, \bibinfo {author} {\bibnamefont {{Krishnan, V.
  Venkatraman}}}, \bibinfo {author} {\bibnamefont {{Verbiest, J. P. W.}}},
  \bibinfo {author} {\bibnamefont {{Wang, J.}}}, \bibinfo {author}
  {\bibnamefont {{Wang, L.}}}, \ and\ \bibinfo {author} {\bibnamefont {{Wu,
  Z.}}},\ }\bibfield  {title} {\enquote {\bibinfo {title} {The second data
  release from the european pulsar timing array - ii. customised pulsar noise
  models for spatially correlated gravitational waves},}\ }\href {\doibase
  10.1051/0004-6361/202346842} {\bibfield  {journal} {\bibinfo  {journal} {A\&
  A}\ }\textbf {\bibinfo {volume} {678}},\ \bibinfo {pages} {A49} (\bibinfo
  {year} {2023}{\natexlab{a}})}\BibitemShut {NoStop}%
\bibitem [{\citenamefont {Agazie}\ \emph {et~al.}(2023)\citenamefont {Agazie}
  \emph {et~al.}}]{NANOGrav_2023}%
  \BibitemOpen
  \bibfield  {author} {\bibinfo {author} {\bibfnamefont {Gabriella}\
  \bibnamefont {Agazie}} \emph {et~al.} (\bibinfo {collaboration} {NANOGrav}),\
  }\bibfield  {title} {\enquote {\bibinfo {title} {{The NANOGrav 15 yr Data
  Set: Evidence for a Gravitational-wave Background}},}\ }\href {\doibase
  10.3847/2041-8213/acdac6} {\bibfield  {journal} {\bibinfo  {journal}
  {Astrophys. J. Lett.}\ }\textbf {\bibinfo {volume} {951}},\ \bibinfo {pages}
  {L8} (\bibinfo {year} {2023})},\ \Eprint {http://arxiv.org/abs/2306.16213}
  {arXiv:2306.16213 [astro-ph.HE]} \BibitemShut {NoStop}%
\bibitem [{\citenamefont {Antoniadis}\ \emph
  {et~al.}(2023{\natexlab{a}})\citenamefont {Antoniadis} \emph
  {et~al.}}]{EPTA_2023}%
  \BibitemOpen
  \bibfield  {author} {\bibinfo {author} {\bibfnamefont {J.}~\bibnamefont
  {Antoniadis}} \emph {et~al.} (\bibinfo {collaboration} {EPTA, InPTA:}),\
  }\bibfield  {title} {\enquote {\bibinfo {title} {{The second data release
  from the European Pulsar Timing Array - III. Search for gravitational wave
  signals}},}\ }\href {\doibase 10.1051/0004-6361/202346844} {\bibfield
  {journal} {\bibinfo  {journal} {Astron. Astrophys.}\ }\textbf {\bibinfo
  {volume} {678}},\ \bibinfo {pages} {A50} (\bibinfo {year}
  {2023}{\natexlab{a}})},\ \Eprint {http://arxiv.org/abs/2306.16214}
  {arXiv:2306.16214 [astro-ph.HE]} \BibitemShut {NoStop}%
\bibitem [{\citenamefont {Reardon}\ \emph {et~al.}(2023)\citenamefont {Reardon}
  \emph {et~al.}}]{PPTA_2023}%
  \BibitemOpen
  \bibfield  {author} {\bibinfo {author} {\bibfnamefont {Daniel~J.}\
  \bibnamefont {Reardon}} \emph {et~al.},\ }\bibfield  {title} {\enquote
  {\bibinfo {title} {{Search for an Isotropic Gravitational-wave Background
  with the Parkes Pulsar Timing Array}},}\ }\href {\doibase
  10.3847/2041-8213/acdd02} {\bibfield  {journal} {\bibinfo  {journal}
  {Astrophys. J. Lett.}\ }\textbf {\bibinfo {volume} {951}},\ \bibinfo {pages}
  {L6} (\bibinfo {year} {2023})},\ \Eprint {http://arxiv.org/abs/2306.16215}
  {arXiv:2306.16215 [astro-ph.HE]} \BibitemShut {NoStop}%
\bibitem [{\citenamefont {{Xu}}\ \emph {et~al.}(2023)\citenamefont {{Xu}},
  \citenamefont {{Chen}}, \citenamefont {{Guo}}, \citenamefont {{Jiang}},
  \citenamefont {{Wang}}, \citenamefont {{Xu}}, \citenamefont {{Xue}},
  \citenamefont {{Nicolas Caballero}}, \citenamefont {{Yuan}}, \citenamefont
  {{Xu}}, \citenamefont {{Wang}}, \citenamefont {{Hao}}, \citenamefont {{Luo}},
  \citenamefont {{Lee}}, \citenamefont {{Han}}, \citenamefont {{Jiang}},
  \citenamefont {{Shen}}, \citenamefont {{Wang}}, \citenamefont {{Wang}},
  \citenamefont {{Xu}}, \citenamefont {{Wu}}, \citenamefont {{Manchester}},
  \citenamefont {{Qian}}, \citenamefont {{Guan}}, \citenamefont {{Huang}},
  \citenamefont {{Sun}},\ and\ \citenamefont {{Zhu}}}]{CPTA}%
  \BibitemOpen
  \bibfield  {author} {\bibinfo {author} {\bibfnamefont {Heng}\ \bibnamefont
  {{Xu}}}, \bibinfo {author} {\bibfnamefont {Siyuan}\ \bibnamefont {{Chen}}},
  \bibinfo {author} {\bibfnamefont {Yanjun}\ \bibnamefont {{Guo}}}, \bibinfo
  {author} {\bibfnamefont {Jinchen}\ \bibnamefont {{Jiang}}}, \bibinfo {author}
  {\bibfnamefont {Bojun}\ \bibnamefont {{Wang}}}, \bibinfo {author}
  {\bibfnamefont {Jiangwei}\ \bibnamefont {{Xu}}}, \bibinfo {author}
  {\bibfnamefont {Zihan}\ \bibnamefont {{Xue}}}, \bibinfo {author}
  {\bibfnamefont {R.}~\bibnamefont {{Nicolas Caballero}}}, \bibinfo {author}
  {\bibfnamefont {Jianping}\ \bibnamefont {{Yuan}}}, \bibinfo {author}
  {\bibfnamefont {Yonghua}\ \bibnamefont {{Xu}}}, \bibinfo {author}
  {\bibfnamefont {Jingbo}\ \bibnamefont {{Wang}}}, \bibinfo {author}
  {\bibfnamefont {Longfei}\ \bibnamefont {{Hao}}}, \bibinfo {author}
  {\bibfnamefont {Jingtao}\ \bibnamefont {{Luo}}}, \bibinfo {author}
  {\bibfnamefont {Kejia}\ \bibnamefont {{Lee}}}, \bibinfo {author}
  {\bibfnamefont {Jinlin}\ \bibnamefont {{Han}}}, \bibinfo {author}
  {\bibfnamefont {Peng}\ \bibnamefont {{Jiang}}}, \bibinfo {author}
  {\bibfnamefont {Zhiqiang}\ \bibnamefont {{Shen}}}, \bibinfo {author}
  {\bibfnamefont {Min}\ \bibnamefont {{Wang}}}, \bibinfo {author}
  {\bibfnamefont {Na}~\bibnamefont {{Wang}}}, \bibinfo {author} {\bibfnamefont
  {Renxin}\ \bibnamefont {{Xu}}}, \bibinfo {author} {\bibfnamefont {Xiangping}\
  \bibnamefont {{Wu}}}, \bibinfo {author} {\bibfnamefont {Richard}\
  \bibnamefont {{Manchester}}}, \bibinfo {author} {\bibfnamefont {Lei}\
  \bibnamefont {{Qian}}}, \bibinfo {author} {\bibfnamefont {Xin}\ \bibnamefont
  {{Guan}}}, \bibinfo {author} {\bibfnamefont {Menglin}\ \bibnamefont
  {{Huang}}}, \bibinfo {author} {\bibfnamefont {Chun}\ \bibnamefont {{Sun}}}, \
  and\ \bibinfo {author} {\bibfnamefont {Yan}\ \bibnamefont {{Zhu}}},\
  }\bibfield  {title} {\enquote {\bibinfo {title} {{Searching for the
  Nano-Hertz Stochastic Gravitational Wave Background with the Chinese Pulsar
  Timing Array Data Release I}},}\ }\href {\doibase 10.1088/1674-4527/acdfa5}
  {\bibfield  {journal} {\bibinfo  {journal} {Research in Astronomy and
  Astrophysics}\ }\textbf {\bibinfo {volume} {23}},\ \bibinfo {eid} {075024}
  (\bibinfo {year} {2023})},\ \Eprint {http://arxiv.org/abs/2306.16216}
  {arXiv:2306.16216 [astro-ph.HE]} \BibitemShut {NoStop}%
\bibitem [{\citenamefont {{Hellings}}\ and\ \citenamefont
  {{Downs}}(1983)}]{HellingsDowns1983}%
  \BibitemOpen
  \bibfield  {author} {\bibinfo {author} {\bibfnamefont {R.~W.}\ \bibnamefont
  {{Hellings}}}\ and\ \bibinfo {author} {\bibfnamefont {G.~S.}\ \bibnamefont
  {{Downs}}},\ }\bibfield  {title} {\enquote {\bibinfo {title} {{Upper limits
  on the isotropic gravitational radiation background from pulsar timing
  analysis.}}}\ }\href {\doibase 10.1086/183954} {\bibfield  {journal}
  {\bibinfo  {journal} {\apjl}\ }\textbf {\bibinfo {volume} {265}},\ \bibinfo
  {pages} {L39--L42} (\bibinfo {year} {1983})}\BibitemShut {NoStop}%
\bibitem [{\citenamefont {{EPTA Collaboration and InPTA Collaboration}}\ \emph
  {et~al.}(2023{\natexlab{b}})\citenamefont {{EPTA Collaboration and InPTA
  Collaboration}}, \citenamefont {{Antoniadis, J.}}, \citenamefont {{Arumugam,
  P.}}, \citenamefont {{Arumugam, S.}}, \citenamefont {{Babak, S.}},
  \citenamefont {{Bagchi, M.}}, \citenamefont {{Bak Nielsen, A.-S.}},
  \citenamefont {{Bassa, C. G.}}, \citenamefont {{Bathula, A.}}, \citenamefont
  {{Berthereau, A.}}, \citenamefont {{Bonetti, M.}}, \citenamefont {{Bortolas,
  E.}}, \citenamefont {{Brook, P. R.}}, \citenamefont {{Burgay, M.}},
  \citenamefont {{Caballero, R. N.}}, \citenamefont {{Chalumeau, A.}},
  \citenamefont {{Champion, D. J.}}, \citenamefont {{Chanlaridis, S.}},
  \citenamefont {{Chen, S.}}, \citenamefont {{Cognard, I.}}, \citenamefont
  {{Dandapat, S.}}, \citenamefont {{Deb, D.}}, \citenamefont {{Desai, S.}},
  \citenamefont {{Desvignes, G.}}, \citenamefont {{Dhanda-Batra, N.}},
  \citenamefont {{Dwivedi, C.}}, \citenamefont {{Falxa, M.}}, \citenamefont
  {{Ferdman, R. D.}}, \citenamefont {{Franchini, A.}}, \citenamefont {{Gair, J.
  R.}}, \citenamefont {{Goncharov, B.}}, \citenamefont {{Gopakumar, A.}},
  \citenamefont {{Graikou, E.}}, \citenamefont {{Grießmeier, J.-M.}},
  \citenamefont {{Guillemot, L.}}, \citenamefont {{Guo, Y. J.}}, \citenamefont
  {{Gupta, Y.}}, \citenamefont {{Hisano, S.}}, \citenamefont {{Hu, H.}},
  \citenamefont {{Iraci, F.}}, \citenamefont {{Izquierdo-Villalba, D.}},
  \citenamefont {{Jang, J.}}, \citenamefont {{Jawor, J.}}, \citenamefont
  {{Janssen, G. H.}}, \citenamefont {{Jessner, A.}}, \citenamefont {{Joshi, B.
  C.}}, \citenamefont {{Kareem, F.}}, \citenamefont {{Karuppusamy, R.}},
  \citenamefont {{Keane, E. F.}}, \citenamefont {{Keith, M. J.}}, \citenamefont
  {{Kharbanda, D.}}, \citenamefont {{Kikunaga, T.}}, \citenamefont {{Kolhe,
  N.}}, \citenamefont {{Kramer, M.}}, \citenamefont {{Krishnakumar, M. A.}},
  \citenamefont {{Lackeos, K.}}, \citenamefont {{Lee, K. J.}}, \citenamefont
  {{Liu, K.}}, \citenamefont {{Liu, Y.}}, \citenamefont {{Lyne, A. G.}},
  \citenamefont {{McKee, J. W.}}, \citenamefont {{Maan, Y.}}, \citenamefont
  {{Main, R. A.}}, \citenamefont {{Mickaliger, M. B.}}, \citenamefont {{Niţu,
  I. C.}}, \citenamefont {{Nobleson, K.}}, \citenamefont {{Paladi, A. K.}},
  \citenamefont {{Parthasarathy, A.}}, \citenamefont {{Perera, B. B. P.}},
  \citenamefont {{Perrodin, D.}}, \citenamefont {{Petiteau, A.}}, \citenamefont
  {{Porayko, N. K.}}, \citenamefont {{Possenti, A.}}, \citenamefont {{Prabu,
  T.}}, \citenamefont {{Quelquejay Leclere, H.}}, \citenamefont {{Rana, P.}},
  \citenamefont {{Samajdar, A.}}, \citenamefont {{Sanidas, S. A.}},
  \citenamefont {{Sesana, A.}}, \citenamefont {{Shaifullah, G.}}, \citenamefont
  {{Singha, J.}}, \citenamefont {{Speri, L.}}, \citenamefont {{Spiewak, R.}},
  \citenamefont {{Srivastava, A.}}, \citenamefont {{Stappers, B. W.}},
  \citenamefont {{Surnis, M.}}, \citenamefont {{Susarla, S. C.}}, \citenamefont
  {{Susobhanan, A.}}, \citenamefont {{Takahashi, K.}}, \citenamefont
  {{Tarafdar, P.}}, \citenamefont {{Theureau, G.}}, \citenamefont {{Tiburzi,
  C.}}, \citenamefont {{van der Wateren, E.}}, \citenamefont {{Vecchio, A.}},
  \citenamefont {{Venkatraman Krishnan, V.}}, \citenamefont {{Verbiest, J. P.
  W.}}, \citenamefont {{Wang, J.}}, \citenamefont {{Wang, L.}},\ and\
  \citenamefont {{Wu, Z.}}}]{EPTA_III_2023}%
  \BibitemOpen
  \bibfield  {author} {\bibinfo {author} {\bibnamefont {{EPTA Collaboration and
  InPTA Collaboration}}}, \bibinfo {author} {\bibnamefont {{Antoniadis, J.}}},
  \bibinfo {author} {\bibnamefont {{Arumugam, P.}}}, \bibinfo {author}
  {\bibnamefont {{Arumugam, S.}}}, \bibinfo {author} {\bibnamefont {{Babak,
  S.}}}, \bibinfo {author} {\bibnamefont {{Bagchi, M.}}}, \bibinfo {author}
  {\bibnamefont {{Bak Nielsen, A.-S.}}}, \bibinfo {author} {\bibnamefont
  {{Bassa, C. G.}}}, \bibinfo {author} {\bibnamefont {{Bathula, A.}}}, \bibinfo
  {author} {\bibnamefont {{Berthereau, A.}}}, \bibinfo {author} {\bibnamefont
  {{Bonetti, M.}}}, \bibinfo {author} {\bibnamefont {{Bortolas, E.}}}, \bibinfo
  {author} {\bibnamefont {{Brook, P. R.}}}, \bibinfo {author} {\bibnamefont
  {{Burgay, M.}}}, \bibinfo {author} {\bibnamefont {{Caballero, R. N.}}},
  \bibinfo {author} {\bibnamefont {{Chalumeau, A.}}}, \bibinfo {author}
  {\bibnamefont {{Champion, D. J.}}}, \bibinfo {author} {\bibnamefont
  {{Chanlaridis, S.}}}, \bibinfo {author} {\bibnamefont {{Chen, S.}}}, \bibinfo
  {author} {\bibnamefont {{Cognard, I.}}}, \bibinfo {author} {\bibnamefont
  {{Dandapat, S.}}}, \bibinfo {author} {\bibnamefont {{Deb, D.}}}, \bibinfo
  {author} {\bibnamefont {{Desai, S.}}}, \bibinfo {author} {\bibnamefont
  {{Desvignes, G.}}}, \bibinfo {author} {\bibnamefont {{Dhanda-Batra, N.}}},
  \bibinfo {author} {\bibnamefont {{Dwivedi, C.}}}, \bibinfo {author}
  {\bibnamefont {{Falxa, M.}}}, \bibinfo {author} {\bibnamefont {{Ferdman, R.
  D.}}}, \bibinfo {author} {\bibnamefont {{Franchini, A.}}}, \bibinfo {author}
  {\bibnamefont {{Gair, J. R.}}}, \bibinfo {author} {\bibnamefont {{Goncharov,
  B.}}}, \bibinfo {author} {\bibnamefont {{Gopakumar, A.}}}, \bibinfo {author}
  {\bibnamefont {{Graikou, E.}}}, \bibinfo {author} {\bibnamefont
  {{Grießmeier, J.-M.}}}, \bibinfo {author} {\bibnamefont {{Guillemot, L.}}},
  \bibinfo {author} {\bibnamefont {{Guo, Y. J.}}}, \bibinfo {author}
  {\bibnamefont {{Gupta, Y.}}}, \bibinfo {author} {\bibnamefont {{Hisano,
  S.}}}, \bibinfo {author} {\bibnamefont {{Hu, H.}}}, \bibinfo {author}
  {\bibnamefont {{Iraci, F.}}}, \bibinfo {author} {\bibnamefont
  {{Izquierdo-Villalba, D.}}}, \bibinfo {author} {\bibnamefont {{Jang, J.}}},
  \bibinfo {author} {\bibnamefont {{Jawor, J.}}}, \bibinfo {author}
  {\bibnamefont {{Janssen, G. H.}}}, \bibinfo {author} {\bibnamefont {{Jessner,
  A.}}}, \bibinfo {author} {\bibnamefont {{Joshi, B. C.}}}, \bibinfo {author}
  {\bibnamefont {{Kareem, F.}}}, \bibinfo {author} {\bibnamefont {{Karuppusamy,
  R.}}}, \bibinfo {author} {\bibnamefont {{Keane, E. F.}}}, \bibinfo {author}
  {\bibnamefont {{Keith, M. J.}}}, \bibinfo {author} {\bibnamefont {{Kharbanda,
  D.}}}, \bibinfo {author} {\bibnamefont {{Kikunaga, T.}}}, \bibinfo {author}
  {\bibnamefont {{Kolhe, N.}}}, \bibinfo {author} {\bibnamefont {{Kramer,
  M.}}}, \bibinfo {author} {\bibnamefont {{Krishnakumar, M. A.}}}, \bibinfo
  {author} {\bibnamefont {{Lackeos, K.}}}, \bibinfo {author} {\bibnamefont
  {{Lee, K. J.}}}, \bibinfo {author} {\bibnamefont {{Liu, K.}}}, \bibinfo
  {author} {\bibnamefont {{Liu, Y.}}}, \bibinfo {author} {\bibnamefont {{Lyne,
  A. G.}}}, \bibinfo {author} {\bibnamefont {{McKee, J. W.}}}, \bibinfo
  {author} {\bibnamefont {{Maan, Y.}}}, \bibinfo {author} {\bibnamefont {{Main,
  R. A.}}}, \bibinfo {author} {\bibnamefont {{Mickaliger, M. B.}}}, \bibinfo
  {author} {\bibnamefont {{Niţu, I. C.}}}, \bibinfo {author} {\bibnamefont
  {{Nobleson, K.}}}, \bibinfo {author} {\bibnamefont {{Paladi, A. K.}}},
  \bibinfo {author} {\bibnamefont {{Parthasarathy, A.}}}, \bibinfo {author}
  {\bibnamefont {{Perera, B. B. P.}}}, \bibinfo {author} {\bibnamefont
  {{Perrodin, D.}}}, \bibinfo {author} {\bibnamefont {{Petiteau, A.}}},
  \bibinfo {author} {\bibnamefont {{Porayko, N. K.}}}, \bibinfo {author}
  {\bibnamefont {{Possenti, A.}}}, \bibinfo {author} {\bibnamefont {{Prabu,
  T.}}}, \bibinfo {author} {\bibnamefont {{Quelquejay Leclere, H.}}}, \bibinfo
  {author} {\bibnamefont {{Rana, P.}}}, \bibinfo {author} {\bibnamefont
  {{Samajdar, A.}}}, \bibinfo {author} {\bibnamefont {{Sanidas, S. A.}}},
  \bibinfo {author} {\bibnamefont {{Sesana, A.}}}, \bibinfo {author}
  {\bibnamefont {{Shaifullah, G.}}}, \bibinfo {author} {\bibnamefont {{Singha,
  J.}}}, \bibinfo {author} {\bibnamefont {{Speri, L.}}}, \bibinfo {author}
  {\bibnamefont {{Spiewak, R.}}}, \bibinfo {author} {\bibnamefont {{Srivastava,
  A.}}}, \bibinfo {author} {\bibnamefont {{Stappers, B. W.}}}, \bibinfo
  {author} {\bibnamefont {{Surnis, M.}}}, \bibinfo {author} {\bibnamefont
  {{Susarla, S. C.}}}, \bibinfo {author} {\bibnamefont {{Susobhanan, A.}}},
  \bibinfo {author} {\bibnamefont {{Takahashi, K.}}}, \bibinfo {author}
  {\bibnamefont {{Tarafdar, P.}}}, \bibinfo {author} {\bibnamefont {{Theureau,
  G.}}}, \bibinfo {author} {\bibnamefont {{Tiburzi, C.}}}, \bibinfo {author}
  {\bibnamefont {{van der Wateren, E.}}}, \bibinfo {author} {\bibnamefont
  {{Vecchio, A.}}}, \bibinfo {author} {\bibnamefont {{Venkatraman Krishnan,
  V.}}}, \bibinfo {author} {\bibnamefont {{Verbiest, J. P. W.}}}, \bibinfo
  {author} {\bibnamefont {{Wang, J.}}}, \bibinfo {author} {\bibnamefont {{Wang,
  L.}}}, \ and\ \bibinfo {author} {\bibnamefont {{Wu, Z.}}},\ }\bibfield
  {title} {\enquote {\bibinfo {title} {The second data release from the
  european pulsar timing array - iii. search for gravitational wave signals},}\
  }\href {\doibase 10.1051/0004-6361/202346844} {\bibfield  {journal} {\bibinfo
   {journal} {A\& A}\ }\textbf {\bibinfo {volume} {678}},\ \bibinfo {pages}
  {A50} (\bibinfo {year} {2023}{\natexlab{b}})}\BibitemShut {NoStop}%
\bibitem [{\citenamefont {Antoniadis}\ \emph
  {et~al.}(2023{\natexlab{b}})\citenamefont {Antoniadis}, \citenamefont
  {Arumugam}, \citenamefont {Arumugam}, \citenamefont {Auclair}, \citenamefont
  {Babak}, \citenamefont {Bagchi}, \citenamefont {Nielsen}, \citenamefont
  {Barausse}, \citenamefont {Bassa}, \citenamefont {Bathula}, \citenamefont
  {Berthereau}, \citenamefont {Bonetti}, \citenamefont {Bortolas},
  \citenamefont {Brook}, \citenamefont {Burgay}, \citenamefont {Caballero},
  \citenamefont {Caprini}, \citenamefont {Chalumeau}, \citenamefont {Champion},
  \citenamefont {Chanlaridis}, \citenamefont {Chen}, \citenamefont {Cognard},
  \citenamefont {Crisostomi}, \citenamefont {Dandapat}, \citenamefont {Deb},
  \citenamefont {Desai}, \citenamefont {Desvignes}, \citenamefont
  {Dhanda-Batra}, \citenamefont {Dwivedi}, \citenamefont {Falxa}, \citenamefont
  {Fastidio}, \citenamefont {Ferdman}, \citenamefont {Franchini}, \citenamefont
  {Gair}, \citenamefont {Goncharov}, \citenamefont {Gopakumar}, \citenamefont
  {Graikou}, \citenamefont {Grießmeier}, \citenamefont {Gualandris},
  \citenamefont {Guillemot}, \citenamefont {Guo}, \citenamefont {Gupta},
  \citenamefont {Hisano}, \citenamefont {Hu}, \citenamefont {Iraci},
  \citenamefont {Izquierdo-Villalba}, \citenamefont {Jang}, \citenamefont
  {Jawor}, \citenamefont {Janssen}, \citenamefont {Jessner}, \citenamefont
  {Joshi}, \citenamefont {Kareem}, \citenamefont {Karuppusamy}, \citenamefont
  {Keane}, \citenamefont {Keith}, \citenamefont {Kharbanda}, \citenamefont
  {Khizriev}, \citenamefont {Kikunaga}, \citenamefont {Kolhe}, \citenamefont
  {Kramer}, \citenamefont {Krishnakumar}, \citenamefont {Lackeos},
  \citenamefont {Lee}, \citenamefont {Liu}, \citenamefont {Liu}, \citenamefont
  {Lyne}, \citenamefont {McKee}, \citenamefont {Maan}, \citenamefont {Main},
  \citenamefont {Mickaliger}, \citenamefont {Middleton}, \citenamefont
  {Neronov}, \citenamefont {Nitu}, \citenamefont {Nobleson}, \citenamefont
  {Paladi}, \citenamefont {Parthasarathy}, \citenamefont {Perera},
  \citenamefont {Perrodin}, \citenamefont {Petiteau}, \citenamefont {Porayko},
  \citenamefont {Possenti}, \citenamefont {Prabu}, \citenamefont {Postnov},
  \citenamefont {Leclere}, \citenamefont {Rana}, \citenamefont {Pol},
  \citenamefont {Samajdar}, \citenamefont {Sanidas}, \citenamefont {Semikoz},
  \citenamefont {Sesana}, \citenamefont {Shaifullah}, \citenamefont {Singha},
  \citenamefont {Smarra}, \citenamefont {Speri}, \citenamefont {Spiewak},
  \citenamefont {Srivastava}, \citenamefont {Stappers}, \citenamefont {Steer},
  \citenamefont {Surnis}, \citenamefont {Susarla}, \citenamefont {Susobhanan},
  \citenamefont {Takahashi}, \citenamefont {Tarafdar}, \citenamefont
  {Theureau}, \citenamefont {Tiburzi}, \citenamefont {Truant}, \citenamefont
  {van~der Wateren}, \citenamefont {Valtolina}, \citenamefont {Vecchio},
  \citenamefont {Krishnan}, \citenamefont {Verbiest}, \citenamefont {Wang},
  \citenamefont {Wang},\ and\ \citenamefont {Wu}}]{EPTA_V_2023}%
  \BibitemOpen
  \bibfield  {author} {\bibinfo {author} {\bibfnamefont {J.}~\bibnamefont
  {Antoniadis}}, \bibinfo {author} {\bibfnamefont {P.}~\bibnamefont
  {Arumugam}}, \bibinfo {author} {\bibfnamefont {S.}~\bibnamefont {Arumugam}},
  \bibinfo {author} {\bibfnamefont {P.}~\bibnamefont {Auclair}}, \bibinfo
  {author} {\bibfnamefont {S.}~\bibnamefont {Babak}}, \bibinfo {author}
  {\bibfnamefont {M.}~\bibnamefont {Bagchi}}, \bibinfo {author} {\bibfnamefont
  {A.~S.~Bak}\ \bibnamefont {Nielsen}}, \bibinfo {author} {\bibfnamefont
  {E.}~\bibnamefont {Barausse}}, \bibinfo {author} {\bibfnamefont {C.~G.}\
  \bibnamefont {Bassa}}, \bibinfo {author} {\bibfnamefont {A.}~\bibnamefont
  {Bathula}}, \bibinfo {author} {\bibfnamefont {A.}~\bibnamefont {Berthereau}},
  \bibinfo {author} {\bibfnamefont {M.}~\bibnamefont {Bonetti}}, \bibinfo
  {author} {\bibfnamefont {E.}~\bibnamefont {Bortolas}}, \bibinfo {author}
  {\bibfnamefont {P.~R.}\ \bibnamefont {Brook}}, \bibinfo {author}
  {\bibfnamefont {M.}~\bibnamefont {Burgay}}, \bibinfo {author} {\bibfnamefont
  {R.~N.}\ \bibnamefont {Caballero}}, \bibinfo {author} {\bibfnamefont
  {C.}~\bibnamefont {Caprini}}, \bibinfo {author} {\bibfnamefont
  {A.}~\bibnamefont {Chalumeau}}, \bibinfo {author} {\bibfnamefont {D.~J.}\
  \bibnamefont {Champion}}, \bibinfo {author} {\bibfnamefont {S.}~\bibnamefont
  {Chanlaridis}}, \bibinfo {author} {\bibfnamefont {S.}~\bibnamefont {Chen}},
  \bibinfo {author} {\bibfnamefont {I.}~\bibnamefont {Cognard}}, \bibinfo
  {author} {\bibfnamefont {M.}~\bibnamefont {Crisostomi}}, \bibinfo {author}
  {\bibfnamefont {S.}~\bibnamefont {Dandapat}}, \bibinfo {author}
  {\bibfnamefont {D.}~\bibnamefont {Deb}}, \bibinfo {author} {\bibfnamefont
  {S.}~\bibnamefont {Desai}}, \bibinfo {author} {\bibfnamefont
  {G.}~\bibnamefont {Desvignes}}, \bibinfo {author} {\bibfnamefont
  {N.}~\bibnamefont {Dhanda-Batra}}, \bibinfo {author} {\bibfnamefont
  {C.}~\bibnamefont {Dwivedi}}, \bibinfo {author} {\bibfnamefont
  {M.}~\bibnamefont {Falxa}}, \bibinfo {author} {\bibfnamefont
  {F.}~\bibnamefont {Fastidio}}, \bibinfo {author} {\bibfnamefont {R.~D.}\
  \bibnamefont {Ferdman}}, \bibinfo {author} {\bibfnamefont {A.}~\bibnamefont
  {Franchini}}, \bibinfo {author} {\bibfnamefont {J.~R.}\ \bibnamefont {Gair}},
  \bibinfo {author} {\bibfnamefont {B.}~\bibnamefont {Goncharov}}, \bibinfo
  {author} {\bibfnamefont {A.}~\bibnamefont {Gopakumar}}, \bibinfo {author}
  {\bibfnamefont {E.}~\bibnamefont {Graikou}}, \bibinfo {author} {\bibfnamefont
  {J.~M.}\ \bibnamefont {Grießmeier}}, \bibinfo {author} {\bibfnamefont
  {A.}~\bibnamefont {Gualandris}}, \bibinfo {author} {\bibfnamefont
  {L.}~\bibnamefont {Guillemot}}, \bibinfo {author} {\bibfnamefont {Y.~J.}\
  \bibnamefont {Guo}}, \bibinfo {author} {\bibfnamefont {Y.}~\bibnamefont
  {Gupta}}, \bibinfo {author} {\bibfnamefont {S.}~\bibnamefont {Hisano}},
  \bibinfo {author} {\bibfnamefont {H.}~\bibnamefont {Hu}}, \bibinfo {author}
  {\bibfnamefont {F.}~\bibnamefont {Iraci}}, \bibinfo {author} {\bibfnamefont
  {D.}~\bibnamefont {Izquierdo-Villalba}}, \bibinfo {author} {\bibfnamefont
  {J.}~\bibnamefont {Jang}}, \bibinfo {author} {\bibfnamefont {J.}~\bibnamefont
  {Jawor}}, \bibinfo {author} {\bibfnamefont {G.~H.}\ \bibnamefont {Janssen}},
  \bibinfo {author} {\bibfnamefont {A.}~\bibnamefont {Jessner}}, \bibinfo
  {author} {\bibfnamefont {B.~C.}\ \bibnamefont {Joshi}}, \bibinfo {author}
  {\bibfnamefont {F.}~\bibnamefont {Kareem}}, \bibinfo {author} {\bibfnamefont
  {R.}~\bibnamefont {Karuppusamy}}, \bibinfo {author} {\bibfnamefont {E.~F.}\
  \bibnamefont {Keane}}, \bibinfo {author} {\bibfnamefont {M.~J.}\ \bibnamefont
  {Keith}}, \bibinfo {author} {\bibfnamefont {D.}~\bibnamefont {Kharbanda}},
  \bibinfo {author} {\bibfnamefont {T.}~\bibnamefont {Khizriev}}, \bibinfo
  {author} {\bibfnamefont {T.}~\bibnamefont {Kikunaga}}, \bibinfo {author}
  {\bibfnamefont {N.}~\bibnamefont {Kolhe}}, \bibinfo {author} {\bibfnamefont
  {M.}~\bibnamefont {Kramer}}, \bibinfo {author} {\bibfnamefont {M.~A.}\
  \bibnamefont {Krishnakumar}}, \bibinfo {author} {\bibfnamefont
  {K.}~\bibnamefont {Lackeos}}, \bibinfo {author} {\bibfnamefont {K.~J.}\
  \bibnamefont {Lee}}, \bibinfo {author} {\bibfnamefont {K.}~\bibnamefont
  {Liu}}, \bibinfo {author} {\bibfnamefont {Y.}~\bibnamefont {Liu}}, \bibinfo
  {author} {\bibfnamefont {A.~G.}\ \bibnamefont {Lyne}}, \bibinfo {author}
  {\bibfnamefont {J.~W.}\ \bibnamefont {McKee}}, \bibinfo {author}
  {\bibfnamefont {Y.}~\bibnamefont {Maan}}, \bibinfo {author} {\bibfnamefont
  {R.~A.}\ \bibnamefont {Main}}, \bibinfo {author} {\bibfnamefont {M.~B.}\
  \bibnamefont {Mickaliger}}, \bibinfo {author} {\bibfnamefont
  {H.}~\bibnamefont {Middleton}}, \bibinfo {author} {\bibfnamefont
  {A.}~\bibnamefont {Neronov}}, \bibinfo {author} {\bibfnamefont {I.~C.}\
  \bibnamefont {Nitu}}, \bibinfo {author} {\bibfnamefont {K.}~\bibnamefont
  {Nobleson}}, \bibinfo {author} {\bibfnamefont {A.~K.}\ \bibnamefont
  {Paladi}}, \bibinfo {author} {\bibfnamefont {A.}~\bibnamefont
  {Parthasarathy}}, \bibinfo {author} {\bibfnamefont {B.~B.~P.}\ \bibnamefont
  {Perera}}, \bibinfo {author} {\bibfnamefont {D.}~\bibnamefont {Perrodin}},
  \bibinfo {author} {\bibfnamefont {A.}~\bibnamefont {Petiteau}}, \bibinfo
  {author} {\bibfnamefont {N.~K.}\ \bibnamefont {Porayko}}, \bibinfo {author}
  {\bibfnamefont {A.}~\bibnamefont {Possenti}}, \bibinfo {author}
  {\bibfnamefont {T.}~\bibnamefont {Prabu}}, \bibinfo {author} {\bibfnamefont
  {K.}~\bibnamefont {Postnov}}, \bibinfo {author} {\bibfnamefont
  {H.~Quelquejay}\ \bibnamefont {Leclere}}, \bibinfo {author} {\bibfnamefont
  {P.}~\bibnamefont {Rana}}, \bibinfo {author} {\bibfnamefont {A.~Roper}\
  \bibnamefont {Pol}}, \bibinfo {author} {\bibfnamefont {A.}~\bibnamefont
  {Samajdar}}, \bibinfo {author} {\bibfnamefont {S.~A.}\ \bibnamefont
  {Sanidas}}, \bibinfo {author} {\bibfnamefont {D.}~\bibnamefont {Semikoz}},
  \bibinfo {author} {\bibfnamefont {A.}~\bibnamefont {Sesana}}, \bibinfo
  {author} {\bibfnamefont {G.}~\bibnamefont {Shaifullah}}, \bibinfo {author}
  {\bibfnamefont {J.}~\bibnamefont {Singha}}, \bibinfo {author} {\bibfnamefont
  {C.}~\bibnamefont {Smarra}}, \bibinfo {author} {\bibfnamefont
  {L.}~\bibnamefont {Speri}}, \bibinfo {author} {\bibfnamefont
  {R.}~\bibnamefont {Spiewak}}, \bibinfo {author} {\bibfnamefont
  {A.}~\bibnamefont {Srivastava}}, \bibinfo {author} {\bibfnamefont {B.~W.}\
  \bibnamefont {Stappers}}, \bibinfo {author} {\bibfnamefont {D.~A.}\
  \bibnamefont {Steer}}, \bibinfo {author} {\bibfnamefont {M.}~\bibnamefont
  {Surnis}}, \bibinfo {author} {\bibfnamefont {S.~C.}\ \bibnamefont {Susarla}},
  \bibinfo {author} {\bibfnamefont {A.}~\bibnamefont {Susobhanan}}, \bibinfo
  {author} {\bibfnamefont {K.}~\bibnamefont {Takahashi}}, \bibinfo {author}
  {\bibfnamefont {P.}~\bibnamefont {Tarafdar}}, \bibinfo {author}
  {\bibfnamefont {G.}~\bibnamefont {Theureau}}, \bibinfo {author}
  {\bibfnamefont {C.}~\bibnamefont {Tiburzi}}, \bibinfo {author} {\bibfnamefont
  {R.~J.}\ \bibnamefont {Truant}}, \bibinfo {author} {\bibfnamefont
  {E.}~\bibnamefont {van~der Wateren}}, \bibinfo {author} {\bibfnamefont
  {S.}~\bibnamefont {Valtolina}}, \bibinfo {author} {\bibfnamefont
  {A.}~\bibnamefont {Vecchio}}, \bibinfo {author} {\bibfnamefont
  {V.~Venkatraman}\ \bibnamefont {Krishnan}}, \bibinfo {author} {\bibfnamefont
  {J.~P.~W.}\ \bibnamefont {Verbiest}}, \bibinfo {author} {\bibfnamefont
  {J.}~\bibnamefont {Wang}}, \bibinfo {author} {\bibfnamefont {L.}~\bibnamefont
  {Wang}}, \ and\ \bibinfo {author} {\bibfnamefont {Z.}~\bibnamefont {Wu}},\
  }\href@noop {} {\enquote {\bibinfo {title} {The second data release from the
  european pulsar timing array: V. implications for massive black holes, dark
  matter and the early universe},}\ } (\bibinfo {year} {2023}{\natexlab{b}}),\
  \Eprint {http://arxiv.org/abs/2306.16227} {arXiv:2306.16227 [astro-ph.CO]}
  \BibitemShut {NoStop}%
\bibitem [{\citenamefont {{Arzoumanian}}\ \emph {et~al.}(2020)\citenamefont
  {{Arzoumanian}}, \citenamefont {{Baker}}, \citenamefont {{Blumer}},
  \citenamefont {{B{\'e}csy}}, \citenamefont {{Brazier}}, \citenamefont
  {{Brook}}, \citenamefont {{Burke-Spolaor}}, \citenamefont {{Chatterjee}},
  \citenamefont {{Chen}}, \citenamefont {{Cordes}}, \citenamefont {{Cornish}},
  \citenamefont {{Crawford}}, \citenamefont {{Cromartie}}, \citenamefont
  {{Decesar}}, \citenamefont {{Demorest}}, \citenamefont {{Dolch}},
  \citenamefont {{Ellis}}, \citenamefont {{Ferrara}}, \citenamefont {{Fiore}},
  \citenamefont {{Fonseca}}, \citenamefont {{Garver-Daniels}}, \citenamefont
  {{Gentile}}, \citenamefont {{Good}}, \citenamefont {{Hazboun}}, \citenamefont
  {{Holgado}}, \citenamefont {{Islo}}, \citenamefont {{Jennings}},
  \citenamefont {{Jones}}, \citenamefont {{Kaiser}}, \citenamefont {{Kaplan}},
  \citenamefont {{Kelley}}, \citenamefont {{Key}}, \citenamefont {{Laal}},
  \citenamefont {{Lam}}, \citenamefont {{Lazio}}, \citenamefont {{Lorimer}},
  \citenamefont {{Luo}}, \citenamefont {{Lynch}}, \citenamefont {{Madison}},
  \citenamefont {{McLaughlin}}, \citenamefont {{Mingarelli}}, \citenamefont
  {{Ng}}, \citenamefont {{Nice}}, \citenamefont {{Pennucci}}, \citenamefont
  {{Pol}}, \citenamefont {{Ransom}}, \citenamefont {{Ray}}, \citenamefont
  {{Shapiro-Albert}}, \citenamefont {{Siemens}}, \citenamefont {{Simon}},
  \citenamefont {{Spiewak}}, \citenamefont {{Stairs}}, \citenamefont
  {{Stinebring}}, \citenamefont {{Stovall}}, \citenamefont {{Sun}},
  \citenamefont {{Swiggum}}, \citenamefont {{Taylor}}, \citenamefont
  {{Turner}}, \citenamefont {{Vallisneri}}, \citenamefont {{Vigeland}},
  \citenamefont {{Witt}},\ and\ \citenamefont {{Nanograv
  Collaboration}}}]{nanograv12.5gwb}%
  \BibitemOpen
  \bibfield  {author} {\bibinfo {author} {\bibfnamefont {Zaven}\ \bibnamefont
  {{Arzoumanian}}}, \bibinfo {author} {\bibfnamefont {Paul~T.}\ \bibnamefont
  {{Baker}}}, \bibinfo {author} {\bibfnamefont {Harsha}\ \bibnamefont
  {{Blumer}}}, \bibinfo {author} {\bibfnamefont {Bence}\ \bibnamefont
  {{B{\'e}csy}}}, \bibinfo {author} {\bibfnamefont {Adam}\ \bibnamefont
  {{Brazier}}}, \bibinfo {author} {\bibfnamefont {Paul~R.}\ \bibnamefont
  {{Brook}}}, \bibinfo {author} {\bibfnamefont {Sarah}\ \bibnamefont
  {{Burke-Spolaor}}}, \bibinfo {author} {\bibfnamefont {Shami}\ \bibnamefont
  {{Chatterjee}}}, \bibinfo {author} {\bibfnamefont {Siyuan}\ \bibnamefont
  {{Chen}}}, \bibinfo {author} {\bibfnamefont {James~M.}\ \bibnamefont
  {{Cordes}}}, \bibinfo {author} {\bibfnamefont {Neil~J.}\ \bibnamefont
  {{Cornish}}}, \bibinfo {author} {\bibfnamefont {Fronefield}\ \bibnamefont
  {{Crawford}}}, \bibinfo {author} {\bibfnamefont {H.~Thankful}\ \bibnamefont
  {{Cromartie}}}, \bibinfo {author} {\bibfnamefont {Megan~E.}\ \bibnamefont
  {{Decesar}}}, \bibinfo {author} {\bibfnamefont {Paul~B.}\ \bibnamefont
  {{Demorest}}}, \bibinfo {author} {\bibfnamefont {Timothy}\ \bibnamefont
  {{Dolch}}}, \bibinfo {author} {\bibfnamefont {Justin~A.}\ \bibnamefont
  {{Ellis}}}, \bibinfo {author} {\bibfnamefont {Elizabeth~C.}\ \bibnamefont
  {{Ferrara}}}, \bibinfo {author} {\bibfnamefont {William}\ \bibnamefont
  {{Fiore}}}, \bibinfo {author} {\bibfnamefont {Emmanuel}\ \bibnamefont
  {{Fonseca}}}, \bibinfo {author} {\bibfnamefont {Nathan}\ \bibnamefont
  {{Garver-Daniels}}}, \bibinfo {author} {\bibfnamefont {Peter~A.}\
  \bibnamefont {{Gentile}}}, \bibinfo {author} {\bibfnamefont {Deborah~C.}\
  \bibnamefont {{Good}}}, \bibinfo {author} {\bibfnamefont {Jeffrey~S.}\
  \bibnamefont {{Hazboun}}}, \bibinfo {author} {\bibfnamefont {A.~Miguel}\
  \bibnamefont {{Holgado}}}, \bibinfo {author} {\bibfnamefont {Kristina}\
  \bibnamefont {{Islo}}}, \bibinfo {author} {\bibfnamefont {Ross~J.}\
  \bibnamefont {{Jennings}}}, \bibinfo {author} {\bibfnamefont {Megan~L.}\
  \bibnamefont {{Jones}}}, \bibinfo {author} {\bibfnamefont {Andrew~R.}\
  \bibnamefont {{Kaiser}}}, \bibinfo {author} {\bibfnamefont {David~L.}\
  \bibnamefont {{Kaplan}}}, \bibinfo {author} {\bibfnamefont {Luke~Zoltan}\
  \bibnamefont {{Kelley}}}, \bibinfo {author} {\bibfnamefont {Joey~Shapiro}\
  \bibnamefont {{Key}}}, \bibinfo {author} {\bibfnamefont {Nima}\ \bibnamefont
  {{Laal}}}, \bibinfo {author} {\bibfnamefont {Michael~T.}\ \bibnamefont
  {{Lam}}}, \bibinfo {author} {\bibfnamefont {T.~Joseph~W.}\ \bibnamefont
  {{Lazio}}}, \bibinfo {author} {\bibfnamefont {Duncan~R.}\ \bibnamefont
  {{Lorimer}}}, \bibinfo {author} {\bibfnamefont {Jing}\ \bibnamefont {{Luo}}},
  \bibinfo {author} {\bibfnamefont {Ryan~S.}\ \bibnamefont {{Lynch}}}, \bibinfo
  {author} {\bibfnamefont {Dustin~R.}\ \bibnamefont {{Madison}}}, \bibinfo
  {author} {\bibfnamefont {Maura~A.}\ \bibnamefont {{McLaughlin}}}, \bibinfo
  {author} {\bibfnamefont {Chiara M.~F.}\ \bibnamefont {{Mingarelli}}},
  \bibinfo {author} {\bibfnamefont {Cherry}\ \bibnamefont {{Ng}}}, \bibinfo
  {author} {\bibfnamefont {David~J.}\ \bibnamefont {{Nice}}}, \bibinfo {author}
  {\bibfnamefont {Timothy~T.}\ \bibnamefont {{Pennucci}}}, \bibinfo {author}
  {\bibfnamefont {Nihan~S.}\ \bibnamefont {{Pol}}}, \bibinfo {author}
  {\bibfnamefont {Scott~M.}\ \bibnamefont {{Ransom}}}, \bibinfo {author}
  {\bibfnamefont {Paul~S.}\ \bibnamefont {{Ray}}}, \bibinfo {author}
  {\bibfnamefont {Brent~J.}\ \bibnamefont {{Shapiro-Albert}}}, \bibinfo
  {author} {\bibfnamefont {Xavier}\ \bibnamefont {{Siemens}}}, \bibinfo
  {author} {\bibfnamefont {Joseph}\ \bibnamefont {{Simon}}}, \bibinfo {author}
  {\bibfnamefont {Ren{\'e}e}\ \bibnamefont {{Spiewak}}}, \bibinfo {author}
  {\bibfnamefont {Ingrid~H.}\ \bibnamefont {{Stairs}}}, \bibinfo {author}
  {\bibfnamefont {Daniel~R.}\ \bibnamefont {{Stinebring}}}, \bibinfo {author}
  {\bibfnamefont {Kevin}\ \bibnamefont {{Stovall}}}, \bibinfo {author}
  {\bibfnamefont {Jerry~P.}\ \bibnamefont {{Sun}}}, \bibinfo {author}
  {\bibfnamefont {Joseph~K.}\ \bibnamefont {{Swiggum}}}, \bibinfo {author}
  {\bibfnamefont {Stephen~R.}\ \bibnamefont {{Taylor}}}, \bibinfo {author}
  {\bibfnamefont {Jacob~E.}\ \bibnamefont {{Turner}}}, \bibinfo {author}
  {\bibfnamefont {Michele}\ \bibnamefont {{Vallisneri}}}, \bibinfo {author}
  {\bibfnamefont {Sarah~J.}\ \bibnamefont {{Vigeland}}}, \bibinfo {author}
  {\bibfnamefont {Caitlin~A.}\ \bibnamefont {{Witt}}}, \ and\ \bibinfo {author}
  {\bibnamefont {{Nanograv Collaboration}}},\ }\bibfield  {title} {\enquote
  {\bibinfo {title} {{The NANOGrav 12.5 yr Data Set: Search for an Isotropic
  Stochastic Gravitational-wave Background}},}\ }\href {\doibase
  10.3847/2041-8213/abd401} {\bibfield  {journal} {\bibinfo  {journal} {\apjl}\
  }\textbf {\bibinfo {volume} {905}},\ \bibinfo {eid} {L34} (\bibinfo {year}
  {2020})},\ \Eprint {http://arxiv.org/abs/2009.04496} {arXiv:2009.04496
  [astro-ph.HE]} \BibitemShut {NoStop}%
\bibitem [{\citenamefont {{Goncharov}}\ \emph
  {et~al.}(2021{\natexlab{b}})\citenamefont {{Goncharov}}, \citenamefont
  {{Shannon}}, \citenamefont {{Reardon}}, \citenamefont {{Hobbs}},
  \citenamefont {{Zic}}, \citenamefont {{Bailes}}, \citenamefont {{Cury{\l}o}},
  \citenamefont {{Dai}}, \citenamefont {{Kerr}}, \citenamefont {{Lower}},
  \citenamefont {{Manchester}}, \citenamefont {{Mandow}}, \citenamefont
  {{Middleton}}, \citenamefont {{Miles}}, \citenamefont {{Parthasarathy}},
  \citenamefont {{Thrane}}, \citenamefont {{Thyagarajan}}, \citenamefont
  {{Xue}}, \citenamefont {{Zhu}}, \citenamefont {{Cameron}}, \citenamefont
  {{Feng}}, \citenamefont {{Luo}}, \citenamefont {{Russell}}, \citenamefont
  {{Sarkissian}}, \citenamefont {{Spiewak}}, \citenamefont {{Wang}},
  \citenamefont {{Wang}}, \citenamefont {{Zhang}},\ and\ \citenamefont
  {{Zhang}}}]{GoncharovShannon2021}%
  \BibitemOpen
  \bibfield  {author} {\bibinfo {author} {\bibfnamefont {Boris}\ \bibnamefont
  {{Goncharov}}}, \bibinfo {author} {\bibfnamefont {R.~M.}\ \bibnamefont
  {{Shannon}}}, \bibinfo {author} {\bibfnamefont {D.~J.}\ \bibnamefont
  {{Reardon}}}, \bibinfo {author} {\bibfnamefont {G.}~\bibnamefont {{Hobbs}}},
  \bibinfo {author} {\bibfnamefont {A.}~\bibnamefont {{Zic}}}, \bibinfo
  {author} {\bibfnamefont {M.}~\bibnamefont {{Bailes}}}, \bibinfo {author}
  {\bibfnamefont {M.}~\bibnamefont {{Cury{\l}o}}}, \bibinfo {author}
  {\bibfnamefont {S.}~\bibnamefont {{Dai}}}, \bibinfo {author} {\bibfnamefont
  {M.}~\bibnamefont {{Kerr}}}, \bibinfo {author} {\bibfnamefont {M.~E.}\
  \bibnamefont {{Lower}}}, \bibinfo {author} {\bibfnamefont {R.~N.}\
  \bibnamefont {{Manchester}}}, \bibinfo {author} {\bibfnamefont
  {R.}~\bibnamefont {{Mandow}}}, \bibinfo {author} {\bibfnamefont
  {H.}~\bibnamefont {{Middleton}}}, \bibinfo {author} {\bibfnamefont {M.~T.}\
  \bibnamefont {{Miles}}}, \bibinfo {author} {\bibfnamefont {A.}~\bibnamefont
  {{Parthasarathy}}}, \bibinfo {author} {\bibfnamefont {E.}~\bibnamefont
  {{Thrane}}}, \bibinfo {author} {\bibfnamefont {N.}~\bibnamefont
  {{Thyagarajan}}}, \bibinfo {author} {\bibfnamefont {X.}~\bibnamefont
  {{Xue}}}, \bibinfo {author} {\bibfnamefont {X.~J.}\ \bibnamefont {{Zhu}}},
  \bibinfo {author} {\bibfnamefont {A.~D.}\ \bibnamefont {{Cameron}}}, \bibinfo
  {author} {\bibfnamefont {Y.}~\bibnamefont {{Feng}}}, \bibinfo {author}
  {\bibfnamefont {R.}~\bibnamefont {{Luo}}}, \bibinfo {author} {\bibfnamefont
  {C.~J.}\ \bibnamefont {{Russell}}}, \bibinfo {author} {\bibfnamefont
  {J.}~\bibnamefont {{Sarkissian}}}, \bibinfo {author} {\bibfnamefont
  {R.}~\bibnamefont {{Spiewak}}}, \bibinfo {author} {\bibfnamefont
  {S.}~\bibnamefont {{Wang}}}, \bibinfo {author} {\bibfnamefont {J.~B.}\
  \bibnamefont {{Wang}}}, \bibinfo {author} {\bibfnamefont {L.}~\bibnamefont
  {{Zhang}}}, \ and\ \bibinfo {author} {\bibfnamefont {S.}~\bibnamefont
  {{Zhang}}},\ }\bibfield  {title} {\enquote {\bibinfo {title} {{On the
  Evidence for a Common-spectrum Process in the Search for the Nanohertz
  Gravitational-wave Background with the Parkes Pulsar Timing Array}},}\ }\href
  {\doibase 10.3847/2041-8213/ac17f4} {\bibfield  {journal} {\bibinfo
  {journal} {\apjl}\ }\textbf {\bibinfo {volume} {917}},\ \bibinfo {eid} {L19}
  (\bibinfo {year} {2021}{\natexlab{b}})},\ \Eprint
  {http://arxiv.org/abs/2107.12112} {arXiv:2107.12112 [astro-ph.HE]}
  \BibitemShut {NoStop}%
\bibitem [{\citenamefont {{Chen}}\ \emph {et~al.}(2021)\citenamefont {{Chen}},
  \citenamefont {{Caballero}}, \citenamefont {{Guo}}, \citenamefont
  {{Chalumeau}}, \citenamefont {{Liu}}, \citenamefont {{Shaifullah}},
  \citenamefont {{Lee}}, \citenamefont {{Babak}}, \citenamefont {{Desvignes}},
  \citenamefont {{Parthasarathy}}, \citenamefont {{Hu}}, \citenamefont {{van
  der Wateren}}, \citenamefont {{Antoniadis}}, \citenamefont {{Bak Nielsen}},
  \citenamefont {{Bassa}}, \citenamefont {{Berthereau}}, \citenamefont
  {{Burgay}}, \citenamefont {{Champion}}, \citenamefont {{Cognard}},
  \citenamefont {{Falxa}}, \citenamefont {{Ferdman}}, \citenamefont {{Freire}},
  \citenamefont {{Gair}}, \citenamefont {{Graikou}}, \citenamefont
  {{Guillemot}}, \citenamefont {{Jang}}, \citenamefont {{Janssen}},
  \citenamefont {{Karuppusamy}}, \citenamefont {{Keith}}, \citenamefont
  {{Kramer}}, \citenamefont {{Liu}}, \citenamefont {{Lyne}}, \citenamefont
  {{Main}}, \citenamefont {{McKee}}, \citenamefont {{Mickaliger}},
  \citenamefont {{Perera}}, \citenamefont {{Perrodin}}, \citenamefont
  {{Petiteau}}, \citenamefont {{Porayko}}, \citenamefont {{Possenti}},
  \citenamefont {{Samajdar}}, \citenamefont {{Sanidas}}, \citenamefont
  {{Sesana}}, \citenamefont {{Speri}}, \citenamefont {{Stappers}},
  \citenamefont {{Theureau}}, \citenamefont {{Tiburzi}}, \citenamefont
  {{Vecchio}}, \citenamefont {{Verbiest}}, \citenamefont {{Wang}},
  \citenamefont {{Wang}},\ and\ \citenamefont {{Xu}}}]{epta2021gwb}%
  \BibitemOpen
  \bibfield  {author} {\bibinfo {author} {\bibfnamefont {S.}~\bibnamefont
  {{Chen}}}, \bibinfo {author} {\bibfnamefont {R.~N.}\ \bibnamefont
  {{Caballero}}}, \bibinfo {author} {\bibfnamefont {Y.~J.}\ \bibnamefont
  {{Guo}}}, \bibinfo {author} {\bibfnamefont {A.}~\bibnamefont {{Chalumeau}}},
  \bibinfo {author} {\bibfnamefont {K.}~\bibnamefont {{Liu}}}, \bibinfo
  {author} {\bibfnamefont {G.}~\bibnamefont {{Shaifullah}}}, \bibinfo {author}
  {\bibfnamefont {K.~J.}\ \bibnamefont {{Lee}}}, \bibinfo {author}
  {\bibfnamefont {S.}~\bibnamefont {{Babak}}}, \bibinfo {author} {\bibfnamefont
  {G.}~\bibnamefont {{Desvignes}}}, \bibinfo {author} {\bibfnamefont
  {A.}~\bibnamefont {{Parthasarathy}}}, \bibinfo {author} {\bibfnamefont
  {H.}~\bibnamefont {{Hu}}}, \bibinfo {author} {\bibfnamefont {E.}~\bibnamefont
  {{van der Wateren}}}, \bibinfo {author} {\bibfnamefont {J.}~\bibnamefont
  {{Antoniadis}}}, \bibinfo {author} {\bibfnamefont {A.~S.}\ \bibnamefont {{Bak
  Nielsen}}}, \bibinfo {author} {\bibfnamefont {C.~G.}\ \bibnamefont
  {{Bassa}}}, \bibinfo {author} {\bibfnamefont {A.}~\bibnamefont
  {{Berthereau}}}, \bibinfo {author} {\bibfnamefont {M.}~\bibnamefont
  {{Burgay}}}, \bibinfo {author} {\bibfnamefont {D.~J.}\ \bibnamefont
  {{Champion}}}, \bibinfo {author} {\bibfnamefont {I.}~\bibnamefont
  {{Cognard}}}, \bibinfo {author} {\bibfnamefont {M.}~\bibnamefont {{Falxa}}},
  \bibinfo {author} {\bibfnamefont {R.~D.}\ \bibnamefont {{Ferdman}}}, \bibinfo
  {author} {\bibfnamefont {P.~C.~C.}\ \bibnamefont {{Freire}}}, \bibinfo
  {author} {\bibfnamefont {J.~R.}\ \bibnamefont {{Gair}}}, \bibinfo {author}
  {\bibfnamefont {E.}~\bibnamefont {{Graikou}}}, \bibinfo {author}
  {\bibfnamefont {L.}~\bibnamefont {{Guillemot}}}, \bibinfo {author}
  {\bibfnamefont {J.}~\bibnamefont {{Jang}}}, \bibinfo {author} {\bibfnamefont
  {G.~H.}\ \bibnamefont {{Janssen}}}, \bibinfo {author} {\bibfnamefont
  {R.}~\bibnamefont {{Karuppusamy}}}, \bibinfo {author} {\bibfnamefont {M.~J.}\
  \bibnamefont {{Keith}}}, \bibinfo {author} {\bibfnamefont {M.}~\bibnamefont
  {{Kramer}}}, \bibinfo {author} {\bibfnamefont {X.~J.}\ \bibnamefont {{Liu}}},
  \bibinfo {author} {\bibfnamefont {A.~G.}\ \bibnamefont {{Lyne}}}, \bibinfo
  {author} {\bibfnamefont {R.~A.}\ \bibnamefont {{Main}}}, \bibinfo {author}
  {\bibfnamefont {J.~W.}\ \bibnamefont {{McKee}}}, \bibinfo {author}
  {\bibfnamefont {M.~B.}\ \bibnamefont {{Mickaliger}}}, \bibinfo {author}
  {\bibfnamefont {B.~B.~P.}\ \bibnamefont {{Perera}}}, \bibinfo {author}
  {\bibfnamefont {D.}~\bibnamefont {{Perrodin}}}, \bibinfo {author}
  {\bibfnamefont {A.}~\bibnamefont {{Petiteau}}}, \bibinfo {author}
  {\bibfnamefont {N.~K.}\ \bibnamefont {{Porayko}}}, \bibinfo {author}
  {\bibfnamefont {A.}~\bibnamefont {{Possenti}}}, \bibinfo {author}
  {\bibfnamefont {A.}~\bibnamefont {{Samajdar}}}, \bibinfo {author}
  {\bibfnamefont {S.~A.}\ \bibnamefont {{Sanidas}}}, \bibinfo {author}
  {\bibfnamefont {A.}~\bibnamefont {{Sesana}}}, \bibinfo {author}
  {\bibfnamefont {L.}~\bibnamefont {{Speri}}}, \bibinfo {author} {\bibfnamefont
  {B.~W.}\ \bibnamefont {{Stappers}}}, \bibinfo {author} {\bibfnamefont
  {G.}~\bibnamefont {{Theureau}}}, \bibinfo {author} {\bibfnamefont
  {C.}~\bibnamefont {{Tiburzi}}}, \bibinfo {author} {\bibfnamefont
  {A.}~\bibnamefont {{Vecchio}}}, \bibinfo {author} {\bibfnamefont {J.~P.~W.}\
  \bibnamefont {{Verbiest}}}, \bibinfo {author} {\bibfnamefont
  {J.}~\bibnamefont {{Wang}}}, \bibinfo {author} {\bibfnamefont
  {L.}~\bibnamefont {{Wang}}}, \ and\ \bibinfo {author} {\bibfnamefont
  {H.}~\bibnamefont {{Xu}}},\ }\bibfield  {title} {\enquote {\bibinfo {title}
  {{Common-red-signal analysis with 24-yr high-precision timing of the European
  Pulsar Timing Array: inferences in the stochastic gravitational-wave
  background search}},}\ }\href {\doibase 10.1093/mnras/stab2833} {\bibfield
  {journal} {\bibinfo  {journal} {\mnras}\ }\textbf {\bibinfo {volume} {508}},\
  \bibinfo {pages} {4970--4993} (\bibinfo {year} {2021})},\ \Eprint
  {http://arxiv.org/abs/2110.13184} {arXiv:2110.13184 [astro-ph.HE]}
  \BibitemShut {NoStop}%
\bibitem [{\citenamefont {{van Haasteren}}\ \emph {et~al.}(2009)\citenamefont
  {{van Haasteren}}, \citenamefont {{Levin}}, \citenamefont {{McDonald}},\ and\
  \citenamefont {{Lu}}}]{vanHaasterenLevin2009}%
  \BibitemOpen
  \bibfield  {author} {\bibinfo {author} {\bibfnamefont {Rutger}\ \bibnamefont
  {{van Haasteren}}}, \bibinfo {author} {\bibfnamefont {Yuri}\ \bibnamefont
  {{Levin}}}, \bibinfo {author} {\bibfnamefont {Patrick}\ \bibnamefont
  {{McDonald}}}, \ and\ \bibinfo {author} {\bibfnamefont {Tingting}\
  \bibnamefont {{Lu}}},\ }\bibfield  {title} {\enquote {\bibinfo {title} {{On
  measuring the gravitational-wave background using Pulsar Timing Arrays}},}\
  }\href {\doibase 10.1111/j.1365-2966.2009.14590.x} {\bibfield  {journal}
  {\bibinfo  {journal} {Monthly Notices of the Royal Astronomical Society}\
  }\textbf {\bibinfo {volume} {395}},\ \bibinfo {pages} {1005--1014} (\bibinfo
  {year} {2009})},\ \Eprint {http://arxiv.org/abs/0809.0791} {arXiv:0809.0791
  [astro-ph]} \BibitemShut {NoStop}%
\bibitem [{\citenamefont {{Lentati}}\ \emph {et~al.}(2014)\citenamefont
  {{Lentati}}, \citenamefont {{Alexander}}, \citenamefont {{Hobson}},
  \citenamefont {{Feroz}}, \citenamefont {{van Haasteren}}, \citenamefont
  {{Lee}},\ and\ \citenamefont {{Shannon}}}]{Lentati_2014}%
  \BibitemOpen
  \bibfield  {author} {\bibinfo {author} {\bibfnamefont {L.}~\bibnamefont
  {{Lentati}}}, \bibinfo {author} {\bibfnamefont {P.}~\bibnamefont
  {{Alexander}}}, \bibinfo {author} {\bibfnamefont {M.~P.}\ \bibnamefont
  {{Hobson}}}, \bibinfo {author} {\bibfnamefont {F.}~\bibnamefont {{Feroz}}},
  \bibinfo {author} {\bibfnamefont {R.}~\bibnamefont {{van Haasteren}}},
  \bibinfo {author} {\bibfnamefont {K.~J.}\ \bibnamefont {{Lee}}}, \ and\
  \bibinfo {author} {\bibfnamefont {R.~M.}\ \bibnamefont {{Shannon}}},\
  }\bibfield  {title} {\enquote {\bibinfo {title} {{TEMPONEST: a Bayesian
  approach to pulsar timing analysis}},}\ }\href {\doibase
  10.1093/mnras/stt2122} {\bibfield  {journal} {\bibinfo  {journal} {Monthly
  Notices of the Royal Astronomical Society}\ }\textbf {\bibinfo {volume}
  {437}},\ \bibinfo {pages} {3004--3023} (\bibinfo {year} {2014})},\ \Eprint
  {http://arxiv.org/abs/1310.2120} {arXiv:1310.2120 [astro-ph.IM]} \BibitemShut
  {NoStop}%
\bibitem [{\citenamefont {{Taylor}}\ \emph {et~al.}(2017)\citenamefont
  {{Taylor}}, \citenamefont {{Lentati}}, \citenamefont {{Babak}}, \citenamefont
  {{Brem}}, \citenamefont {{Gair}}, \citenamefont {{Sesana}},\ and\
  \citenamefont {{Vecchio}}}]{Taylor_2017}%
  \BibitemOpen
  \bibfield  {author} {\bibinfo {author} {\bibfnamefont {S.~R.}\ \bibnamefont
  {{Taylor}}}, \bibinfo {author} {\bibfnamefont {L.}~\bibnamefont {{Lentati}}},
  \bibinfo {author} {\bibfnamefont {S.}~\bibnamefont {{Babak}}}, \bibinfo
  {author} {\bibfnamefont {P.}~\bibnamefont {{Brem}}}, \bibinfo {author}
  {\bibfnamefont {J.~R.}\ \bibnamefont {{Gair}}}, \bibinfo {author}
  {\bibfnamefont {A.}~\bibnamefont {{Sesana}}}, \ and\ \bibinfo {author}
  {\bibfnamefont {A.}~\bibnamefont {{Vecchio}}},\ }\bibfield  {title} {\enquote
  {\bibinfo {title} {{All correlations must die: Assessing the significance of
  a stochastic gravitational-wave background in pulsar timing arrays}},}\
  }\href {\doibase 10.1103/PhysRevD.95.042002} {\bibfield  {journal} {\bibinfo
  {journal} {\prd}\ }\textbf {\bibinfo {volume} {95}},\ \bibinfo {eid} {042002}
  (\bibinfo {year} {2017})},\ \Eprint {http://arxiv.org/abs/1606.09180}
  {arXiv:1606.09180 [astro-ph.IM]} \BibitemShut {NoStop}%
\bibitem [{\citenamefont {Ellis}\ \emph {et~al.}(2020)\citenamefont {Ellis},
  \citenamefont {Vallisneri}, \citenamefont {Taylor},\ and\ \citenamefont
  {Baker}}]{enterprise}%
  \BibitemOpen
  \bibfield  {author} {\bibinfo {author} {\bibfnamefont {Justin~A.}\
  \bibnamefont {Ellis}}, \bibinfo {author} {\bibfnamefont {Michele}\
  \bibnamefont {Vallisneri}}, \bibinfo {author} {\bibfnamefont {Stephen~R.}\
  \bibnamefont {Taylor}}, \ and\ \bibinfo {author} {\bibfnamefont {Paul~T.}\
  \bibnamefont {Baker}},\ }\href {\doibase 10.5281/zenodo.4059815} {\enquote
  {\bibinfo {title} {Enterprise: Enhanced numerical toolbox enabling a robust
  pulsar inference suite},}\ }\bibinfo {howpublished} {Zenodo} (\bibinfo {year}
  {2020})\BibitemShut {NoStop}%
\bibitem [{\citenamefont {Taylor}\ \emph {et~al.}(2021)\citenamefont {Taylor},
  \citenamefont {Baker}, \citenamefont {Hazboun}, \citenamefont {Simon},\ and\
  \citenamefont {Vigeland}}]{enterprise_ext}%
  \BibitemOpen
  \bibfield  {author} {\bibinfo {author} {\bibfnamefont {Stephen~R.}\
  \bibnamefont {Taylor}}, \bibinfo {author} {\bibfnamefont {Paul~T.}\
  \bibnamefont {Baker}}, \bibinfo {author} {\bibfnamefont {Jeffrey~S.}\
  \bibnamefont {Hazboun}}, \bibinfo {author} {\bibfnamefont {Joseph}\
  \bibnamefont {Simon}}, \ and\ \bibinfo {author} {\bibfnamefont {Sarah~J.}\
  \bibnamefont {Vigeland}},\ }\href
  {https://github.com/nanograv/enterprise_extensions} {\enquote {\bibinfo
  {title} {enterprise{\_}extensions},}\ } (\bibinfo {year} {2021}),\ \bibinfo
  {note} {v2.3.3}\BibitemShut {NoStop}%
\bibitem [{\citenamefont {Ellis}\ and\ \citenamefont {van
  Haasteren}(2017)}]{justin_ellis_2017_1037579}%
  \BibitemOpen
  \bibfield  {author} {\bibinfo {author} {\bibfnamefont {Justin}\ \bibnamefont
  {Ellis}}\ and\ \bibinfo {author} {\bibfnamefont {Rutger}\ \bibnamefont {van
  Haasteren}},\ }\href {\doibase 10.5281/zenodo.1037579} {\enquote {\bibinfo
  {title} {jellis18/ptmcmcsampler: Official release},}\ } (\bibinfo {year}
  {2017})\BibitemShut {NoStop}%
\bibitem [{\citenamefont {Mitridate}(2023)}]{andrea_mitridate_2023}%
  \BibitemOpen
  \bibfield  {author} {\bibinfo {author} {\bibfnamefont {Andrea}\ \bibnamefont
  {Mitridate}},\ }\bibfield  {title} {\enquote {\bibinfo {title} {Ptarcade},}\
  }\href {\doibase 10.5281/zenodo.7876430} {\  (\bibinfo {year} {2023}),\
  10.5281/zenodo.7876430}\BibitemShut {NoStop}%
\bibitem [{\citenamefont {Mitridate}\ \emph {et~al.}(2023)\citenamefont
  {Mitridate}, \citenamefont {Wright}, \citenamefont {von Eckardstein},
  \citenamefont {Schr\"oder}, \citenamefont {Nay}, \citenamefont {Olum},
  \citenamefont {Schmitz},\ and\ \citenamefont {Trickle}}]{Mitridate:2023oar}%
  \BibitemOpen
  \bibfield  {author} {\bibinfo {author} {\bibfnamefont {Andrea}\ \bibnamefont
  {Mitridate}}, \bibinfo {author} {\bibfnamefont {David}\ \bibnamefont
  {Wright}}, \bibinfo {author} {\bibfnamefont {Richard}\ \bibnamefont {von
  Eckardstein}}, \bibinfo {author} {\bibfnamefont {Tobias}\ \bibnamefont
  {Schr\"oder}}, \bibinfo {author} {\bibfnamefont {Jonathan}\ \bibnamefont
  {Nay}}, \bibinfo {author} {\bibfnamefont {Ken}\ \bibnamefont {Olum}},
  \bibinfo {author} {\bibfnamefont {Kai}\ \bibnamefont {Schmitz}}, \ and\
  \bibinfo {author} {\bibfnamefont {Tanner}\ \bibnamefont {Trickle}},\
  }\bibfield  {title} {\enquote {\bibinfo {title} {{PTArcade}},}\ }\href@noop
  {} {\  (\bibinfo {year} {2023})},\ \Eprint {http://arxiv.org/abs/2306.16377}
  {arXiv:2306.16377 [hep-ph]} \BibitemShut {NoStop}%
\bibitem [{\citenamefont {Verbiest}\ \emph {et~al.}(2012)\citenamefont
  {Verbiest}, \citenamefont {Weisberg}, \citenamefont {Chael}, \citenamefont
  {Lee},\ and\ \citenamefont {Lorimer}}]{Verbiest_2012}%
  \BibitemOpen
  \bibfield  {author} {\bibinfo {author} {\bibfnamefont {J.~P.~W.}\
  \bibnamefont {Verbiest}}, \bibinfo {author} {\bibfnamefont {J.~M.}\
  \bibnamefont {Weisberg}}, \bibinfo {author} {\bibfnamefont {A.~A.}\
  \bibnamefont {Chael}}, \bibinfo {author} {\bibfnamefont {K.~J.}\ \bibnamefont
  {Lee}}, \ and\ \bibinfo {author} {\bibfnamefont {D.~R.}\ \bibnamefont
  {Lorimer}},\ }\bibfield  {title} {\enquote {\bibinfo {title} {On pulsar
  distance measurements and their uncertainties},}\ }\href {\doibase
  10.1088/0004-637X/755/1/39} {\bibfield  {journal} {\bibinfo  {journal} {The
  Astrophysical Journal}\ }\textbf {\bibinfo {volume} {755}},\ \bibinfo {pages}
  {39} (\bibinfo {year} {2012})}\BibitemShut {NoStop}%
\bibitem [{\citenamefont {chan Hwang}\ \emph {et~al.}(2024)\citenamefont {chan
  Hwang}, \citenamefont {Jeong}, \citenamefont {Noh},\ and\ \citenamefont
  {Smarra}}]{Hwang_2024}%
  \BibitemOpen
  \bibfield  {author} {\bibinfo {author} {\bibfnamefont {Jai}\ \bibnamefont
  {chan Hwang}}, \bibinfo {author} {\bibfnamefont {Donghui}\ \bibnamefont
  {Jeong}}, \bibinfo {author} {\bibfnamefont {Hyerim}\ \bibnamefont {Noh}}, \
  and\ \bibinfo {author} {\bibfnamefont {Clemente}\ \bibnamefont {Smarra}},\
  }\bibfield  {title} {\enquote {\bibinfo {title} {Pulsar timing array
  signature from oscillating metric perturbations due to ultra-light axion},}\
  }\href {\doibase 10.1088/1475-7516/2024/02/014} {\bibfield  {journal}
  {\bibinfo  {journal} {Journal of Cosmology and Astroparticle Physics}\
  }\textbf {\bibinfo {volume} {2024}},\ \bibinfo {pages} {014} (\bibinfo {year}
  {2024})}\BibitemShut {NoStop}%
\bibitem [{\citenamefont {{Hazboun}}\ \emph {et~al.}(2019)\citenamefont
  {{Hazboun}}, \citenamefont {{Romano}},\ and\ \citenamefont
  {{Smith}}}]{HazbounRomano2019}%
  \BibitemOpen
  \bibfield  {author} {\bibinfo {author} {\bibfnamefont {Jeffrey~S.}\
  \bibnamefont {{Hazboun}}}, \bibinfo {author} {\bibfnamefont {Joseph~D.}\
  \bibnamefont {{Romano}}}, \ and\ \bibinfo {author} {\bibfnamefont
  {Tristan~L.}\ \bibnamefont {{Smith}}},\ }\bibfield  {title} {\enquote
  {\bibinfo {title} {{Realistic sensitivity curves for pulsar timing
  arrays}},}\ }\href {\doibase 10.1103/PhysRevD.100.104028} {\bibfield
  {journal} {\bibinfo  {journal} {\prd}\ }\textbf {\bibinfo {volume} {100}},\
  \bibinfo {eid} {104028} (\bibinfo {year} {2019})},\ \Eprint
  {http://arxiv.org/abs/1907.04341} {arXiv:1907.04341 [gr-qc]} \BibitemShut
  {NoStop}%
\bibitem [{\citenamefont {{Blandford}}\ \emph {et~al.}(1984)\citenamefont
  {{Blandford}}, \citenamefont {{Narayan}},\ and\ \citenamefont
  {{Romani}}}]{BlandfordNarayan1984}%
  \BibitemOpen
  \bibfield  {author} {\bibinfo {author} {\bibfnamefont {R.}~\bibnamefont
  {{Blandford}}}, \bibinfo {author} {\bibfnamefont {R.}~\bibnamefont
  {{Narayan}}}, \ and\ \bibinfo {author} {\bibfnamefont {R.~W.}\ \bibnamefont
  {{Romani}}},\ }\bibfield  {title} {\enquote {\bibinfo {title} {{Arrival-time
  analysis for a millisecond pulsar.}}}\ }\href {\doibase 10.1007/BF02714466}
  {\bibfield  {journal} {\bibinfo  {journal} {Journal of Astrophysics and
  Astronomy}\ }\textbf {\bibinfo {volume} {5}},\ \bibinfo {pages} {369--388}
  (\bibinfo {year} {1984})}\BibitemShut {NoStop}%
\bibitem [{\citenamefont {Ramani}\ \emph {et~al.}(2020)\citenamefont {Ramani},
  \citenamefont {Trickle},\ and\ \citenamefont {Zurek}}]{Ramani_2020}%
  \BibitemOpen
  \bibfield  {author} {\bibinfo {author} {\bibfnamefont {Harikrishnan}\
  \bibnamefont {Ramani}}, \bibinfo {author} {\bibfnamefont {Tanner}\
  \bibnamefont {Trickle}}, \ and\ \bibinfo {author} {\bibfnamefont
  {Kathryn~M.}\ \bibnamefont {Zurek}},\ }\bibfield  {title} {\enquote {\bibinfo
  {title} {Observability of dark matter substructure with pulsar timing
  correlations},}\ }\href {\doibase 10.1088/1475-7516/2020/12/033} {\bibfield
  {journal} {\bibinfo  {journal} {Journal of Cosmology and Astroparticle
  Physics}\ }\textbf {\bibinfo {volume} {2020}},\ \bibinfo {pages} {033}
  (\bibinfo {year} {2020})}\BibitemShut {NoStop}%
\bibitem [{\citenamefont {Unal}\ \emph {et~al.}(2022)\citenamefont {Unal},
  \citenamefont {Urban},\ and\ \citenamefont {Kovetz}}]{Unal_2022}%
  \BibitemOpen
  \bibfield  {author} {\bibinfo {author} {\bibfnamefont {Caner}\ \bibnamefont
  {Unal}}, \bibinfo {author} {\bibfnamefont {Federico~R.}\ \bibnamefont
  {Urban}}, \ and\ \bibinfo {author} {\bibfnamefont {Ely~D.}\ \bibnamefont
  {Kovetz}},\ }\href {\doibase https://doi.org/10.48550/arXiv.2209.02741}
  {\enquote {\bibinfo {title} {Probing ultralight scalar, vector and tensor
  dark matter with pulsar timing arrays},}\ } (\bibinfo {year} {2022}),\
  \Eprint {http://arxiv.org/abs/2209.02741} {arXiv:2209.02741 [astro-ph.CO]}
  \BibitemShut {NoStop}%
\bibitem [{\citenamefont {Abney}\ \emph {et~al.}(1996)\citenamefont {Abney},
  \citenamefont {Epstein},\ and\ \citenamefont {Olinto}}]{Abney_1996}%
  \BibitemOpen
  \bibfield  {author} {\bibinfo {author} {\bibfnamefont {Mark}\ \bibnamefont
  {Abney}}, \bibinfo {author} {\bibfnamefont {Richard~I.}\ \bibnamefont
  {Epstein}}, \ and\ \bibinfo {author} {\bibfnamefont {Angela~V.}\ \bibnamefont
  {Olinto}},\ }\bibfield  {title} {\enquote {\bibinfo {title} {Observational
  constraints on the internal structure and dynamics of the vela pulsar},}\
  }\href {\doibase 10.1086/310171} {\bibfield  {journal} {\bibinfo  {journal}
  {The Astrophysical Journal}\ }\textbf {\bibinfo {volume} {466}},\ \bibinfo
  {pages} {L91} (\bibinfo {year} {1996})}\BibitemShut {NoStop}%
\bibitem [{\citenamefont {Mendes}(2015)}]{Mendes:2014ufa}%
  \BibitemOpen
  \bibfield  {author} {\bibinfo {author} {\bibfnamefont {Raissa F.~P.}\
  \bibnamefont {Mendes}},\ }\bibfield  {title} {\enquote {\bibinfo {title}
  {{Possibility of setting a new constraint to scalar-tensor theories}},}\
  }\href {\doibase 10.1103/PhysRevD.91.064024} {\bibfield  {journal} {\bibinfo
  {journal} {Phys. Rev. D}\ }\textbf {\bibinfo {volume} {91}},\ \bibinfo
  {pages} {064024} (\bibinfo {year} {2015})},\ \Eprint
  {http://arxiv.org/abs/1412.6789} {arXiv:1412.6789 [gr-qc]} \BibitemShut
  {NoStop}%
\bibitem [{\citenamefont {Anderson}\ and\ \citenamefont
  {Yunes}(2017)}]{Anderson:2017phb}%
  \BibitemOpen
  \bibfield  {author} {\bibinfo {author} {\bibfnamefont {David}\ \bibnamefont
  {Anderson}}\ and\ \bibinfo {author} {\bibfnamefont {Nicol\'as}\ \bibnamefont
  {Yunes}},\ }\bibfield  {title} {\enquote {\bibinfo {title} {{Solar System
  constraints on massless scalar-tensor gravity with positive coupling constant
  upon cosmological evolution of the scalar field}},}\ }\href {\doibase
  10.1103/PhysRevD.96.064037} {\bibfield  {journal} {\bibinfo  {journal} {Phys.
  Rev. D}\ }\textbf {\bibinfo {volume} {96}},\ \bibinfo {pages} {064037}
  (\bibinfo {year} {2017})},\ \Eprint {http://arxiv.org/abs/1705.06351}
  {arXiv:1705.06351 [gr-qc]} \BibitemShut {NoStop}%
\bibitem [{\citenamefont {Janssen}\ \emph {et~al.}(2015)\citenamefont {Janssen}
  \emph {et~al.}}]{Janssen:2014dka}%
  \BibitemOpen
  \bibfield  {author} {\bibinfo {author} {\bibfnamefont {Gemma}\ \bibnamefont
  {Janssen}} \emph {et~al.},\ }\bibfield  {title} {\enquote {\bibinfo {title}
  {{Gravitational wave astronomy with the SKA}},}\ }\href {\doibase
  10.22323/1.215.0037} {\bibfield  {journal} {\bibinfo  {journal} {PoS}\
  }\textbf {\bibinfo {volume} {AASKA14}},\ \bibinfo {pages} {037} (\bibinfo
  {year} {2015})},\ \Eprint {http://arxiv.org/abs/1501.00127} {arXiv:1501.00127
  [astro-ph.IM]} \BibitemShut {NoStop}%
\bibitem [{\citenamefont {Kim}(2023)}]{Kim:2023pkx}%
  \BibitemOpen
  \bibfield  {author} {\bibinfo {author} {\bibfnamefont {Hyungjin}\
  \bibnamefont {Kim}},\ }\bibfield  {title} {\enquote {\bibinfo {title}
  {{Gravitational interaction of ultralight dark matter with
  interferometers}},}\ }\href {\doibase 10.1088/1475-7516/2023/12/018}
  {\bibfield  {journal} {\bibinfo  {journal} {JCAP}\ }\textbf {\bibinfo
  {volume} {12}},\ \bibinfo {pages} {018} (\bibinfo {year} {2023})},\ \Eprint
  {http://arxiv.org/abs/2306.13348} {arXiv:2306.13348 [hep-ph]} \BibitemShut
  {NoStop}%
\bibitem [{\citenamefont {Akmal}\ \emph {et~al.}(1998)\citenamefont {Akmal},
  \citenamefont {Pandharipande},\ and\ \citenamefont {Ravenhall}}]{Akmal_1998}%
  \BibitemOpen
  \bibfield  {author} {\bibinfo {author} {\bibfnamefont {A.}~\bibnamefont
  {Akmal}}, \bibinfo {author} {\bibfnamefont {V.~R.}\ \bibnamefont
  {Pandharipande}}, \ and\ \bibinfo {author} {\bibfnamefont {D.~G.}\
  \bibnamefont {Ravenhall}},\ }\bibfield  {title} {\enquote {\bibinfo {title}
  {Equation of state of nucleon matter and neutron star structure},}\ }\href
  {\doibase 10.1103/PhysRevC.58.1804} {\bibfield  {journal} {\bibinfo
  {journal} {Phys. Rev. C}\ }\textbf {\bibinfo {volume} {58}},\ \bibinfo
  {pages} {1804--1828} (\bibinfo {year} {1998})}\BibitemShut {NoStop}%
\bibitem [{\citenamefont {Müther}\ \emph {et~al.}(1987)\citenamefont
  {Müther}, \citenamefont {Prakash},\ and\ \citenamefont
  {Ainsworth}}]{Muther_1987}%
  \BibitemOpen
  \bibfield  {author} {\bibinfo {author} {\bibfnamefont {H.}~\bibnamefont
  {Müther}}, \bibinfo {author} {\bibfnamefont {M.}~\bibnamefont {Prakash}}, \
  and\ \bibinfo {author} {\bibfnamefont {T.L.}\ \bibnamefont {Ainsworth}},\
  }\bibfield  {title} {\enquote {\bibinfo {title} {The nuclear symmetry energy
  in relativistic brueckner-hartree-fock calculations},}\ }\href {\doibase
  https://doi.org/10.1016/0370-2693(87)91611-X} {\bibfield  {journal} {\bibinfo
   {journal} {Physics Letters B}\ }\textbf {\bibinfo {volume} {199}},\ \bibinfo
  {pages} {469--474} (\bibinfo {year} {1987})}\BibitemShut {NoStop}%
\bibitem [{\citenamefont {Douchin}\ and\ \citenamefont
  {Haensel}(2001)}]{Douchin_2001}%
  \BibitemOpen
  \bibfield  {author} {\bibinfo {author} {\bibfnamefont {F.}~\bibnamefont
  {Douchin}}\ and\ \bibinfo {author} {\bibfnamefont {P.}~\bibnamefont
  {Haensel}},\ }\bibfield  {title} {\enquote {\bibinfo {title} {A unified
  equation of state of dense matter and neutron star structure},}\ }\href
  {\doibase 10.1051/0004-6361:20011402} {\bibfield  {journal} {\bibinfo
  {journal} {Astronomy \& Astrophysics}\ }\textbf {\bibinfo {volume} {380}},\
  \bibinfo {pages} {151–167} (\bibinfo {year} {2001})}\BibitemShut {NoStop}%
\bibitem [{\citenamefont {Miller}\ \emph {et~al.}(2019)\citenamefont {Miller},
  \citenamefont {Lamb}, \citenamefont {Dittmann}, \citenamefont {Bogdanov},
  \citenamefont {Arzoumanian}, \citenamefont {Gendreau}, \citenamefont
  {Guillot}, \citenamefont {Harding}, \citenamefont {Ho}, \citenamefont
  {Lattimer}, \citenamefont {Ludlam}, \citenamefont {Mahmoodifar},
  \citenamefont {Morsink}, \citenamefont {Ray}, \citenamefont {Strohmayer},
  \citenamefont {Wood}, \citenamefont {Enoto}, \citenamefont {Foster},
  \citenamefont {Okajima}, \citenamefont {Prigozhin},\ and\ \citenamefont
  {Soong}}]{Miller_2019}%
  \BibitemOpen
  \bibfield  {author} {\bibinfo {author} {\bibfnamefont {M.~C.}\ \bibnamefont
  {Miller}}, \bibinfo {author} {\bibfnamefont {F.~K.}\ \bibnamefont {Lamb}},
  \bibinfo {author} {\bibfnamefont {A.~J.}\ \bibnamefont {Dittmann}}, \bibinfo
  {author} {\bibfnamefont {S.}~\bibnamefont {Bogdanov}}, \bibinfo {author}
  {\bibfnamefont {Z.}~\bibnamefont {Arzoumanian}}, \bibinfo {author}
  {\bibfnamefont {K.~C.}\ \bibnamefont {Gendreau}}, \bibinfo {author}
  {\bibfnamefont {S.}~\bibnamefont {Guillot}}, \bibinfo {author} {\bibfnamefont
  {A.~K.}\ \bibnamefont {Harding}}, \bibinfo {author} {\bibfnamefont
  {W.~C.~G.}\ \bibnamefont {Ho}}, \bibinfo {author} {\bibfnamefont {J.~M.}\
  \bibnamefont {Lattimer}}, \bibinfo {author} {\bibfnamefont {R.~M.}\
  \bibnamefont {Ludlam}}, \bibinfo {author} {\bibfnamefont {S.}~\bibnamefont
  {Mahmoodifar}}, \bibinfo {author} {\bibfnamefont {S.~M.}\ \bibnamefont
  {Morsink}}, \bibinfo {author} {\bibfnamefont {P.~S.}\ \bibnamefont {Ray}},
  \bibinfo {author} {\bibfnamefont {T.~E.}\ \bibnamefont {Strohmayer}},
  \bibinfo {author} {\bibfnamefont {K.~S.}\ \bibnamefont {Wood}}, \bibinfo
  {author} {\bibfnamefont {T.}~\bibnamefont {Enoto}}, \bibinfo {author}
  {\bibfnamefont {R.}~\bibnamefont {Foster}}, \bibinfo {author} {\bibfnamefont
  {T.}~\bibnamefont {Okajima}}, \bibinfo {author} {\bibfnamefont
  {G.}~\bibnamefont {Prigozhin}}, \ and\ \bibinfo {author} {\bibfnamefont
  {Y.}~\bibnamefont {Soong}},\ }\bibfield  {title} {\enquote {\bibinfo {title}
  {Psr j0030+0451 mass and radius from nicer data and implications for the
  properties of neutron star matter},}\ }\href {\doibase
  10.3847/2041-8213/ab50c5} {\bibfield  {journal} {\bibinfo  {journal} {The
  Astrophysical Journal Letters}\ }\textbf {\bibinfo {volume} {887}},\ \bibinfo
  {pages} {L24} (\bibinfo {year} {2019})}\BibitemShut {NoStop}%
\bibitem [{\citenamefont {Mata~Sánchez}\ \emph {et~al.}(2020)\citenamefont
  {Mata~Sánchez}, \citenamefont {Istrate}, \citenamefont {van Kerkwijk},
  \citenamefont {Breton},\ and\ \citenamefont {Kaplan}}]{Sanchez_2020}%
  \BibitemOpen
  \bibfield  {author} {\bibinfo {author} {\bibfnamefont {D}~\bibnamefont
  {Mata~Sánchez}}, \bibinfo {author} {\bibfnamefont {A~G}\ \bibnamefont
  {Istrate}}, \bibinfo {author} {\bibfnamefont {M~H}\ \bibnamefont {van
  Kerkwijk}}, \bibinfo {author} {\bibfnamefont {R~P}\ \bibnamefont {Breton}}, \
  and\ \bibinfo {author} {\bibfnamefont {D~L}\ \bibnamefont {Kaplan}},\
  }\bibfield  {title} {\enquote {\bibinfo {title} {{PSR J1012+5307: a
  millisecond pulsar with an extremely low-mass white dwarf companion}},}\
  }\href {\doibase 10.1093/mnras/staa983} {\bibfield  {journal} {\bibinfo
  {journal} {Monthly Notices of the Royal Astronomical Society}\ }\textbf
  {\bibinfo {volume} {494}},\ \bibinfo {pages} {4031--4042} (\bibinfo {year}
  {2020})},\ \Eprint
  {http://arxiv.org/abs/https://academic.oup.com/mnras/article-pdf/494/3/4031/33149104/staa983.pdf}
  {https://academic.oup.com/mnras/article-pdf/494/3/4031/33149104/staa983.pdf}
  \BibitemShut {NoStop}%
\bibitem [{\citenamefont {{Chen, W.-C.}}\ and\ \citenamefont {{Panei, J.
  A.}}(2011)}]{Chen_2011}%
  \BibitemOpen
  \bibfield  {author} {\bibinfo {author} {\bibnamefont {{Chen, W.-C.}}}\ and\
  \bibinfo {author} {\bibnamefont {{Panei, J. A.}}},\ }\bibfield  {title}
  {\enquote {\bibinfo {title} {The progenitor of binary millisecond radio
  pulsar psr j1713+0747},}\ }\href {\doibase 10.1051/0004-6361/201014833}
  {\bibfield  {journal} {\bibinfo  {journal} {A\& A}\ }\textbf {\bibinfo
  {volume} {527}},\ \bibinfo {pages} {A128} (\bibinfo {year}
  {2011})}\BibitemShut {NoStop}%
\bibitem [{\citenamefont {Antoniadis}\ \emph {et~al.}(2012)\citenamefont
  {Antoniadis}, \citenamefont {van Kerkwijk}, \citenamefont {Koester},
  \citenamefont {Freire}, \citenamefont {Wex}, \citenamefont {Tauris},
  \citenamefont {Kramer},\ and\ \citenamefont {Bassa}}]{Antoniadis_2012}%
  \BibitemOpen
  \bibfield  {author} {\bibinfo {author} {\bibfnamefont {J.}~\bibnamefont
  {Antoniadis}}, \bibinfo {author} {\bibfnamefont {M.~H.}\ \bibnamefont {van
  Kerkwijk}}, \bibinfo {author} {\bibfnamefont {D.}~\bibnamefont {Koester}},
  \bibinfo {author} {\bibfnamefont {P.~C.~C.}\ \bibnamefont {Freire}}, \bibinfo
  {author} {\bibfnamefont {N.}~\bibnamefont {Wex}}, \bibinfo {author}
  {\bibfnamefont {T.~M.}\ \bibnamefont {Tauris}}, \bibinfo {author}
  {\bibfnamefont {M.}~\bibnamefont {Kramer}}, \ and\ \bibinfo {author}
  {\bibfnamefont {C.~G.}\ \bibnamefont {Bassa}},\ }\bibfield  {title} {\enquote
  {\bibinfo {title} {{The relativistic pulsar–white dwarf binary PSR
  J1738+0333 – I. Mass determination and evolutionary history}},}\ }\href
  {\doibase 10.1111/j.1365-2966.2012.21124.x} {\bibfield  {journal} {\bibinfo
  {journal} {Monthly Notices of the Royal Astronomical Society}\ }\textbf
  {\bibinfo {volume} {423}},\ \bibinfo {pages} {3316--3327} (\bibinfo {year}
  {2012})},\ \Eprint
  {http://arxiv.org/abs/https://academic.oup.com/mnras/article-pdf/423/4/3316/4895633/mnras0423-3316.pdf}
  {https://academic.oup.com/mnras/article-pdf/423/4/3316/4895633/mnras0423-3316.pdf}
  \BibitemShut {NoStop}%
\bibitem [{\citenamefont {Jacoby}\ \emph {et~al.}(2005)\citenamefont {Jacoby},
  \citenamefont {Hotan}, \citenamefont {Bailes}, \citenamefont {Ord},\ and\
  \citenamefont {Kulkarni}}]{Jacoby_2005}%
  \BibitemOpen
  \bibfield  {author} {\bibinfo {author} {\bibfnamefont {B.~A.}\ \bibnamefont
  {Jacoby}}, \bibinfo {author} {\bibfnamefont {A.}~\bibnamefont {Hotan}},
  \bibinfo {author} {\bibfnamefont {M.}~\bibnamefont {Bailes}}, \bibinfo
  {author} {\bibfnamefont {S.}~\bibnamefont {Ord}}, \ and\ \bibinfo {author}
  {\bibfnamefont {S.~R.}\ \bibnamefont {Kulkarni}},\ }\bibfield  {title}
  {\enquote {\bibinfo {title} {The mass of a millisecond pulsar},}\ }\href
  {\doibase 10.1086/449311} {\bibfield  {journal} {\bibinfo  {journal} {The
  Astrophysical Journal}\ }\textbf {\bibinfo {volume} {629}},\ \bibinfo {pages}
  {L113} (\bibinfo {year} {2005})}\BibitemShut {NoStop}%
\end{thebibliography}%

\end{document}